\documentclass[%
 aip,
 pof,
 reprint,%
]{revtex4-1}

\usepackage{amsmath}
\usepackage{graphicx}
\usepackage{dcolumn}
\usepackage{bm}

\usepackage[utf8]{inputenc}
\usepackage[T1]{fontenc}
\usepackage{mathptmx}
\usepackage{etoolbox}
\usepackage{xcolor}
\usepackage{ulem}

\makeatletter
\def\@email#1#2{%
 \endgroup
 \patchcmd{\titleblock@produce}
  {\frontmatter@RRAPformat}
  {\frontmatter@RRAPformat{\produce@RRAP{*#1\href{mailto:#2}{#2}}}\frontmatter@RRAPformat}
  {}{}
}%
\makeatother
\begin{document}

\preprint{AIP/123-QED}

\title{The effect of tilt on turbulent thermal convection for a heated soap bubble} 



\author{Xiao-Qiu He}

\author{Yong-Liang Xiong}%
\email{xylcfd@hust.edu.cn}
\altaffiliation[Also at ]{Hubei Key Laboratory of Engineering Structural Analysis and Safety Assessment, Wuhan 430074, PR China}
\affiliation{School of Aerospace Engineering, Huazhong University of Science and Technology, Wuhan 430074, PR China}%

\author{Andrew D. Bragg}
\affiliation{Department of Civil and Environmental Engineering, Duke University, Durham, NC 27708, USA}%

\author{Patrick Fischer}
\affiliation{Institut de Math{\'e}matiques de Bordeaux (IMB), Universit{\'e} de Bordeaux, CNRS UMR 5251, France}

\author{Hamid Kellay}
\affiliation{Laboratoire Ondes et Mati\'ere d'Aquitaine (LOMA), Universit\'e de Bordeaux, France}

\date{\today}

\begin{abstract}
    We use direct numerical simulation (DNS) to explore
    the effect of tilt on two-dimensional turbulent thermal convection on a half-soap bubble that is heated at its equator.
    In the DNS, the bubble is tilted by an angle $\delta\in[0^{\circ},90^{\circ}]$,
    the Rayleigh number is varied between $Ra\in[3\times10^6, 3\times10^9]$, and the Prandlt number is fixed at $Pr=7$. 
    The DNS reveals two qualitatively different flow regimes: 
    the dynamic plume regime (DPR) and the stable plume regime (SPR).
    In the DPR, small dynamic plumes constantly emerge from random locations on the equator and dissipate on the bubble.
    In the SPR, the flow is dominated by a single large and stable plume rising from the lower edge of the bubble.
    The scaling behaviour of the Nusselt number $Nu$ and 
    Reynolds number $Re$ are different in these two regimes, 
    with $Nu\propto Ra^{0.3}$ for the DPR and $Nu\propto Ra^{0.24}$ for the SPR.
    Concerning $Re$, the scaling in the DPR lies between $Re\propto Ra^{0.48}$ and $Re\propto Ra^{0.53}$ depending on $Ra$ and $\delta$,
    while in the SPR, the scaling lies between $Re\propto Ra^{0.44}$ and $Re\propto Ra^{0.45}$ depending on $\delta$.
    The turbulent thermal and kinetic energy dissipation rates 
    ($\epsilon_{T^{\prime}}$ and $\epsilon_{u^{\prime}}$, respectively) are also very different in the DPR and SPR.
    The probability density functions (PDF) of the normalized $\log\epsilon_{T^{\prime}}$ and $\log\epsilon_{u^{\prime}}$ are close to a Gaussian PDF
    for small fluctuations, but deviate considerably from a Gaussian at large fluctuations in the DPR.
    In the SPR, the PDFs of normalized $\log\epsilon_{T^{\prime}}$ and $\log\epsilon_{u^{\prime}}$ deviate considerably from a Gaussian PDF even for small values.
    The globally averaged thermal energy dissipation rate due to the mean temperature field was
    shown to exhibit the scaling $\langle\epsilon_{\langle T\rangle}\rangle_{\mathcal{B}}\propto Ra^{-0.23}$ in the DPR,
    and $\langle\epsilon_{\langle T\rangle}\rangle_{\mathcal{B}}\propto Ra^{-0.28}$ in the SPR.
    The globally averaged kinetic energy dissipation rate due to the mean velocity field
    is shown to exhibit the scaling $\langle\epsilon_{\langle u\rangle}\rangle_{\mathcal{B}}\propto Ra^{-0.47}$ in the DPR (the exponent reduces from $0.47$ to $0.43$ as $\delta$ is increased up to $30^\circ$).
    In the SPR, the behavior changes considerably to $\langle\epsilon_{\langle u\rangle}\rangle_{\mathcal{B}}\propto Ra^{-0.27}$.
    For the turbulent dissipation rates, the results indicate the scaling $\langle\epsilon_{T^{\prime}}\rangle_{\mathcal{B}}\propto Ra^{-0.18}$ and 
    $\langle\epsilon_{u^{\prime}}\rangle_{\mathcal{B}}\propto Ra^{-0.29}$ in the DPR.
    However, the dependencies of $\langle\epsilon_{T^{\prime}}\rangle_{\mathcal{B}}$ and $\langle\epsilon_{u^{\prime}}\rangle_{\mathcal{B}}$
    on $Ra$ cannot be described by power-laws in the SPR.

\end{abstract}

\pacs{}

\maketitle 

\section{Introduction}
Turbulent thermal convection is ubiquitous in nature and plays a significant role in large scale flows on the Earth,
such as the cyclones in the atmosphere and the circulation of the deep oceans\citep{LohseXia2010}. Convective flows are also vital for a great number 
of industrial applications, for example cooling systems on chip-boards\cite{Guo2022POF}. The fluid motion in turbulent thermal convection is driven by buoyancy which arises due to temperature gradients imposed by boundary conditions\citep{AhlersGrossmann-32}.
In these flows, the buoyancy force typically injects energy into the large flow structures, and this energy is then (on average) transferred to smaller
scales by the energy cascade, and is finally dissipated at the smallest scales \cite{XuXi2019POF}. 
The rate of energy dissipation regulates both the global energy balances and the local fluctuations of flow quantities\cite{HeChingTong2011POF,Petschel2015JFM,Petschel2013PRL},
and studying its behavior provides insights into the fundamental properties of turbulent convective flows \cite{ZhangZhouSun2017JFM,AnnaRonald2021POF}.

\subsection{Dissipation in RBC}
Rayleigh-B\'enard convection (RBC) is the canonical model system for fundamental studies of
turbulent thermal convection \citep{AhlersGrossmann-32,Stevens2013EJMB}, and the physical mechanisms and flow properties of RBC have been studied extensively in recent decades \cite{ZhangZhouSun2017JFM}.
For a given flow configuration, 
the dynamics of RBC is controlled by two non-dimensional parameters, 
the Rayleigh number $Ra$ and the Prandlt number $Pr$.
$Ra$ is defined as the non-dimensional heating temperature as:
\begin{equation}
	Ra = \frac{{g}\beta {T}_c {H}^3}{{\nu}{\kappa}},
\end{equation}
where ${T}_c$ is the difference between the temperature at the upper and lower boundaries,
${H}$ is the distance between the upper and lower boundaries,
${g}$ denotes the norm of the gravity acceleration, 
${\beta}$ the coefficient of thermal expansion, 
${\nu}$ the kinetic viscosity,  
and ${\kappa}$ the thermal diffusivity.
$Pr$ is defined as the ratio of the momentum diffusivity to the thermal diffusivity as:
\begin{equation}
	Pr = \frac{{\nu}}{{\kappa}}.
\end{equation}
The resulting flow properties of RBC in terms of the global heat flux and nature of the fluid motion are measured by the Nusselt number $Nu$ and Reynold number $Re$, respectively.
$Nu$ is the non-dimensional heat flux defined as:
\begin{equation}
	Nu = \frac{Q}{\lambda\frac{{T}_c}{{H}}},
\end{equation}
where $\lambda$ denotes the thermal conductivity of the fluid and $Q$ is the heat flux through the fluid.
$Re$ is defined as:
\begin{equation}
	Re = \frac{{u}_c {H}}{{\nu}},
\end{equation}
where ${u}_c$ denotes a characteristic velocity of the flow, e.g. the root-mean-square velocity.
A crucial open topic in RBC is to understand how the control parameters $Ra$ and $Pr$ determine the 
response parameters $Nu$ and $Re$, which are emergent properties of the RBC flow.
Central to understanding this is to understand the behavior of the thermal and kinetic dissipation rates which are defined as
\begin{equation}
	\epsilon_{{T}} = {\kappa}\|\boldsymbol{\nabla}{T}\|^2,
\end{equation}
and
\begin{equation}
	\epsilon_{{u}} = 
    \frac{1}{2}{\nu}\|{\left( \boldsymbol{\nabla}{\boldsymbol{u}} +\boldsymbol{\nabla}{\boldsymbol{u}}^\top \right)}\|^2,
\end{equation}
where ${T}$ is the temperature field and ${\boldsymbol{u}}$ the fluid velocity.

In RBC, the following exact relations can be derived\cite{Siggia1990PRA,Siggia1994ARFM}:
\begin{equation}
	\langle\epsilon_{{T}}\rangle_{V} = {\kappa}\frac{({T}_c)^2}{{H}^2}Nu,
	\label{eq:exactEpsilonT}
\end{equation}
\begin{equation}
	\langle\epsilon_{{u}}\rangle_{V} = \frac{{\nu}^{3}}{{H}^4}\left(Nu-1\right)\frac{Ra}{Pr^{2}},
	\label{eq:exactEpsilonU}
\end{equation}
where the operator $\langle\cdot\rangle_{V}$ denotes a volume average.
These exact relations connect the controlling parameters $Ra, Pr$ to the response parameters $Nu$ via $\epsilon_{{T}}$ and $\epsilon_{{u}}$. 
In addition, they are also foundational to the famous Grossmann-Lohse(GL) theory\cite{GrossmannLohse-135,GrossmannLohse2001PRL,GrossmannLohse2002PhysRevE,GrossmannLohse2004POF}.
In the original scenario proposed by the GL theory, $\langle\epsilon_{{T}}\rangle_{V}$ and $\langle\epsilon_{{u}}\rangle_{V}$ are decomposed into the contributions 
due to the boundaries and bulk respectively\cite{GrossmannLohse-135,GrossmannLohse2001PRL,GrossmannLohse2002PhysRevE}.
Later, Grossmann and Lohse extend the physical pictures of heat transport and include the contribution of plumes\cite{GrossmannLohse2004POF}.
In macroscopic perspective, the GL theory successfully reveals the mathematical form of $Nu$ and $Re$ as functions of $Ra$ and $Pr$.
However, understanding the microscopic physical mechanisms of heat transport and flow dynamics still needs the insights offered 
by a deep and detailed investigation of $\epsilon_{T}$ and $\epsilon_{u}$\cite{HeTong2009PRE,ZhangZhouSun2017JFM,XuXi2019POF}.

Measuring $\epsilon_{{u}}$ requires the simultaneous measurement of all 5 (for incompressible flow) components of strain-rate tensor,
and $\epsilon_{{T}}$ requires the simultaneous measurements all 3 components of $\boldsymbol{\nabla}T$.
Therefore, it is very challenging to measure $\epsilon_{{u}}$ and $\epsilon_{{T}}$ in experiments, and only recently has
this been done.
For $1\times10^9\leq Ra\leq1\times10^{10}$, 
He et al.\cite{HeTongXia2007PRL} achieved the first successful experimental measurement of $\epsilon_{{T}}$ 
using a local temperature gradient probe in a cylinderical convection cell.
In their results, the spatial and temporal average thermal dissipation rate
$\langle\epsilon_{{T}}\rangle$ is decomposed into two components $\epsilon_{\langle T\rangle}={\kappa}\|\boldsymbol{\nabla}\langle{T}\rangle\|^2$ 
and $\langle\epsilon_{{T}^{\prime}}\rangle=\langle\epsilon_{{T}}\rangle-\epsilon_{\langle{T}\rangle}$ 
where $\langle\cdot\rangle$ denotes a spatially local temporal average\citep{HeTongXia2007PRL}.
Their results showed that $\langle\epsilon_{{T}^{\prime}}\rangle$ is dominant in the central region of the flow and therefore that the plumes makes a crucial contribution to the total dissipation in this region\citep{HeTongXia2007PRL}.
By contrast, the contribution from $\epsilon_{\langle{T}\rangle}$ is dominant in the thermal boundary layers \citep{HeTongXia2007PRL}.
The probability density function (PDF) of $\epsilon_{{T}^{\prime}}$ measured in the experiments\cite{HeTongXia2007PRL,HeTong2009PRE} are well described
by stretched exponential functions. In the central region of the flow or near the side walls, the PDFs of normalized $\log\epsilon_{{T}^{\prime}}$ are well described by a Gaussian PDF 
for relatively small values of the normalized variable\citep{HeTong2009PRE}.

Concerning $\epsilon_{{u}}$, Ni et al.\cite{NiHuangXia2011PRL} measured its temporal and volume average in the center of a convection cell using particle image velocimetry (PIV). 
Their results validate the crucial assumption made by the GL theory, namely that the flow volume averaged dissipation is dominated by the contribution from the the boundary layers\cite{NiHuangXia2011PRL}.
Recently, Chilla et al.\cite{Chilla2022EPL} utilized correlation image velocimetry (CIV) technology and Fluorinert $FC770$ as the working fluid 
in order to measure $\epsilon_{{u}}$ with $Ra$ up to $2\times10^{12}$.
They found that power-law dependence of $\epsilon_{{u}}$ on $Re$ is $\epsilon_{{u}}\propto Re^{\frac{5}{2}}$ for laminar convection 
and $\epsilon_{{u}}\propto Re^{3}$ for turbulent convection\cite{Chilla2022EPL}.

Verzicco et al.\cite{VerziccoCamussi2003JFM,Verzicco2003EPJB} calculated $\epsilon_{{T}}$ and $\epsilon_{{u}}$ from the three dimensional temperature
and velocity field obtained from DNS and also found that the dominant contribution to the globally average dissipation comes from the boundary layer, confirming the hypothesis of the GL theory\cite{VerziccoCamussi2003JFM,Verzicco2003EPJB}.
Shishkina et al.\cite{ShishkinaWagner2006JFM,ShishkinaWagner2007POF,ShishkinaWagner2008JFM} used $\epsilon_{{T}}$ to develop a method 
for the extraction of plumes from the bulk flow in RBC using DNS data.
For $10^7\leq Ra\leq10^9$ and $Pr=0.7$, Emran and Schumacher\cite{EmranSchumacher2008JFM} found that 
the PDF of normalized $\log\epsilon_T$ deviates from a Guassian distribution at large values of the normalized variable. 
It was also shown that the PDFs of $\epsilon_{{T}}$ can be fitted by stretched exponential functions\cite{EmranSchumacher2008JFM}.
Zhang et al.\cite{ZhangZhouSun2017JFM} systematically studied the statistics of $\epsilon_{{T}}$ and $\epsilon_{{u}}$ 
in a two-dimensional square convection cell for $10^6\leq Ra\leq 10^{10}$.
They obtained PDFs for $\epsilon_{{T}}$ and $\epsilon_{{u}}$ that were very similar to those of Emran and Schumacher\cite{EmranSchumacher2008JFM}.
However, they also found deviations from the GL theory with respect to contributions to the dissipation from the central flow region of RBC\citep{ZhangZhouSun2017JFM}.
Xu et al.\cite{XuXi2019POF} studied the statistics of $\epsilon_{{T}}$ for RBC with very low $Pr=0.025$ and obtained similar results to those of 
Zhang et al.\cite{ZhangZhouSun2017JFM} and Emran\cite{EmranSchumacher2008JFM}, suggesting that at least some of the normalized statistical properties of $\epsilon_{{T}}$ are independent of $Pr$ over
the range spanned by these studies. In addition, Shashwat et al.\cite{Shashwat2019POF} studied the scaling of the thermal dissipation, both 
averaged only inside the boundary layer $\langle\epsilon_{{T}}\rangle_{BL}$ and only inside the bulk $\langle\epsilon_{{T}}\rangle_{BK}$,
as a function of $Ra$ and for a a wide range of $Pr$.
They again found that $\langle\epsilon_{{T}}\rangle_{BL}$ is much larger than $\langle\epsilon_{{T}}\rangle_{BK}$\cite{Shashwat2019POF}, in line with previous studies and the GL theory\cite{VerziccoCamussi2003JFM,Verzicco2003EPJB}.
They also found that a stretched exponential function accurately describes the PDFs of $\epsilon_{{T}}$ measured in both the boundary layer and the bulk\cite{Shashwat2019POF}.

\subsection{Tilted RBC}
Since geopotential lines rarely coincide with the surface of the earth\citep{Hideo1984}, 
most buoyancy-driven flows in nature are subject to a non-vertical mean temperature gradient \cite{bejan2013convection}.
Examples of where this is important are for mantle plumes\citep{Taylor1995,Wortel2000,Condie2008} and atmospheric circulations\citep{Emanuel1994}.
It can also be of importance in engineering applications\citep{Madanan2019}.
The impact of this non-vertical mean temperature gradient can be explored in a canonical setting by inclining the RBC flow by an angle $\delta$ that is varied, so that the mean temperature gradient is misaligned with gravity (with $\delta=0^\circ$ denoting the non-tilted case). 

Ahlers et al.\cite{Ahlers2006} showed in experiments that the large scale circulations (LSC) are accelerated when $\delta$ is small but finite,
and in a rectangular cell, the shape of the LSC is modified due to the increase of $\delta$ in DNS\citep{GuoZhou-127}.
In the numerical study of Wang et al\cite{Wang2018,WangXia-126}, the LSC transform from the double rolls to a single roll when $\delta$ is increased, and increasing
$\delta$ can also lead to the reversal of the LSC\citep{WangXia-126}.
$Re$ and $Nu$ are also influenced by $\delta$.
In DNS, Guo et al.\cite{GuoZhou-127} found that as $\delta$ is increased from $0^\circ$ to $90^\circ$, $Re$ first increases then drops after reaching a maximum, 
while $Nu$ decreases monotonically, with a maximum decrease of $18\%$. 
By means of experiments, Wei et al.\cite{WeiXia-172} measured $Re$ as a function of $Ra$ for $0.5^{\circ}\leq\delta\leq 3.4^{\circ}$ and found the
scaling $Re\sim Ra^{0.43}$ (or $Re\sim Ra^{0.55}$ depending on the definition of $Re$) independent of $\delta$ for these small inclination angles.
The experimental study of Ahlers et al.\citep{Ahlers2006} showed that for small $\delta$, $Nu\sim Re^{\frac{1}{3}}$, with only slight variations with $\delta$.
By means of DNS, Shishkina et al.\cite{Shishkina2016} and Zwirner et al.\cite{Zwirner2018} considered a wide range of $Ra$, $Pr$ and $\delta$.
The results demonstrated that $Nu$ depends on $\delta$ in a complicated, non-monotonic way when $\delta$ is varied over a large range\cite{Shishkina2016,Zwirner2018}.
Recently, with help of both DNS and experiments, Zhang et al.\cite{Zhang2021} studied tilted RBC systematically and 
to elucidate how the misalignment of the mean temperature gradient with gravity influences the flow.
In their study, $Ra$ is decomposed into a vertical Rayleigh number $Ra_V$ and a horizontal Rayleigh number $Ra_H$, and $Nu$ is 
also decomposed into a vertical Nusselt number $Nu_V$ and a horizontal Nusselt number $Nu_H$.
By taking the effect of the misalignment into consideration, 
Zhang et al.\cite{Zhang2021} extended the classical GL theory and predicted $Nu_V$ as a function of $Re$, $Pr$ and $\delta$.

\subsection{Soap bubble}

While the classical RBC setup has been the subject of intense investigation, in many naturally occurring contexts the thermal convection takes place in curved or spherical geometries. Understanding the 
influence of this curved geometry on the thermal convection is therefore of great importance for geophysics and astrophysics\citep{kellay2017hydrodynamics}.
A canonical setup for exploring this is to consider turbulent thermal convection on a half soap bubble that is heated at its equator, and this was first
studied experimentally by Kellay\cite{kellay2017hydrodynamics}. Since the thickness of the soap film is negligible compared to the radius of the bubble, 
the turbulent flow on the bubble corresponds to quasi two-dimensional turbulence on a hemispherical surface\cite{Kellay_2002,kellay2017hydrodynamics}. The
experiments revealed that on the bubble there form large, persistent and isolated vortices\cite{Seychelles2008,Seychelles2010,Meuel2018} which are similar to typhoons or cyclones that occur in the atmosphere\cite{Meuel2012,Meuel2013}.
Indeed, several studies revealed important quantitative similarities of the trajectories and intensities of these vorticites on the bubble with those of cyclones in nature \cite{Seychelles2008,Seychelles2010,Meuel2012,Meuel2013,Meuel2018}.
In fact, the trajectories of the cyclones are successfully predicted by a method first developed to describe that of the vortex on the bubble\cite{Meuel2012}.
DNS of the half soap bubble were first performed by Xiong et al.\cite{Xiong2012}, and Bruneau et al.\cite{Bruneau2018} used the DNS to show that 
the scaling behaviour of $Re$ and $Nu$ are very similar to that in standard RBC, with the DNS yielding $Nu\sim Ra^{0.30}$ and $Nu\sim Ra^{0.49}$. 
He et al.\cite{He2021} further extended the model to investigate the impact of bubble rotation on the convective flow and showed that $Nu$ is not effected by even strong rotation,
while $Re$ decreases considerably with increasing rotation \citep{He2021}. 

An important open issue is how the convective flow on the soap bubble is affected by inclining the bubble, analogous to the tilted RBC discussed earlier. The impact of the tilting could be different from that for standard RBC because the curved surface on the bubble leads to a spatial dependence of the alignment of gravity with the flow direction.

\subsection{Organization of the paper}

The aim of our study is to fill this gap by investigating the effect of tilt on the thermal convection of the soap bubble flow.
Special focus on the thermal and kinetic dissipation fields due to the key role these play in governing the properties of the convective flow.
In section 2 we introduce the governing equations and the energy budgets. In section 3, the results of the DNS are presented and discussed. Conclusions are then drawn in section 4.

\section{Method}
\subsection{Governing Equations}
\begin{figure}
	\centering
    \includegraphics[width = 0.6\textwidth]{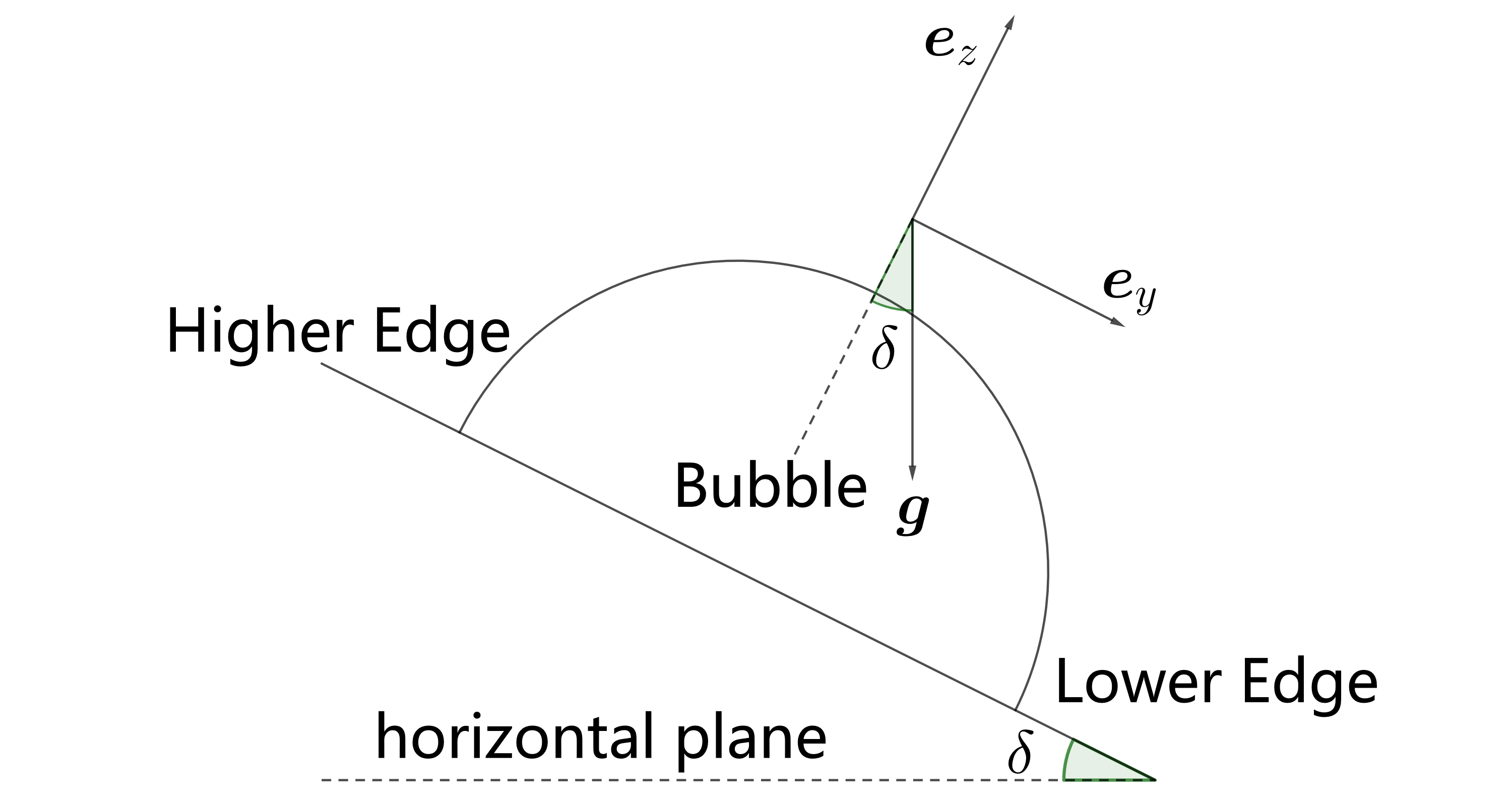}
    \caption{Illustration of the half-soap bubble and the Cartesian coordinate system used. }
  \label{fig:0_bubbleCoordinateSystem}
\end{figure}
In our study, a half-soap bubble of radius ${R}$ is mounted on a base plane which keeps the equator of the bubble at constant temperature ${T}_0$, as shown in figure \ref{fig:0_bubbleCoordinateSystem}.
A three dimensional Cartesian coordinate system is used whose origin is located at the center of the bubble, and is defined by the unit vector $\boldsymbol{e}_y$ that
is parallel to the base plane, $\boldsymbol{e}_z$ that is orthogonal to the base plane, and $\boldsymbol{e}_x$ is defined such that $\boldsymbol{e}_z=\boldsymbol{e}_x\times \boldsymbol{e}_y$.
In this system, an arbitrary vector ${\boldsymbol{a}}$ is represented in terms of its Cartesian components as $\boldsymbol{a}=a_x\boldsymbol{e}_x+a_y\boldsymbol{e}_y+a_z\boldsymbol{e}_z$. 
The base plane is tilted by an angle $\delta\in[0,\pi/2]$, and since the gravitational acceleration vector $\boldsymbol{g}$ is fixed we have $\boldsymbol{e}_g\equiv\boldsymbol{g}/\|\boldsymbol{g}\|= (\boldsymbol{e}_y\sin\delta-\boldsymbol{e}_z\cos\delta)$.

The thickness of the soap film is negligible compared to the radius of the bubble,
and hence the bubble may be approximated as a two-dimensional hemispherical surface.
The bubble is heated at the equator and cools through contact with the surrounding colder air.
The variation of the mass density ${\rho}$ due to the temperature ${T}$ is assumed under
the Boussinesq approximation to be ${\rho}={\rho}_0\left[1-{\beta}({T}-{T}_0)\right]$, where ${\beta}$ is the thermal expansion coefficient,
${T}_0$ is the reference temperature on the equator, ${\rho}_0$ is the massive density of the fluid when $T=T_{0}$.
For this model, the governing equations for the flow are given by the Boussinesq-Navier-Stokes system:
\begin{equation}
	\boldsymbol{\nabla}\cdot{\boldsymbol{u}}=0,
\end{equation}
\begin{equation}
	\partial_t{\boldsymbol{u}} + ({\boldsymbol{u}}\cdot\nabla){\boldsymbol{u}}
	= -\frac{1}{{\rho}_0}\boldsymbol{\nabla}{p} + {\nu}\nabla\cdot\left( \boldsymbol{\nabla}{\boldsymbol{u}} +\boldsymbol{\nabla}{\boldsymbol{u}}^\top \right) -{\beta}({T}-{T}_0){\boldsymbol{g}} - {F}{\boldsymbol{u}},
	\label{eq:momentumEq}
\end{equation}
\begin{equation}
	\partial_t{T} +({\boldsymbol{u}}\cdot\nabla){T}
     = {\kappa}\nabla^2{T} - {S}{T},
\end{equation}
where ${\boldsymbol{u}}$ is the fluid velocity and ${p}$ is the pressure field that includes the hydrostatic contribution.

The terms involving ${S}$ and ${F}$ are the external cooling and friction terms, respectively, which model the heat exchange and friction due to the
the cold air surrounding the bubble. These terms are required in order for the DNS to attain non-trivial steady-state regimes, and are discussed in details in the study of Bruneau et al.\cite{Bruneau2018} and He et al.\cite{He2021}. Analogous terms are also routinely used 
when performing DNS of two-dimension turbulence on flat geometries \cite{Boffetta2012}.

The equations can be non-dimensonlized using the radius of the bubble ${R}$, the initial temperature difference between the equator and the North pole $\Delta{T}$, and the free fall velocity ${u}_c=\sqrt{g\beta\Delta{T} {R}}$ leading to (for notational simplicity, the non-dimensional independent variables are not indicated by a ``hat'' symbol, and all variables are to be understood as non-dimensional hereafter)
\begin{equation}
	\boldsymbol{\nabla}\cdot\boldsymbol{u} = 0,
\end{equation}
\begin{equation}
	\partial_t\boldsymbol{u} + (\boldsymbol{u}\cdot\nabla)\boldsymbol{u}
    =-\boldsymbol{\nabla}p + 
	\frac{1}{\sqrt{Ra/Pr}}\nabla^2\boldsymbol{u}
	- \frac{(T-T_0)}{T_0}\boldsymbol{e}_g - F\boldsymbol{u},
	\label{eq:nonDimMomentEq}
\end{equation}
\begin{equation}
 \partial_t T + (\boldsymbol{u}\cdot\nabla)T
    = \frac{1}{\sqrt{Ra Pr}}\nabla^2 T
	- ST,
	\label{eq:T}
\end{equation}
where the Rayleigh number and Prandlt number are defined as,
\begin{equation}
	Ra = \frac{{g}{\beta}\Delta{T}{R}^3}{{\nu}{\kappa}},
\end{equation}
\begin{equation}
	Pr = \frac{{\kappa}}{{\nu}}.
\end{equation}
In our DNS we use $S=0.06$ and $F=0.06$, which are the values that have already been shown to be suitable in previous studies\cite{Bruneau2018,He2021}. The boundary conditions used are $\boldsymbol{u}=0$ and $T_{0}=1$ on the bubble equator.

\subsection{Tilt leads to stable stratification for the hemispherical flow}\label{Tilt_analysis}

The tilt of the bubble by an angle $\delta$ affects the flow dynamics only through its influence on the buoyancy term. To consider this influence, it is convenient to introduce a spherical coordinate system with coordinates $(r,\theta,\phi)$, and basis vectors $\boldsymbol{e}_r(\theta,\phi)$, $\boldsymbol{e}_\theta(\theta,\phi)$, $\boldsymbol{e}_\phi(\theta,\phi)$. For the hemisphere, the polar coordinate is restricted to $\theta\in[0,\pi/2]$ and is measured from the $\boldsymbol{e}_z$ axis, while the azimuthal coordinate $\phi\in[0,2\pi)$ is measured from the $\boldsymbol{e}_x$ axis. For our two-dimensional flow on the bubble surface, the motion is confined to $r=R$, and there is no flow in the radial direction $\boldsymbol{e}_r(\theta,\phi)$. The unit vector $\boldsymbol{e}_\theta(\theta,\phi)$ depends on the coordinates as 
\begin{equation}
   \boldsymbol{e}_\theta(\theta,\phi) = \boldsymbol{e}_x\cos\theta \cos\phi +  \boldsymbol{e}_y\cos\theta \sin\phi  - \boldsymbol{e}_z\sin\theta,
\end{equation}
and therefore when the flow equations are projected onto the direction $\boldsymbol{e}_\theta(\theta,\phi)$, the buoyancy force projected along this direction (denoted by $B_\theta$) becomes 
\begin{equation}
\begin{split}
B_\theta= -\frac{(T-T_0)}{T_0}\Big(\sin\delta\cos\theta \sin\phi +\cos\delta\sin\theta\Big).
\end{split}
\end{equation}
For the no tilt case $\delta=0$ we have $B_\theta= -((T-T_0)/T_0)\sin\theta$. Since $\sin\theta\geq 0$ on the interval $\theta\in[0,\pi/2]$, then $B_\theta$ will act to accelerate the fluid in the $-\boldsymbol{e}_\theta$ direction in regions where the temperature anomaly is positive, $(T-T_0)/T_0>0$. This means that fluid particles that are heated near the equator will accelerate towards the North pole, corresponding to convection. For $\delta=\pi/2$ then $B_\theta= -((T-T_0)/T_0)\cos\theta \sin\phi$. In this case, while $\cos\theta\geq 0$ on the interval $\theta\in[0,\pi/2]$, $\sin\phi$ changes sign on the interval $\phi\in[0,2\pi)$. Due to this, on the lower side of the hemisphere corresponding to $y>0$ and $\phi\in[0,\pi)$, $B_\theta$ will act to accelerate the fluid in the $-\boldsymbol{e}_\theta$ direction when $(T-T_0)/T_0>0$, but for $\phi\in(\pi,2\pi)$, $B_\theta$ will act to accelerate the fluid in the $+\boldsymbol{e}_\theta$ direction when $(T-T_0)/T_0>0$. It means that when $(T-T_0)/T_0>0$, then for $\phi\in[0,\pi)$, $B_\theta$ will lead to convective motion towards the North pole, while for $\phi\in(\pi,2\pi)$, $B_\theta$ will act to stabilize and stratify the flow. Therefore, while for $\delta=0$, heating at the equator generates buoyancy forces leading to convection and (for sufficiently large $Ra$) turbulence over the entire surface of the bubble\cite{He2021}, for $\delta=\pi/2$, buoyancy forces leading to convection and turbulence can arise for $\phi\in[0,\pi)$, whereas for $\phi\in(\pi,2\pi)$ the buoyancy forces will quench the turbulence and stratify the flow. Note also that since $B_\theta= -((T-T_0)/T_0)\cos\theta \sin\phi$, then the buoyancy forces that produce convection in the region $\phi\in[0,\pi)$ will be strongest near $\phi=\pi/2$. Hence, for $\delta=\pi/2$, we would expect to see the strongest convection and turbulence near the lower edge of the bubble.

For intermediate $\delta$, the tilt will lead to stratification at points where the inequality $\sin\delta\sin\theta \sin\phi +\cos\delta\cos\theta<0$ is satisfied, and this can only be satisfied for $\phi\in(\pi,2\pi)$ and in the region $\theta\in[0,-\tan^{-1}(\tan\delta\sin\phi))$.

\section{Direct Numerical Simulations}

The numerical simulations are conducted by the homebrew code which is introduced in details in previous studies\cite{Bruneau2018,He2021}.
Here we give a brief overview of the numerical methods used in the DNS.
The nondimensional governing equations are solved numerically in a computational space which is accomplished by the stereographic projection.
In the computational space, the geometry of the bubble is a plane circle where the stagger grid is employed for discretization 
and the penalty method is employed for the implementation of the boundary conditions.
The temporal derivatives are approached by the second-order Gear scheme and the non-linear terms are handled by a third-order Murman-like scheme.
The mesh sensitivity are checked and the resolution of $1024\times1024$ and $2048\times2048$ is choosed for the different $Ra$.

The table \ref{tab:case} lists the parameters for all the cases of DNS considered in this study.
$F$ and $S$ are fixed to $0.06$ as in previous studies\cite{Bruneau2018,He2021,Meuel2018}.
$Pr$ is fixed to $7$ since the soap concentration in the water is very low.
The Rayleigh number is varied in the range of $Ra\in[3\times10^6, 3\times10^9]$,
and the full range of tilting angles (in degrees) $\delta\in[0^{\circ},90^{\circ}]$ is explored.
\begin{table}
\caption{The configuration of non-dimensional coefficients for all the cases considered in this study}
\label{tab:case}
\begin{ruledtabular}
    \begin{tabular}{c|c|ccccccc}
        $Ra$                	&$Pr$   &\multicolumn{4}{c}{$\delta$} \\
        \hline
        $3\times10^9$ 			&$7$    &$0^{\circ}$  &$30^{\circ}$  &$60^{\circ}$  &$90^{\circ}$		\\
        $3\times10^8$ 			&$7$    &$0^{\circ}$  &$30^{\circ}$  &$60^{\circ}$  &$90^{\circ}$		\\
        $3\times10^7$ 			&$7$    &$0^{\circ}$  &$30^{\circ}$  &$60^{\circ}$  &$90^{\circ}$		\\
        $3\times10^6$ 			&$7$    &$0^{\circ}$  &$30^{\circ}$  &$60^{\circ}$  &$90^{\circ}$		\\
    \end{tabular}  
\end{ruledtabular}
\end{table}

\section{Results \& Discussion}
\subsection{the Phenomenological Observations}
\begin{figure*}
	\centering
	\includegraphics[width = 0.24\textwidth]{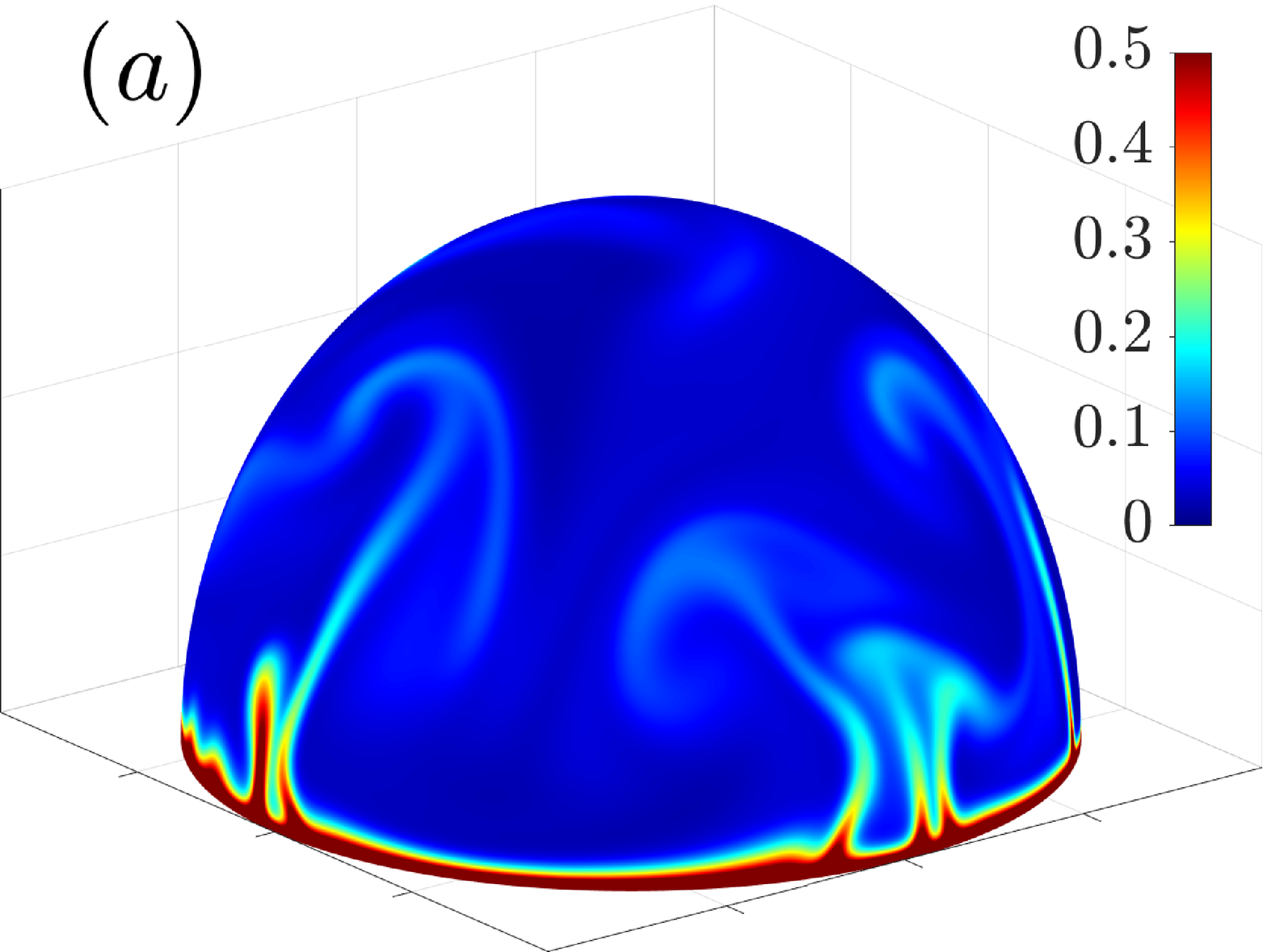}
	\includegraphics[width = 0.24\textwidth]{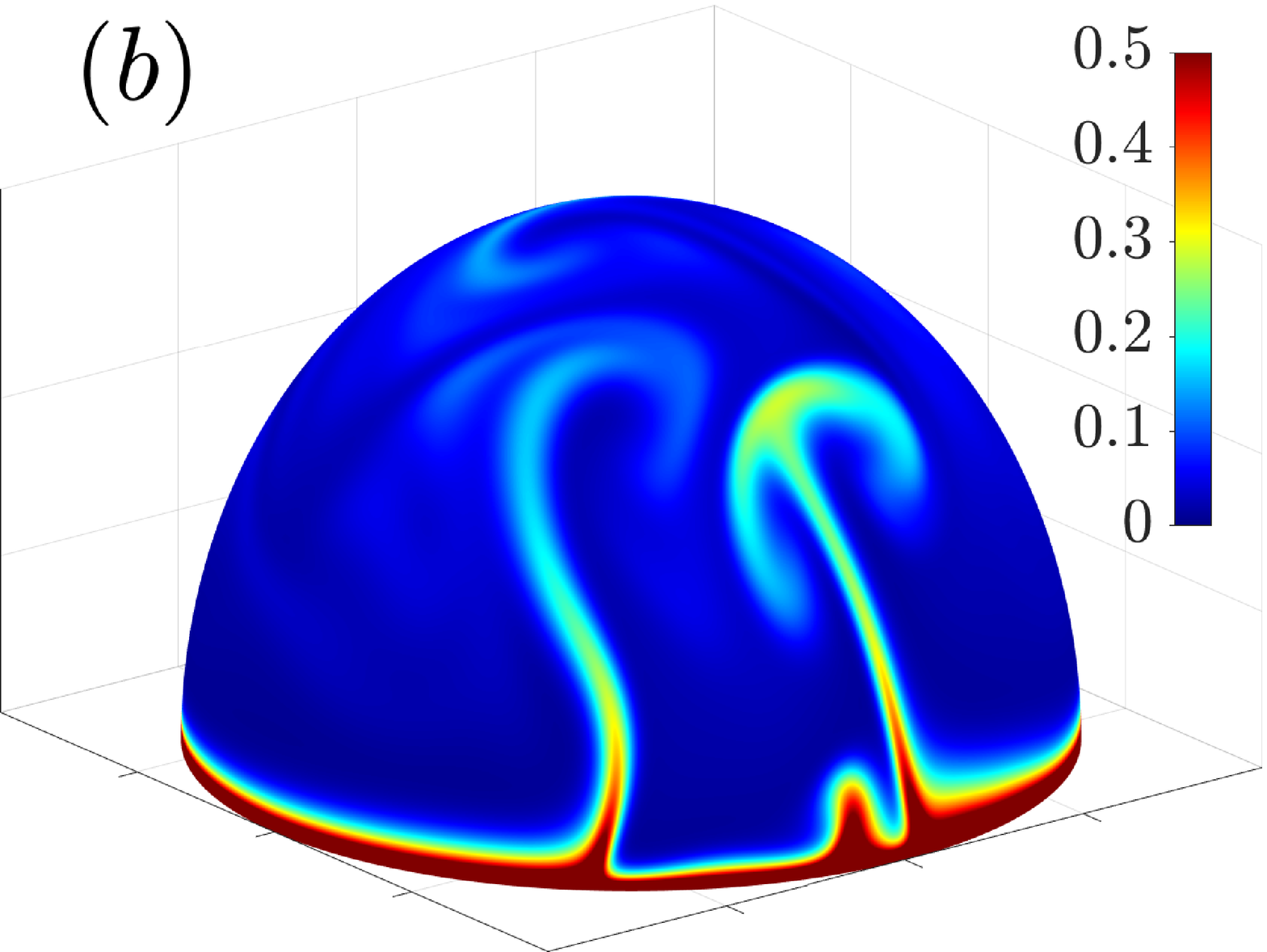}
	\includegraphics[width = 0.24\textwidth]{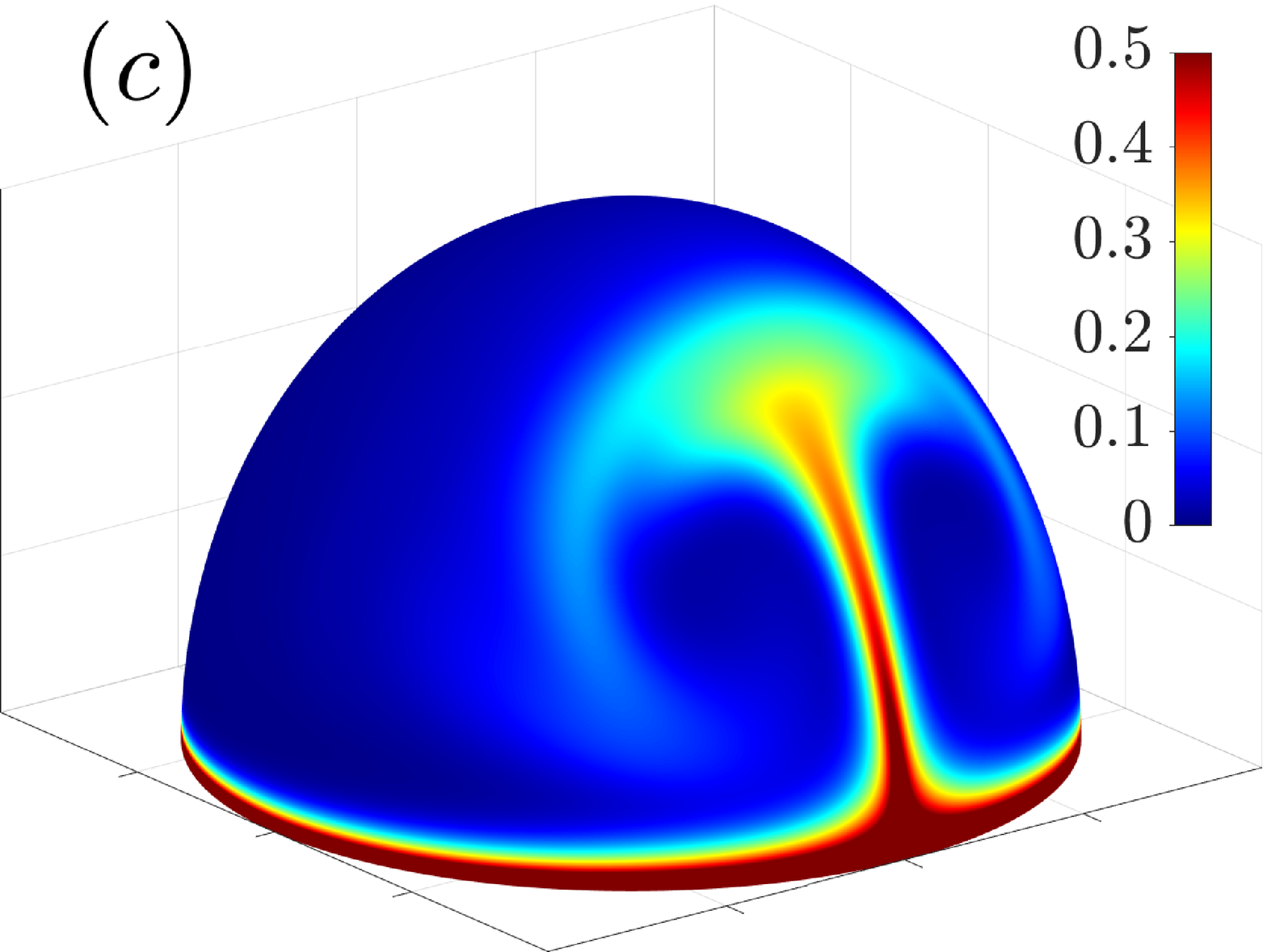}
	\includegraphics[width = 0.24\textwidth]{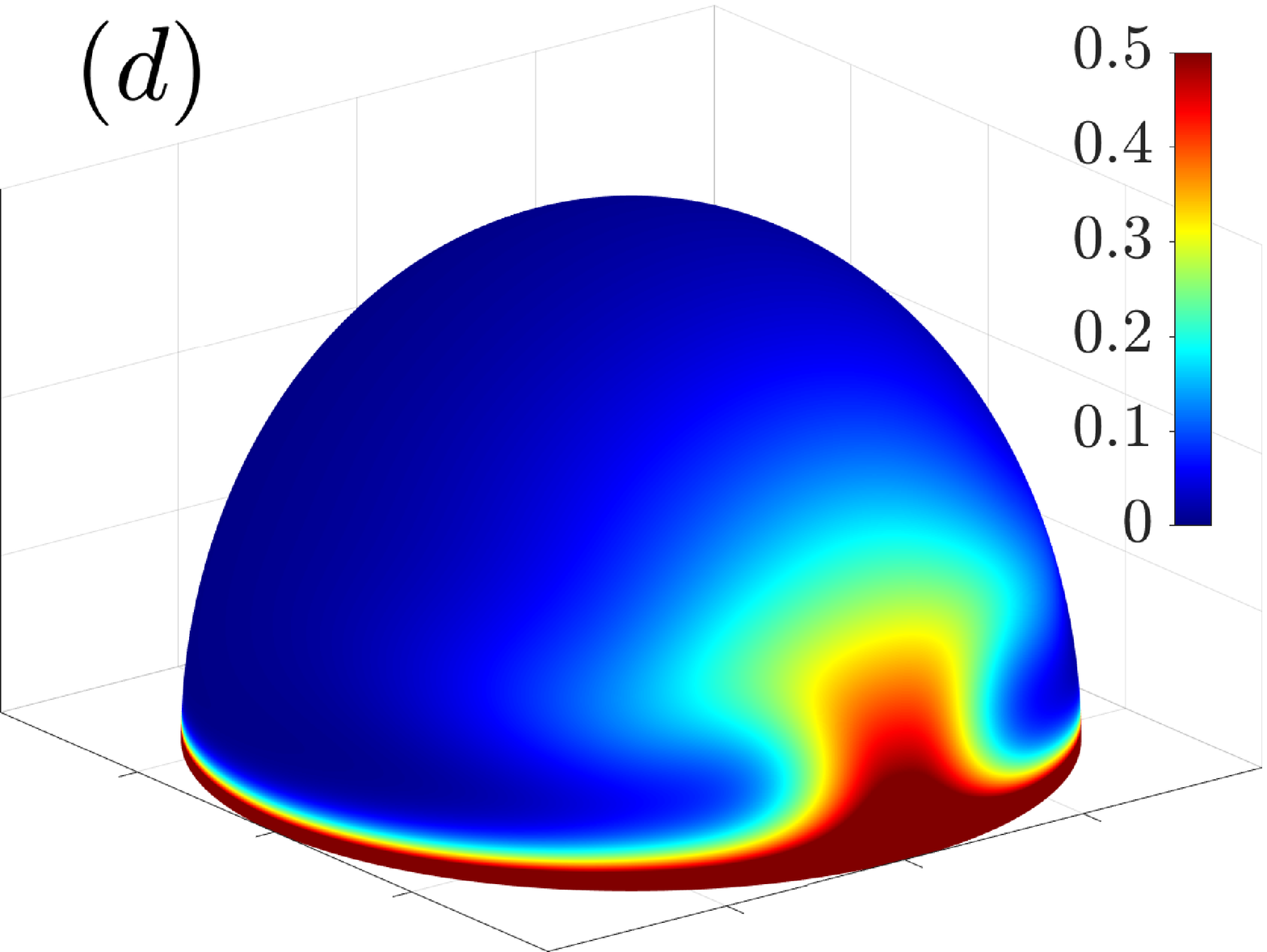}
	\includegraphics[width = 0.24\textwidth]{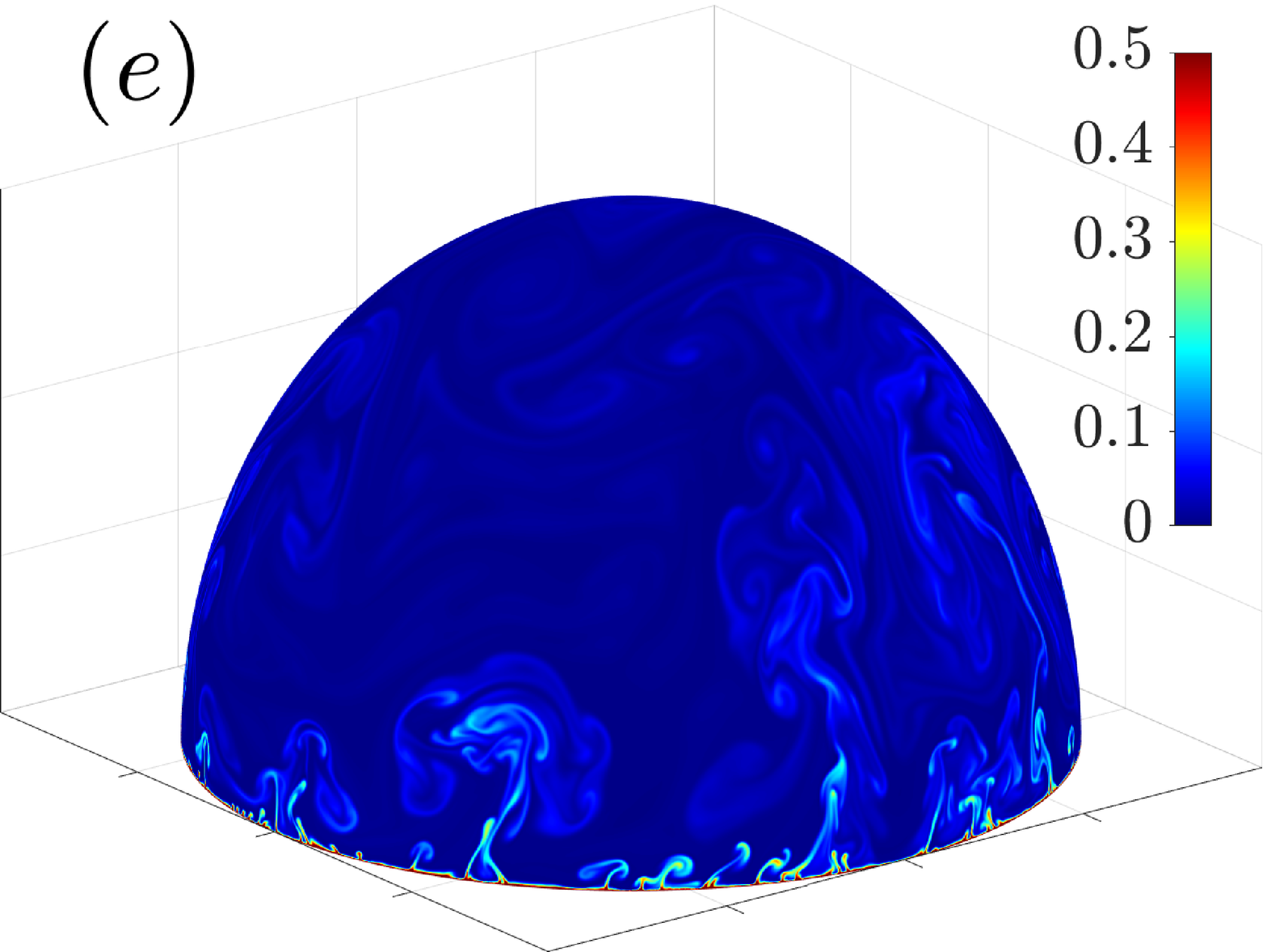}
	\includegraphics[width = 0.24\textwidth]{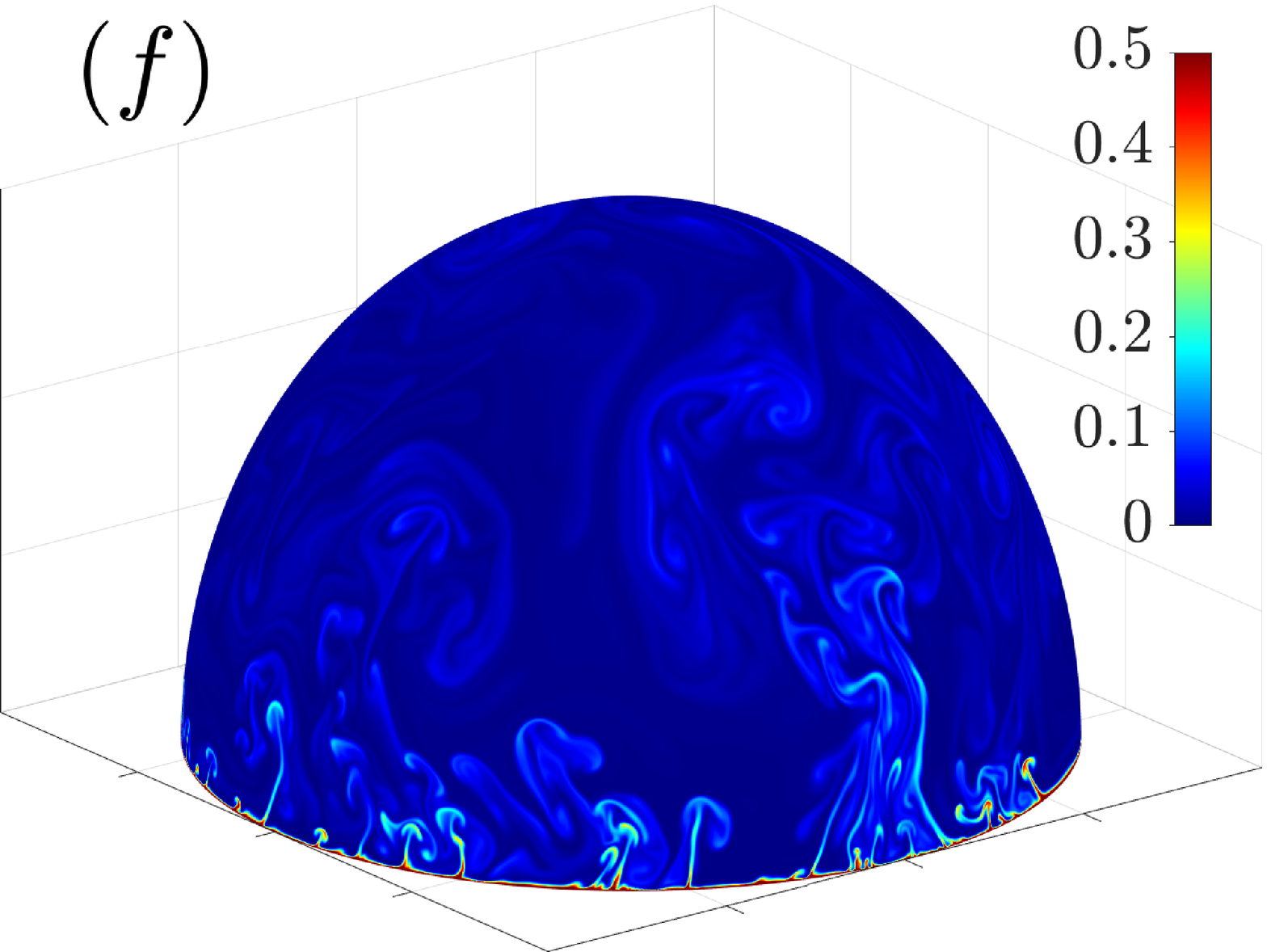}
	\includegraphics[width = 0.24\textwidth]{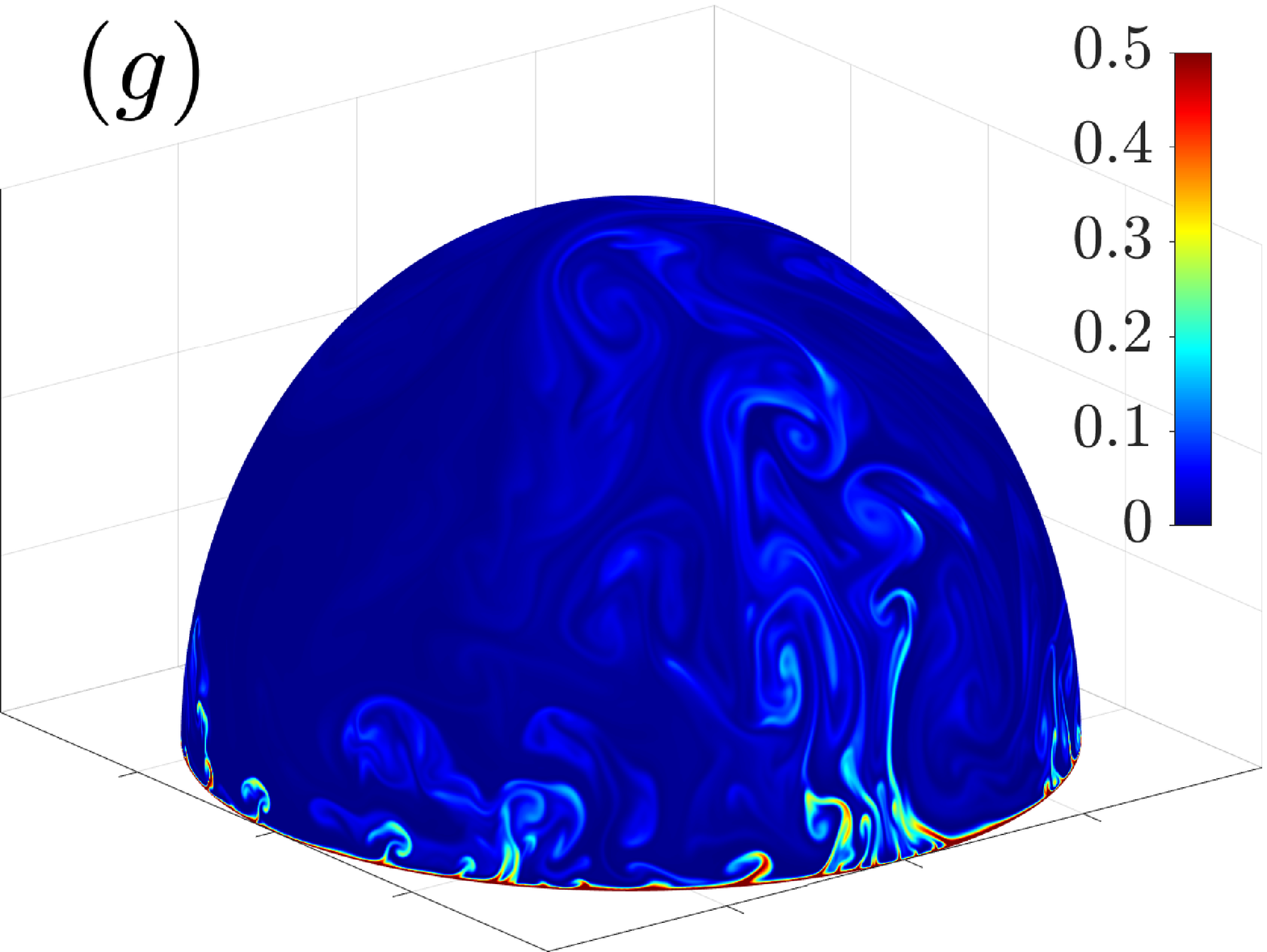}
	\includegraphics[width = 0.24\textwidth]{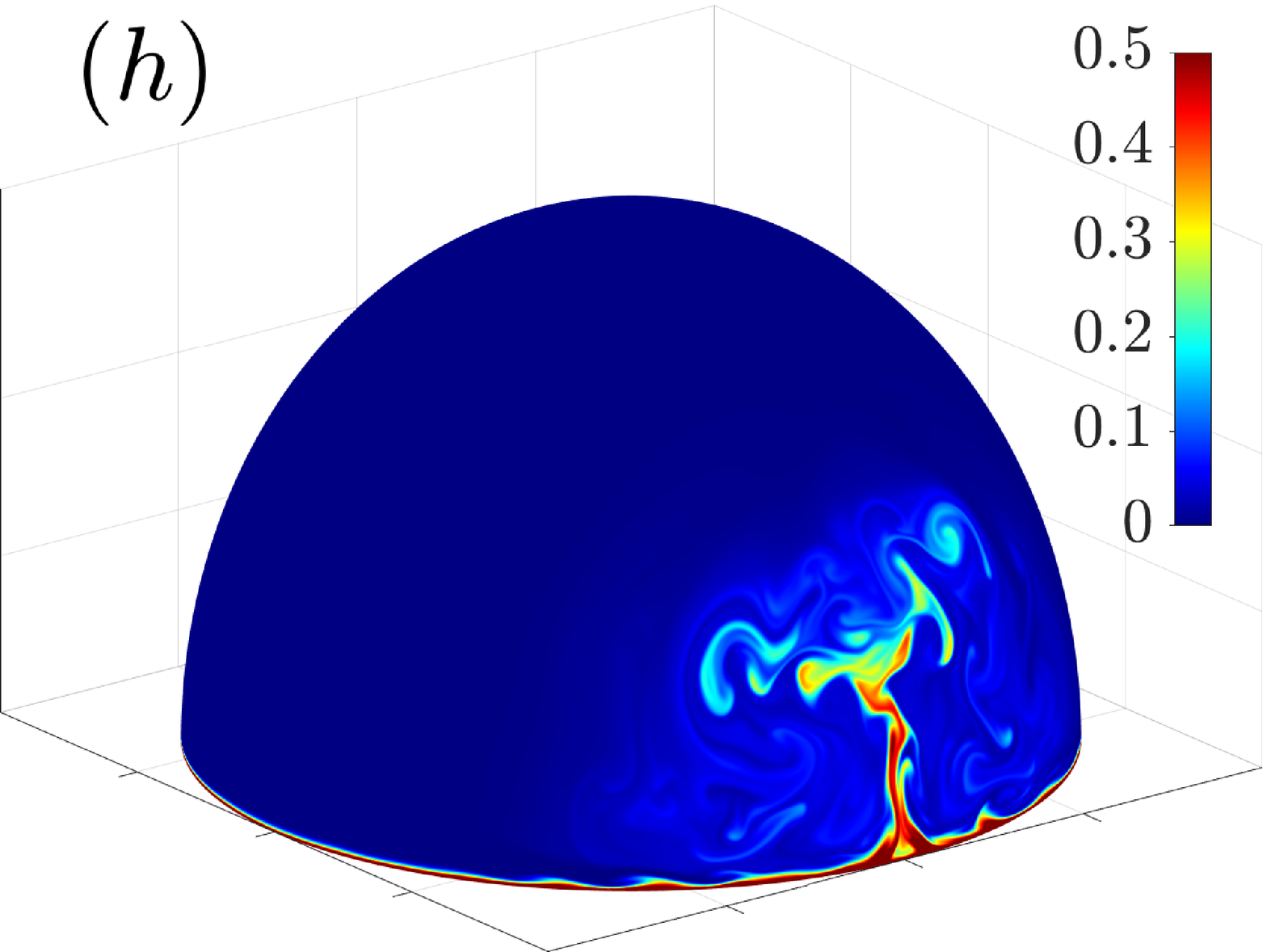}
	\caption{Snapshots of the temperature fields for $Ra=3\times10^6$(the upper line: $(a)$ to $(d)$) and 
    $Ra=3\times10^9$(the lower line: $(e)$ to $(h)$) 
    with $\delta=0^{\circ},30^{\circ},60^{\circ},90^{\circ}$(from left to right).} 
	\label{fig:flow3D}
\end{figure*}

\begin{figure*}
	\centering
	\includegraphics[width = 0.24\textwidth]{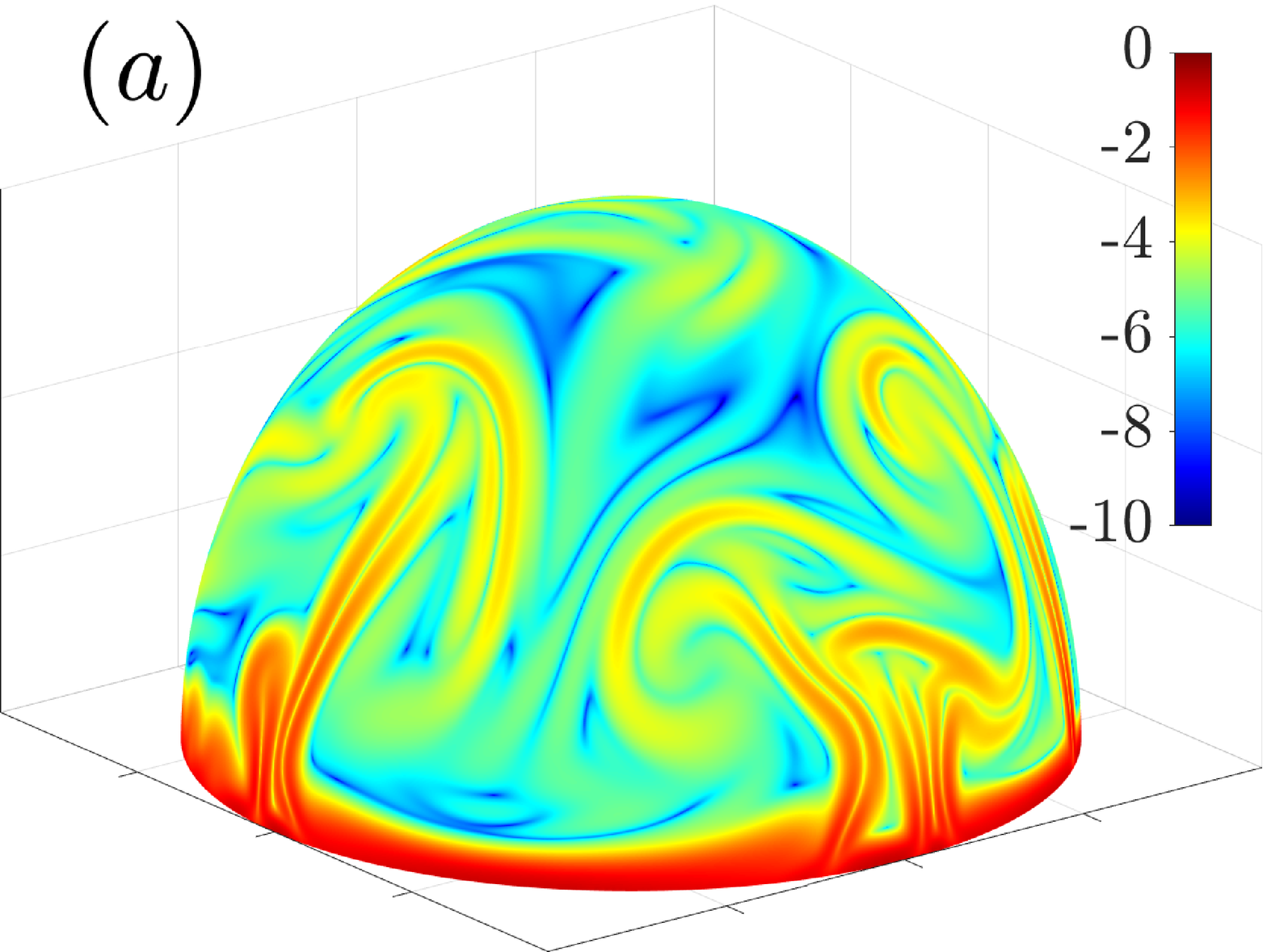}
	\includegraphics[width = 0.24\textwidth]{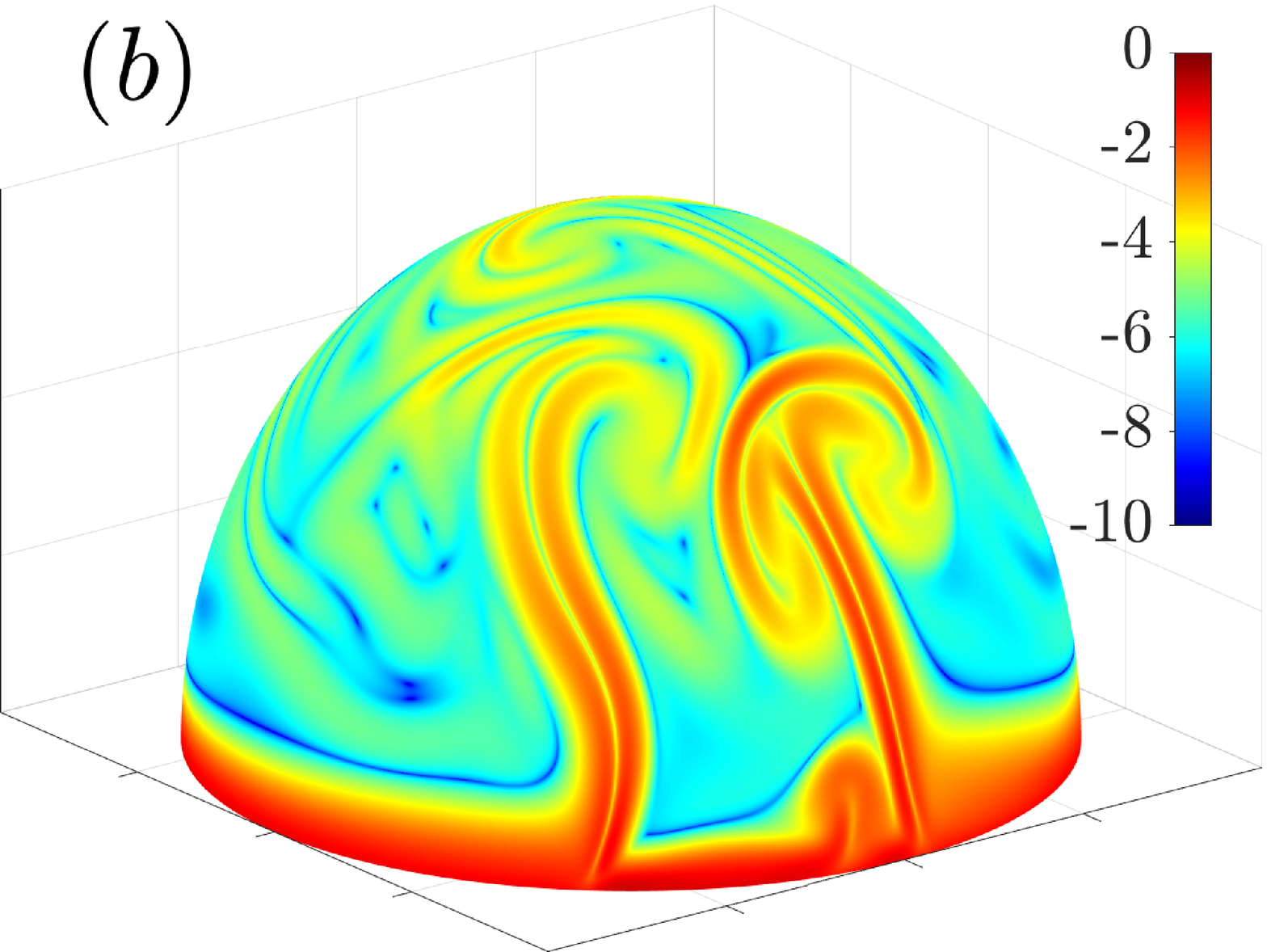}
	\includegraphics[width = 0.24\textwidth]{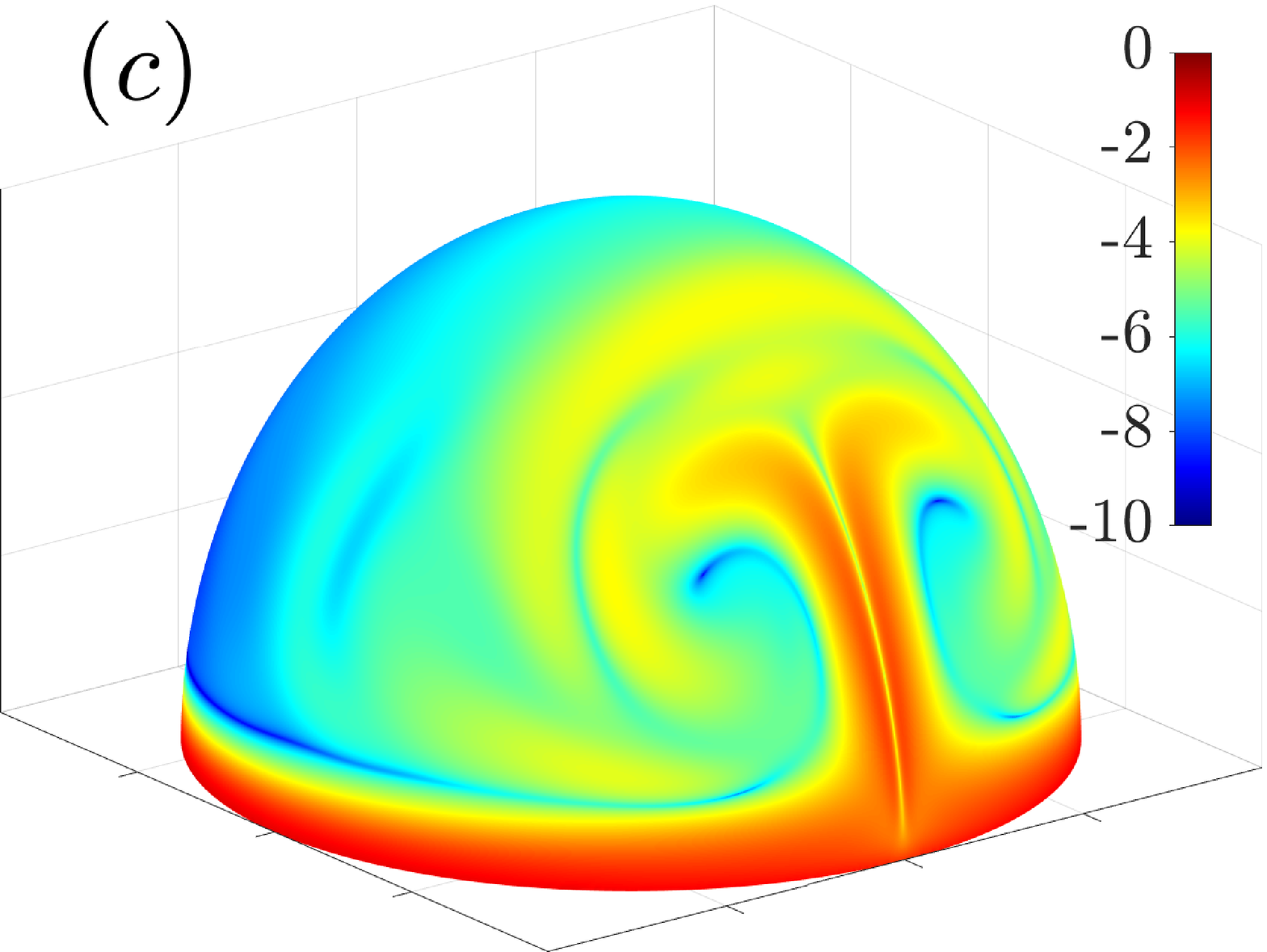}
	\includegraphics[width = 0.24\textwidth]{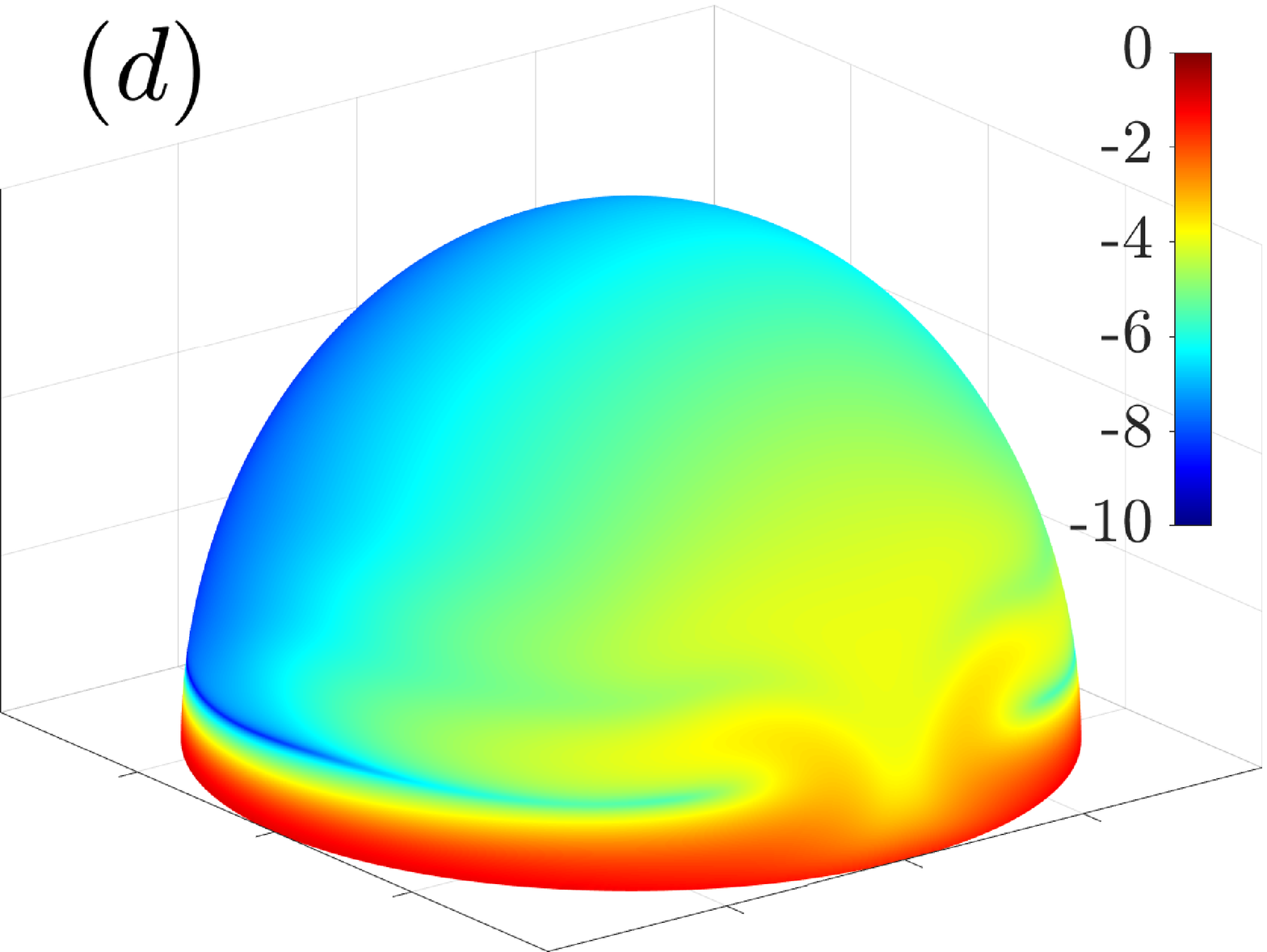}
	\includegraphics[width = 0.24\textwidth]{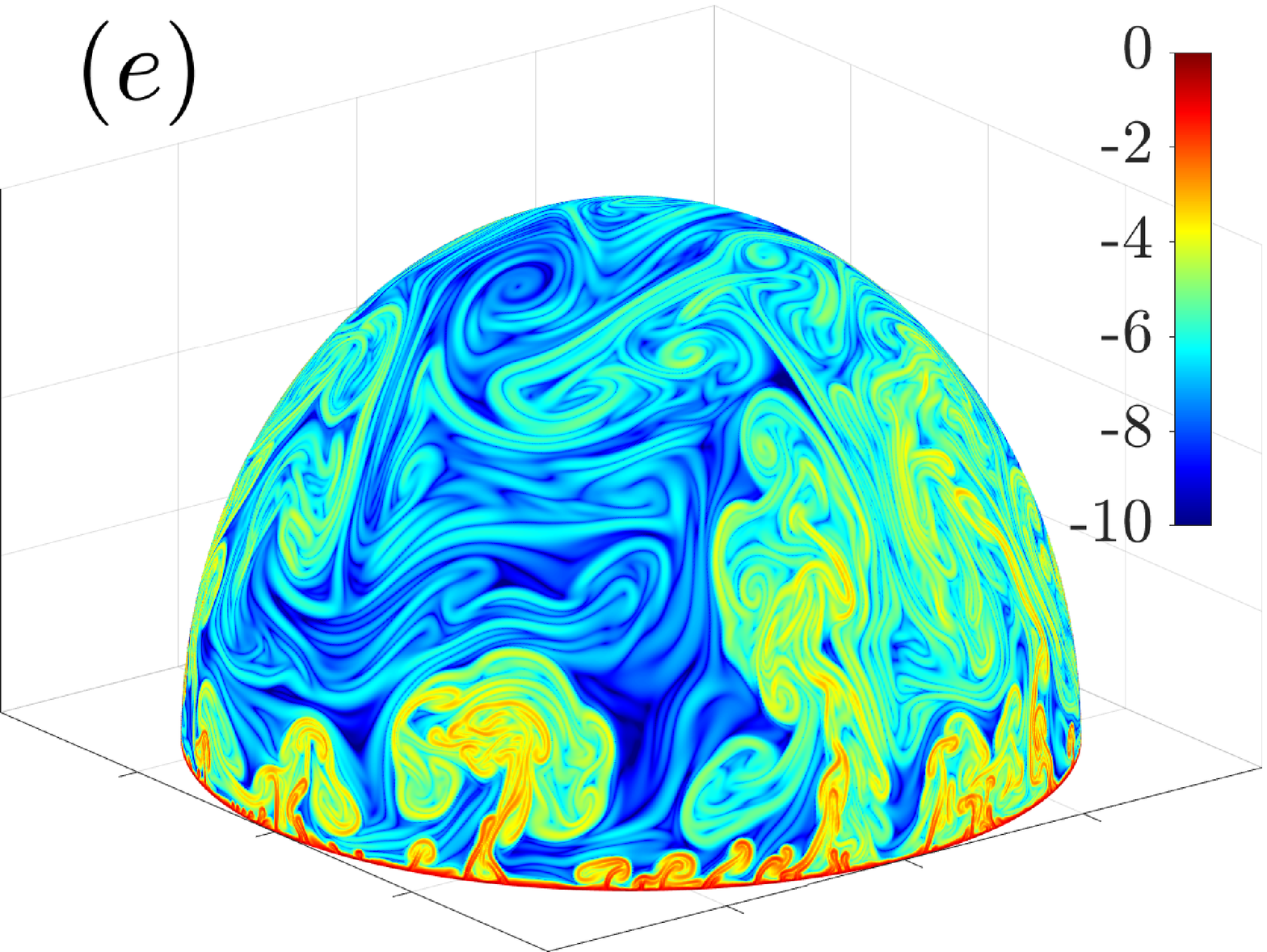}
	\includegraphics[width = 0.24\textwidth]{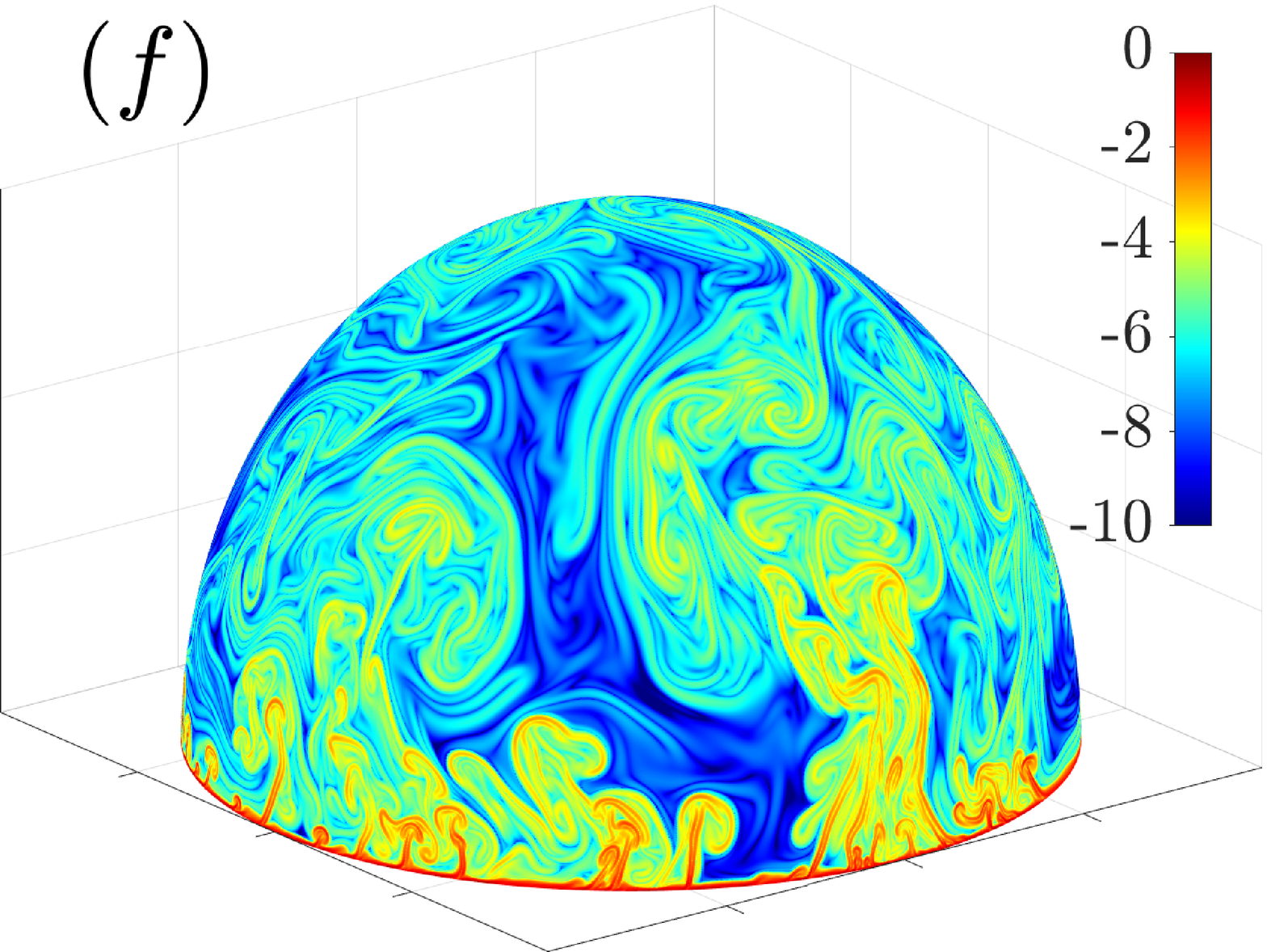}
	\includegraphics[width = 0.24\textwidth]{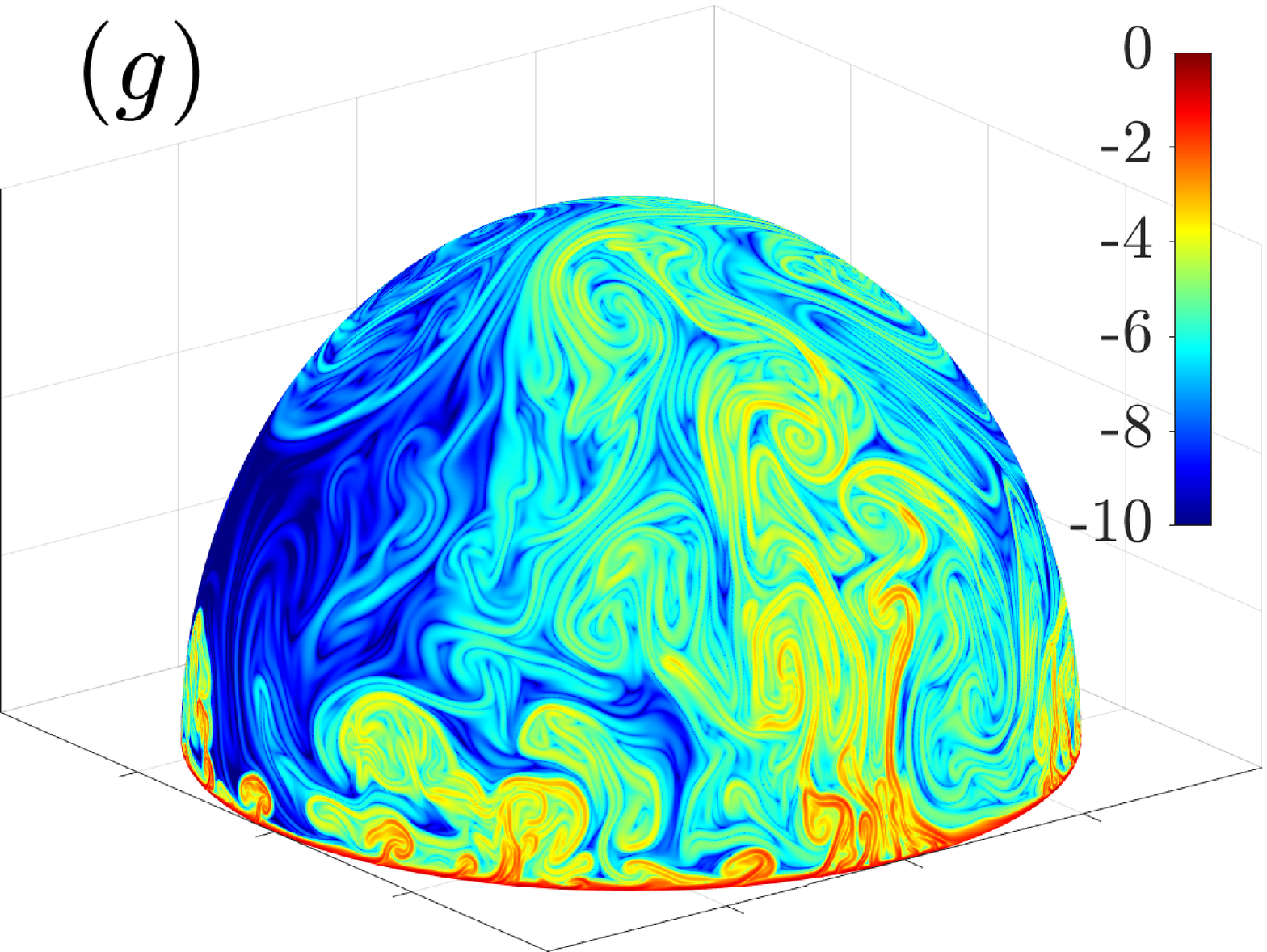}
	\includegraphics[width = 0.24\textwidth]{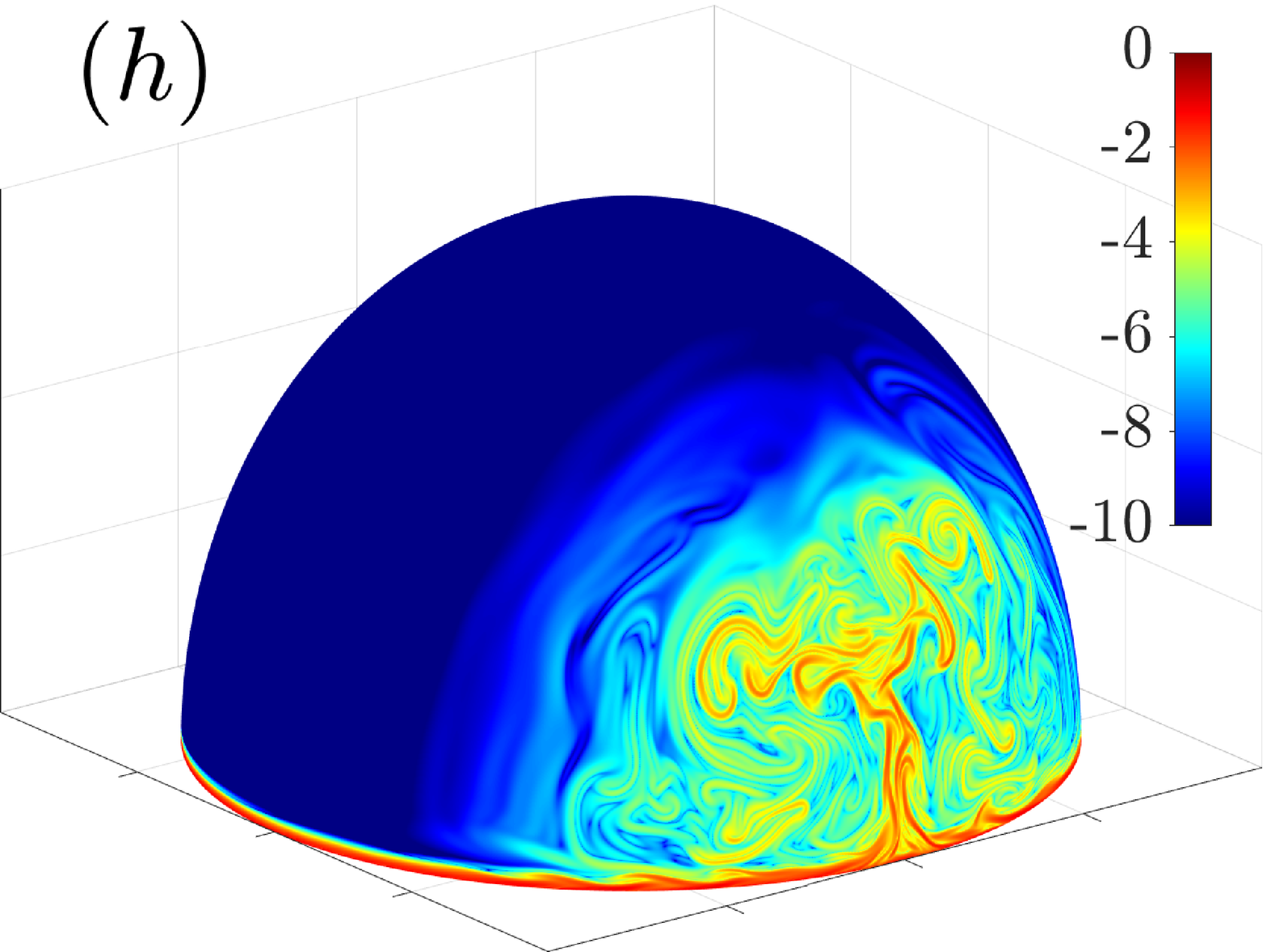}
	\caption{Snapshots of the logarithm of the thermal energy dissipation rate $\log_{10}\left(\epsilon_{T}\right)$
    for $Ra=3\times10^6$(the upper line: $(a)$ to $(d)$) and 
    $Ra=3\times10^9$(the lower line: $(e)$ to $(h)$) 
    with $\delta=0^{\circ},30^{\circ},60^{\circ},90^{\circ}$(from left to right).} 
	\label{fig:epsilonT}
\end{figure*}

\begin{figure*}
	\centering
	\includegraphics[width = 0.24\textwidth]{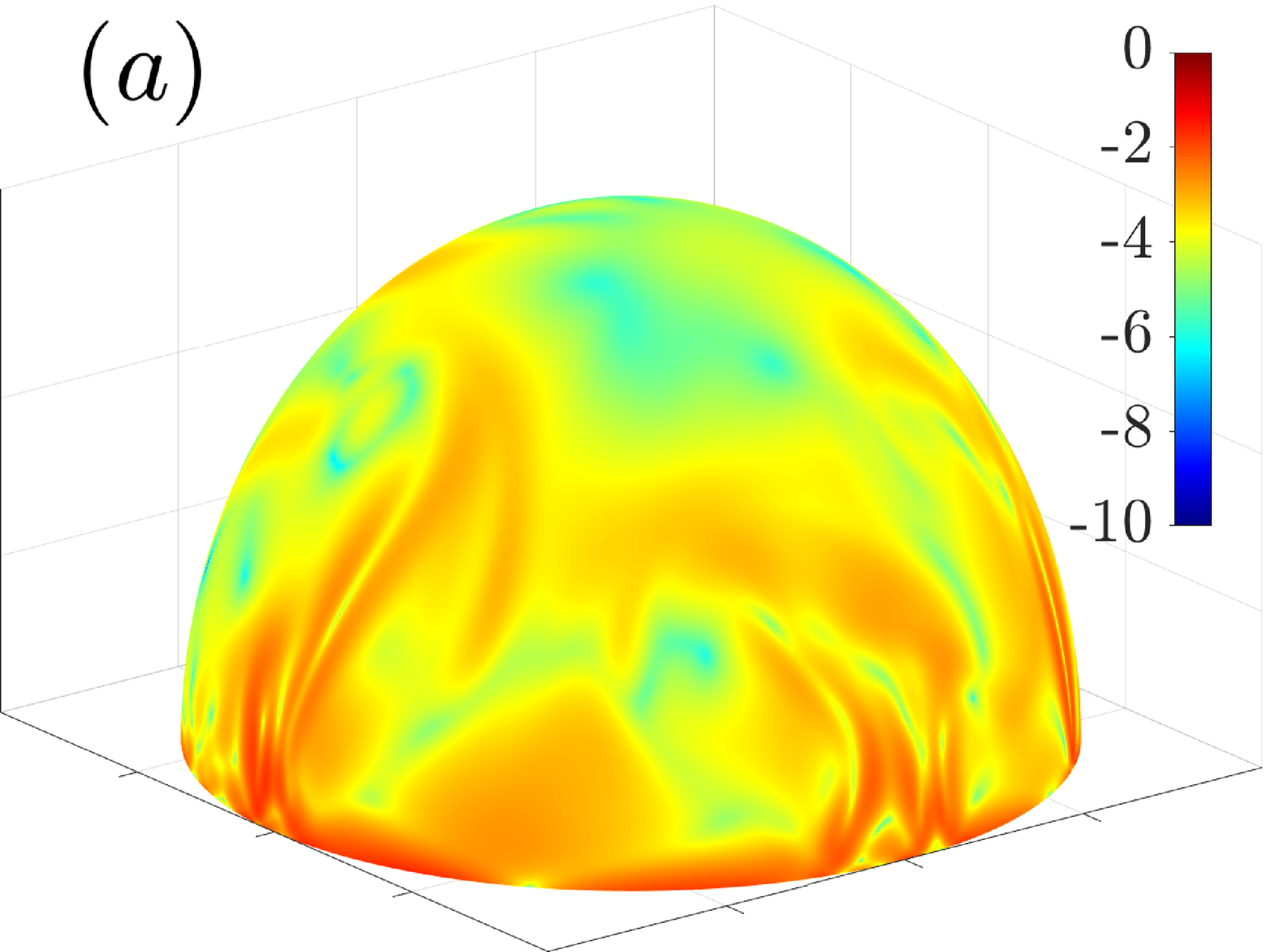}
	\includegraphics[width = 0.24\textwidth]{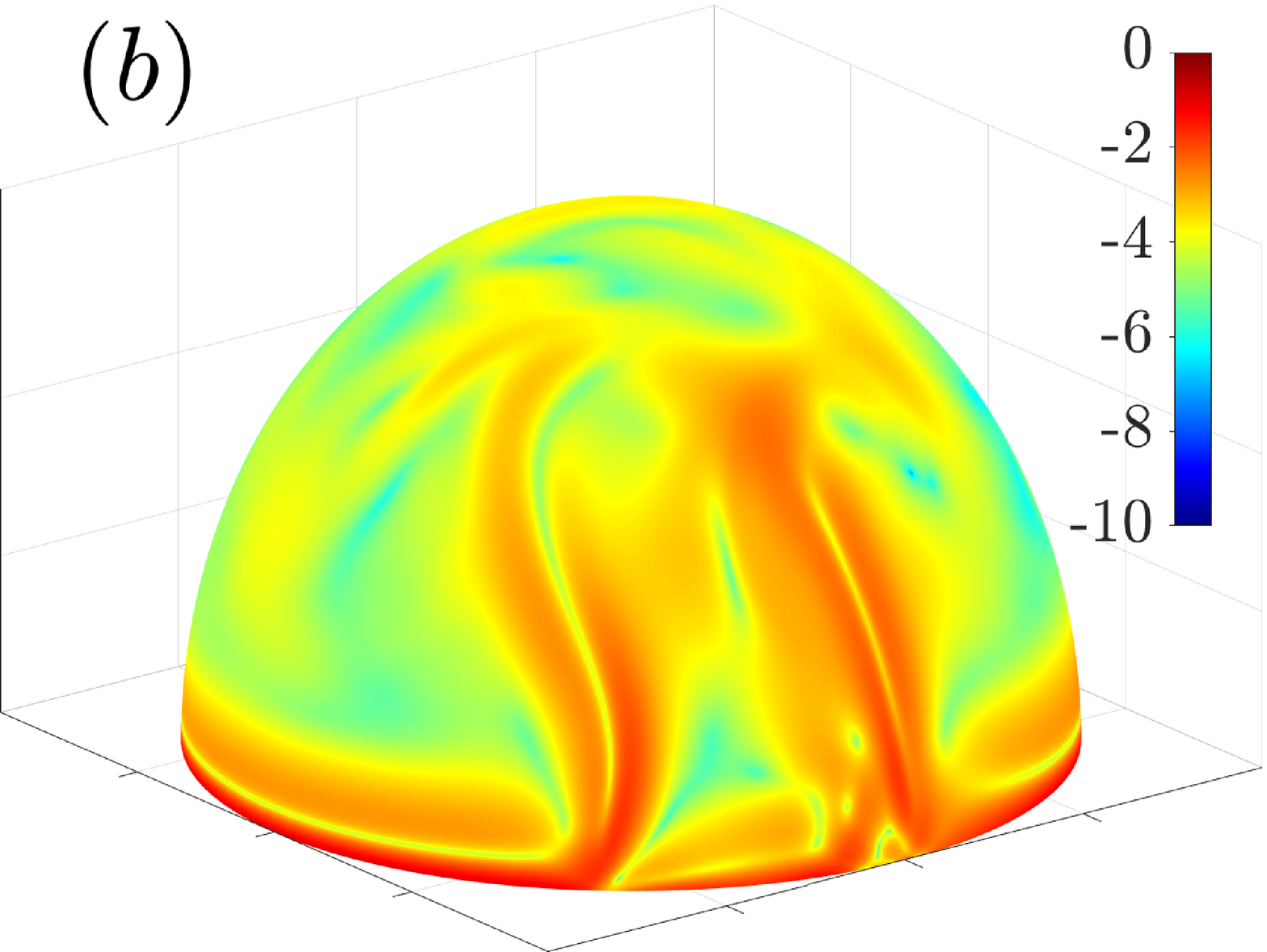}
	\includegraphics[width = 0.24\textwidth]{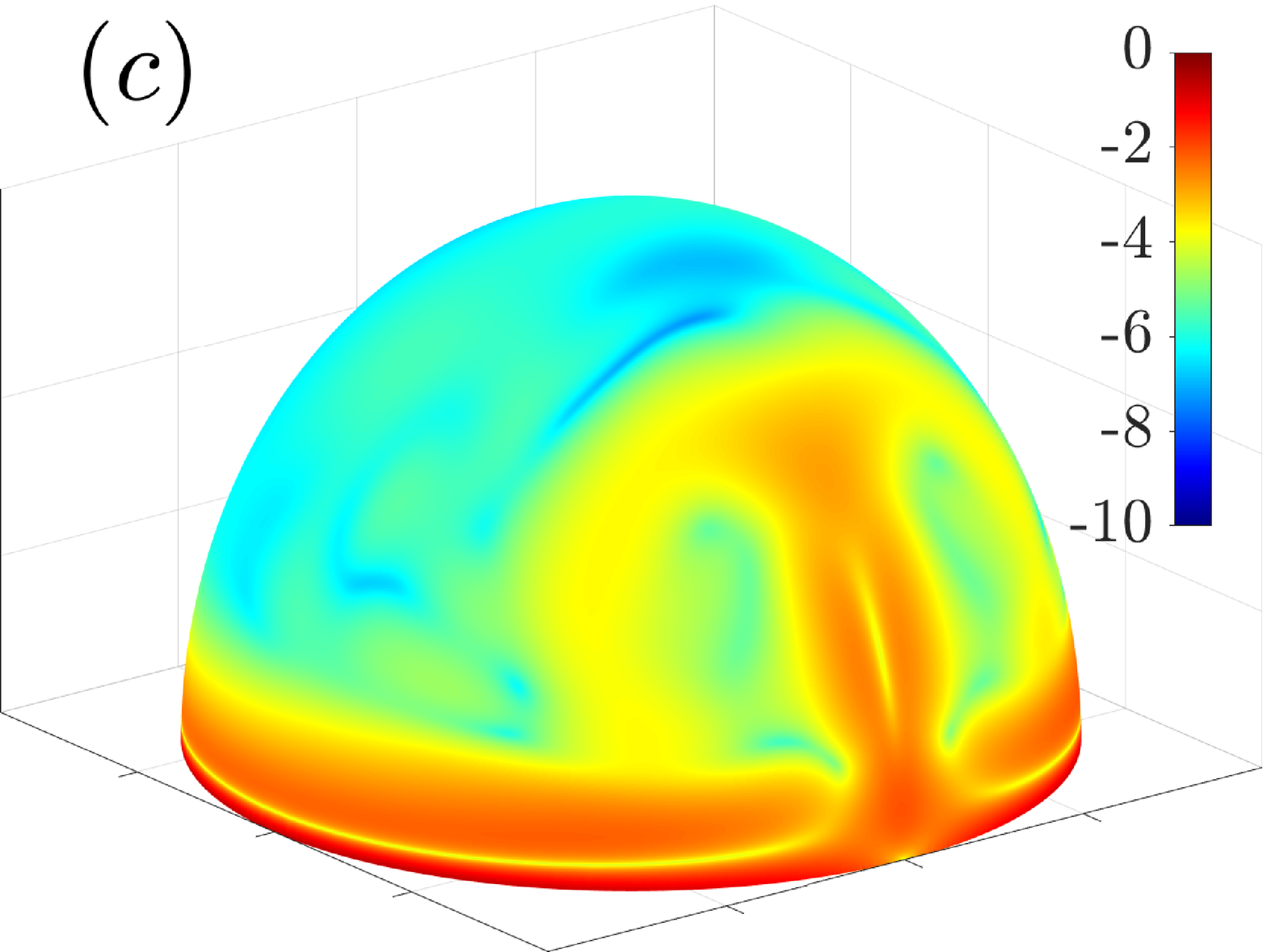}
	\includegraphics[width = 0.24\textwidth]{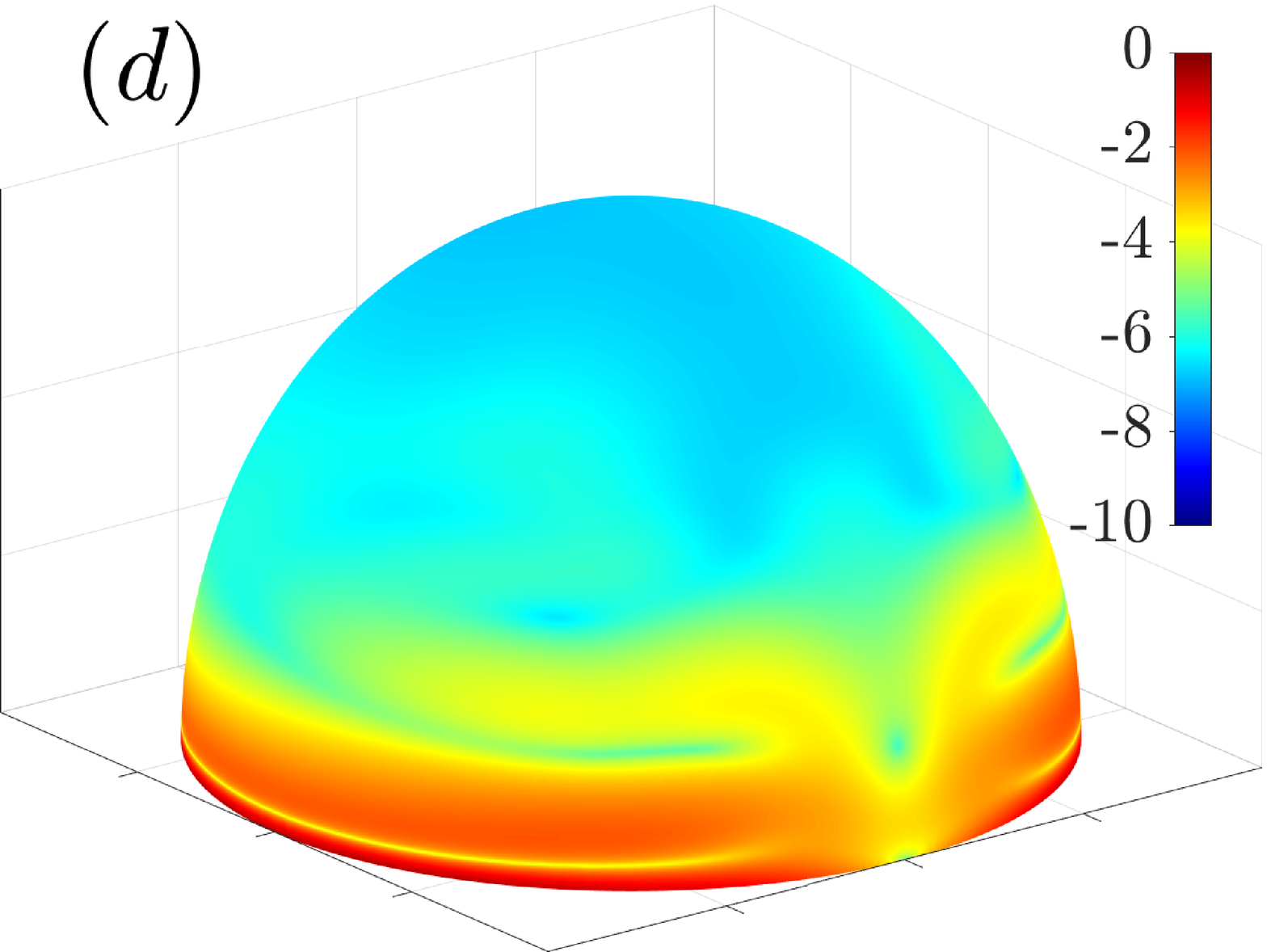}
	\includegraphics[width = 0.24\textwidth]{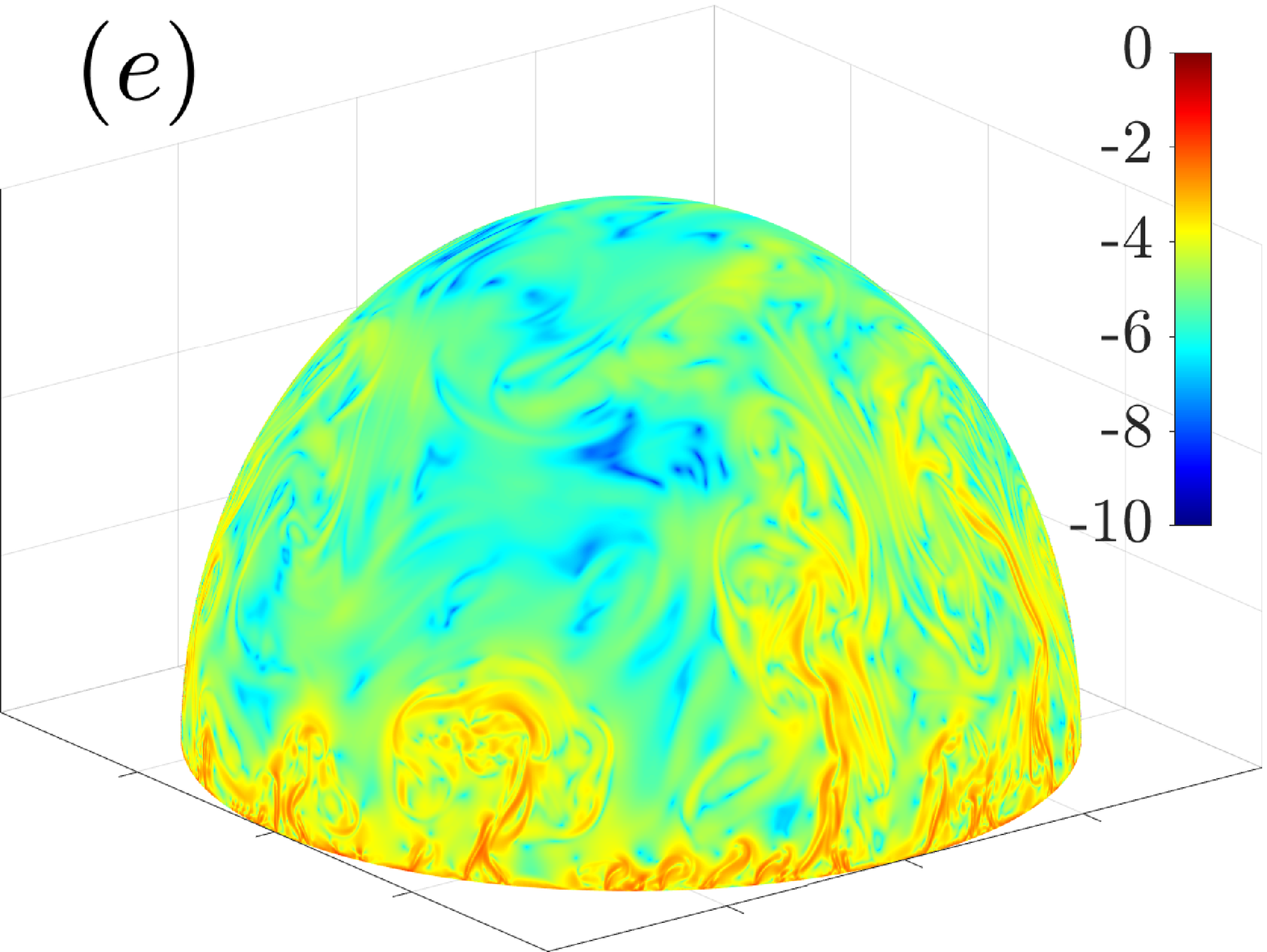}
	\includegraphics[width = 0.24\textwidth]{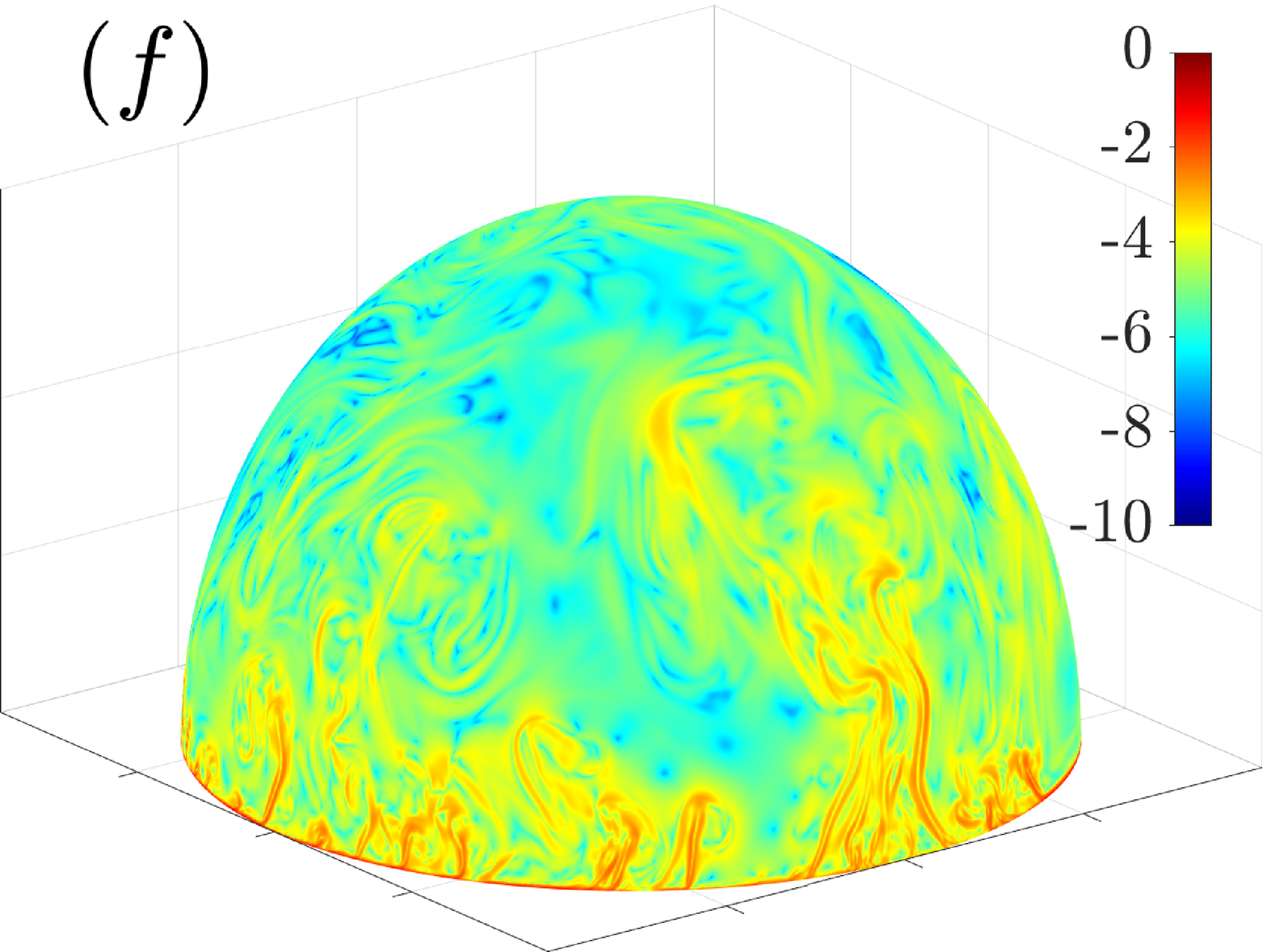}
	\includegraphics[width = 0.24\textwidth]{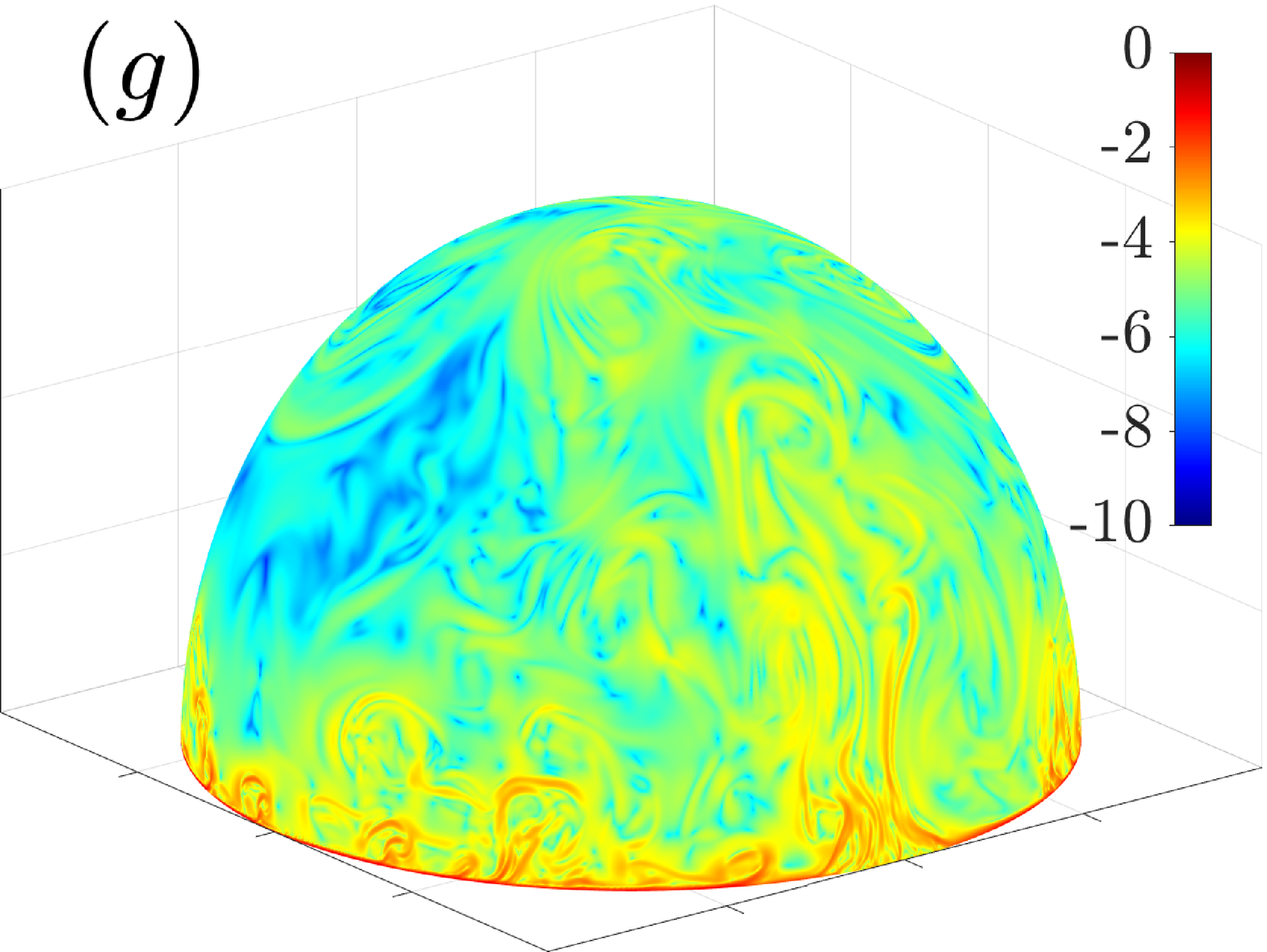}
	\includegraphics[width = 0.24\textwidth]{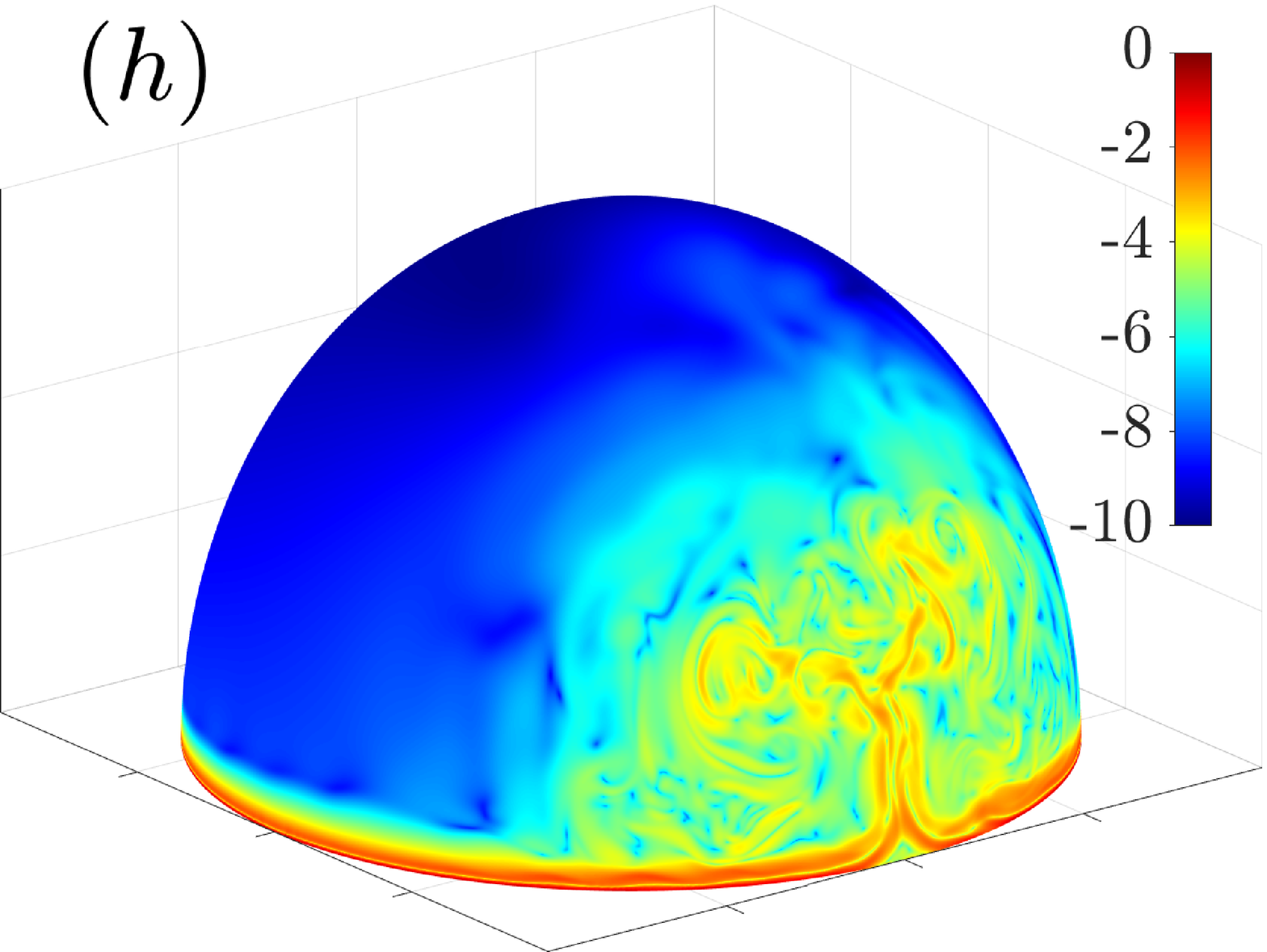}
	\caption{Snapshots of the logarithm of the kinetic energy dissipation rate $\log_{10}\left(\epsilon_{u}\right)$ 
    for $Ra=3\times10^6$(the upper line: $(a)$ to $(d)$) and 
    $Ra=3\times10^9$(the lower line: $(e)$ to $(h)$) 
    with $\delta=0^{\circ},30^{\circ},60^{\circ},90^{\circ}$(from left to right).} 
	\label{fig:epsilonU}
\end{figure*}

We begin by making qualitative observations on the effect of $\delta$ on the behavior of instantaneous flow fields on the bubble.
Figure \ref{fig:flow3D} illustrates typical snapshots of the temperature field $T$ on the surface of the bubble 
for different Rayleigh numbers $Ra$ and tilt angles $\delta$.
The upper line in figure \ref{fig:flow3D} is for $Ra=3\times10^6$, and the lower line in figure \ref{fig:flow3D} is for $Ra=3\times10^9$.
From left to the right, $\delta$ increases from $0^{\circ}$ to $90^{\circ}$.
For the same time instant, figures \ref{fig:epsilonT} and \ref{fig:epsilonU} illustrate the corresponding fields of 
the logarithmic thermal energy dissipation $\log_{10}\left(\epsilon_{T}\right)$ and 
the logarithmic kinetic energy dissipation $\log_{10}\left(\epsilon_{u}\right)$, respectively.

Figure \ref{fig:flow3D} shows that for $\delta=0^\circ$, the flow only features plumes
and does not contain large scale circulations which are usually seen in Rayleigh-B\'enard convection.
This is because for the bubble, there is no cold boundary as there is in in Rayleigh-B\'enard convection, but only a hot boundary at the equator.
As $Ra$ is increased, the plumes become more filamented and smaller.
These flow patterns appearing here for $\delta=0^\circ$ are qualitatively similar to those observed in experiments\cite{Seychelles2008,Seychelles2010,Meuel2012,Meuel2013,Meuel2018} 
and DNS\cite{Xiong2012,Bruneau2018,Meuel2018,He2021}. 

On the other hand, the observed flow patterns go through a dramatic change as $\delta$ is increased from $0^{\circ}$ to $90^{\circ}$.
When $\delta$ is relatively small, e.g. $30^{\circ}$, then the flow patterns are 
very similar to those for $\delta=0^\circ$, with dynamic plumes detaching from the boundary layer at random locations on the equator, and the plumes dissipate
as time proceeds.
We refer to this regime as the dynamic plumes regime (DPR).
When $\delta$ is sufficiently large, however, the flow is dominated by a stable large plume that
rises from the lower edge of the bubble and is persistent in time.
We refer to this as the stable plume regime(SPR).
It should be noted, however, that the stable plume appears as soon as $\delta>0$, however it is relatively weak and blends in 
with the dynamic plumes that dominate in the DPR.

The transition of the flow patterns from the DPR to the SPR as $\delta$ is increased can be understood in terms of the analysis in \S\ref{Tilt_analysis}, where we showed that as $\delta$ is increased, convection will be suppressed in the upper half of the bubble where $\phi\in[\pi,2\pi)$, and that for the lower half where $\phi\in[0,\pi)$, the convection will be strongest near the lower edge at $\phi=\pi/2$.
 
\begin{figure}
	\includegraphics[width = 0.6\textwidth]{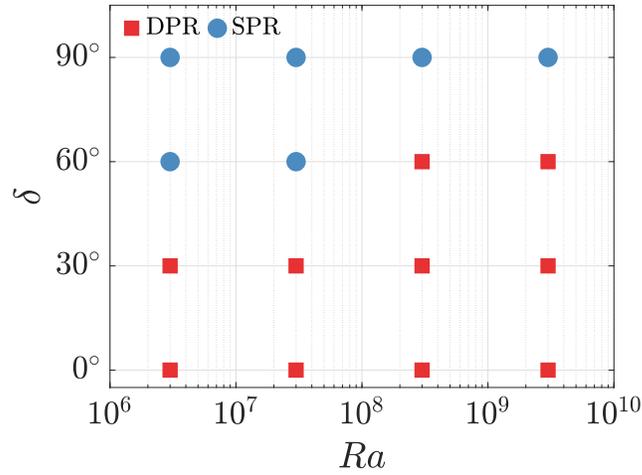}
	\caption{Schematic to illustrate how the flow state depends on $Ra$ and $\delta$.}
	\label{fig:flowState}
\end{figure}

The snapshots of the temperature fields
also reveal that the threshold angle for the flow to transition from the DPR to the SPR depends on $Ra$.
For $Ra=3\times10^6$, the flow is in the SPR for $\delta\gtrsim 60^{\circ}$, while for $Ra=3\times10^9$, the flow is still in the DPR for $\delta=60^{\circ}$ but has transitioned to the SPR at $\delta=90^{\circ}$.
Figure \ref{fig:flowState} illustrates how the flow state depends on $\delta$ and $Ra$. The figure shows that for $\delta\leq 30^\circ$, the flow remains in the DPR for each $Ra$.
For $\delta=60^{\circ}$, the cases with $Ra=3\times10^6$ and $Ra=3\times10^7$ have transitioned into the SPR while the cases with higher $Ra$ remain in the DPR.
For $\delta=90^{\circ}$, however, all of the cases are in the SPR.

Comparing figures \ref{fig:epsilonT} and \ref{fig:epsilonU} with figure \ref{fig:flow3D} shows that the plumes are closely associated with 
regions of large thermal and kinetic energy dissipation field, similar to what is observed in Rayleigh-B\'enard convection\cite{ZhangZhouSun2017JFM,EmranSchumacher2008JFM,XuXi2019POF,ShishkinaWagner2007POF}.
This indicates that the plumes play a key role in the dissipation of thermal and kinetic energy on the bubble, and also suggests that the dissipation rates for these
two fields will be coupled. To explore this, we define the correlation coefficient between $\epsilon_{T}$ and $\epsilon_{u}$ as
\begin{equation}
    \langle c(\boldsymbol{x})\rangle = 
    \frac{\langle(\epsilon_{T}-\langle\epsilon_{T}\rangle)(\epsilon_{u}-\langle\epsilon_{u}\rangle)\rangle}{\sqrt{\langle(\epsilon_{T}-\langle\epsilon_{T}\rangle)^2\rangle}\sqrt{\langle(\epsilon_{u}-\langle\epsilon_{u}\rangle)^2\rangle}},
\end{equation}
where $\langle\cdot\rangle$ here donates a time average at a given location $\boldsymbol{x}$ on the bubble surface.
\begin{figure*}
	\centering
	\includegraphics[width = 0.24\textwidth]{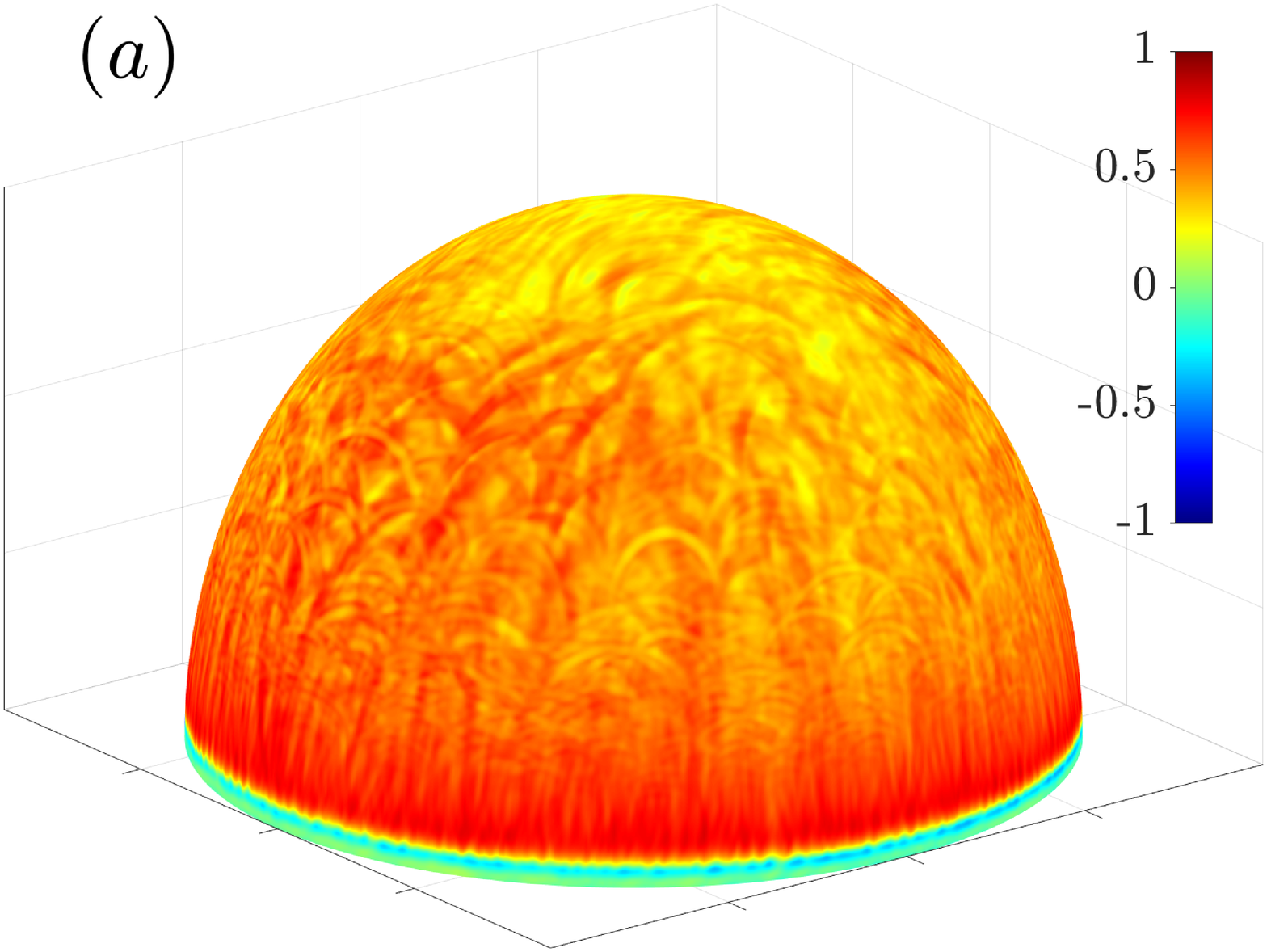}
	\includegraphics[width = 0.24\textwidth]{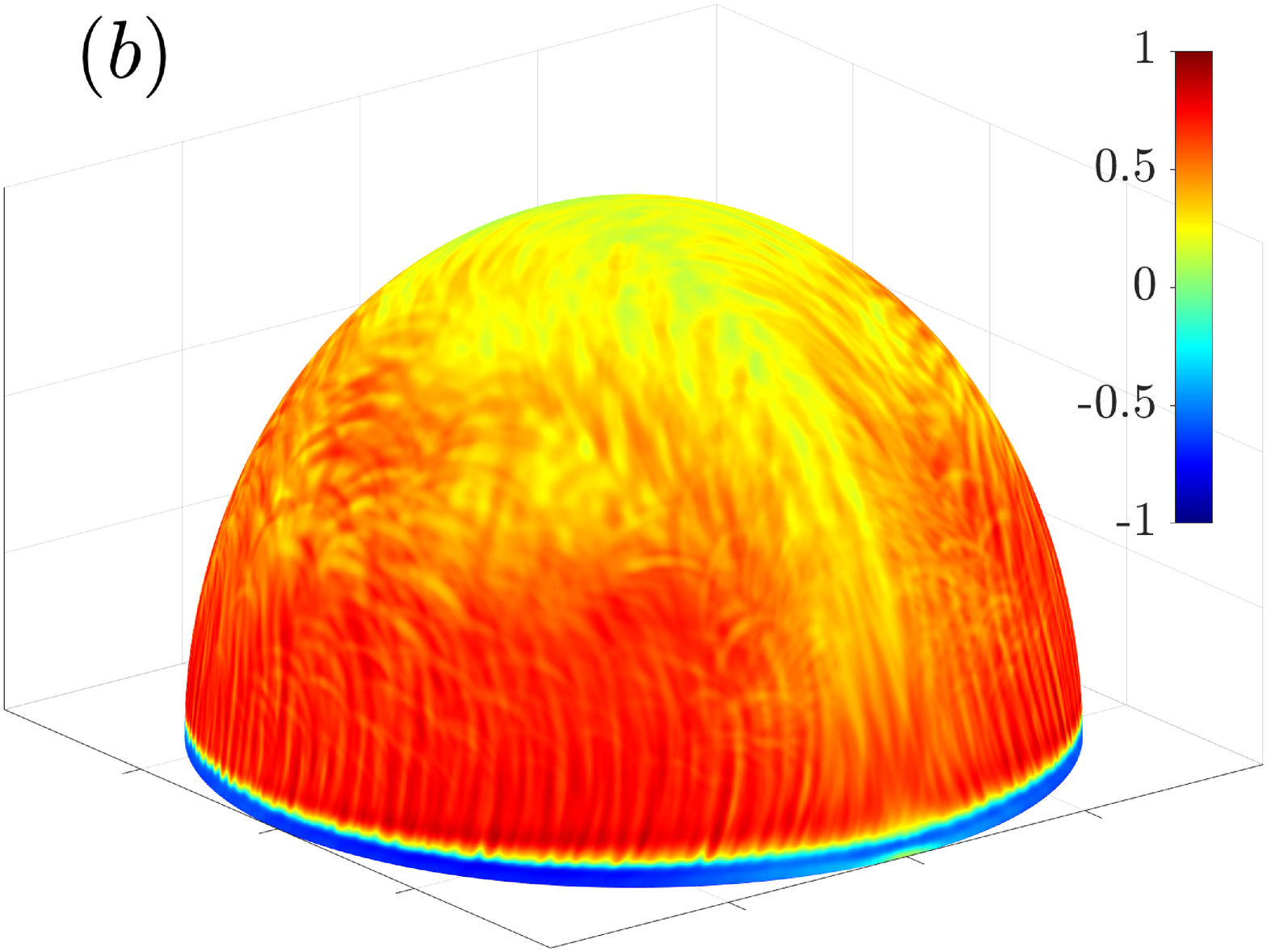}
	\includegraphics[width = 0.24\textwidth]{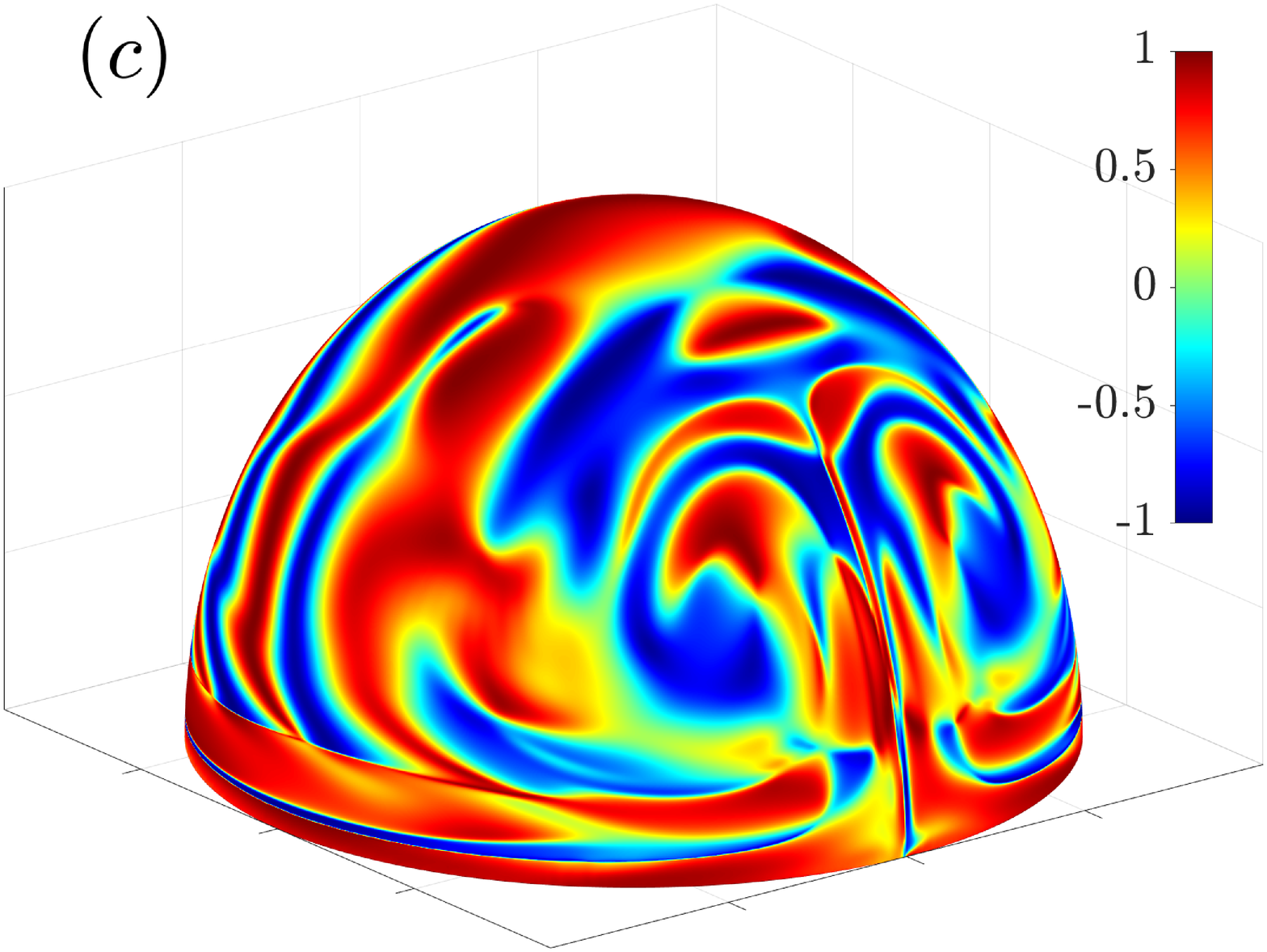}
	\includegraphics[width = 0.24\textwidth]{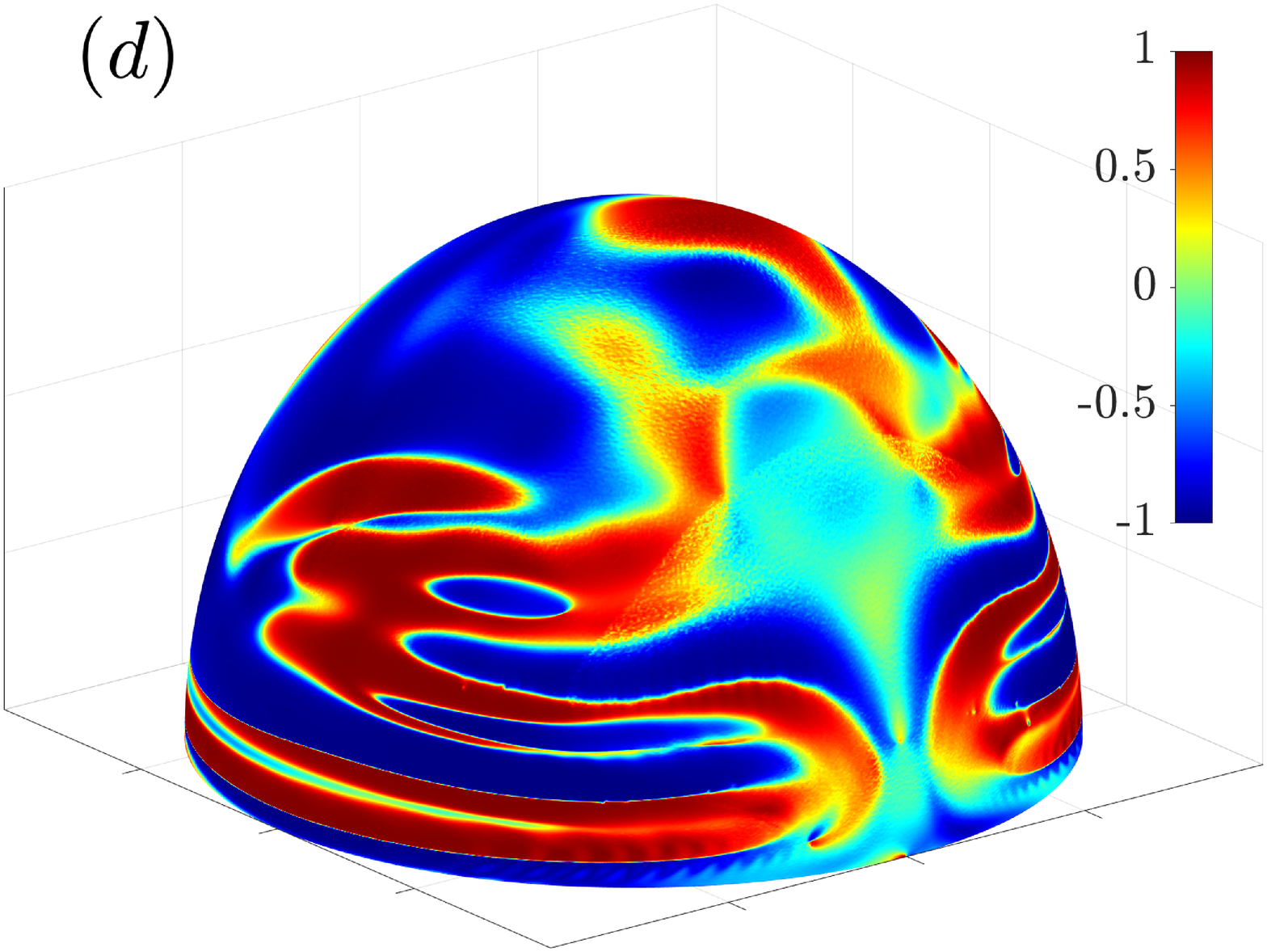}
	\includegraphics[width = 0.24\textwidth]{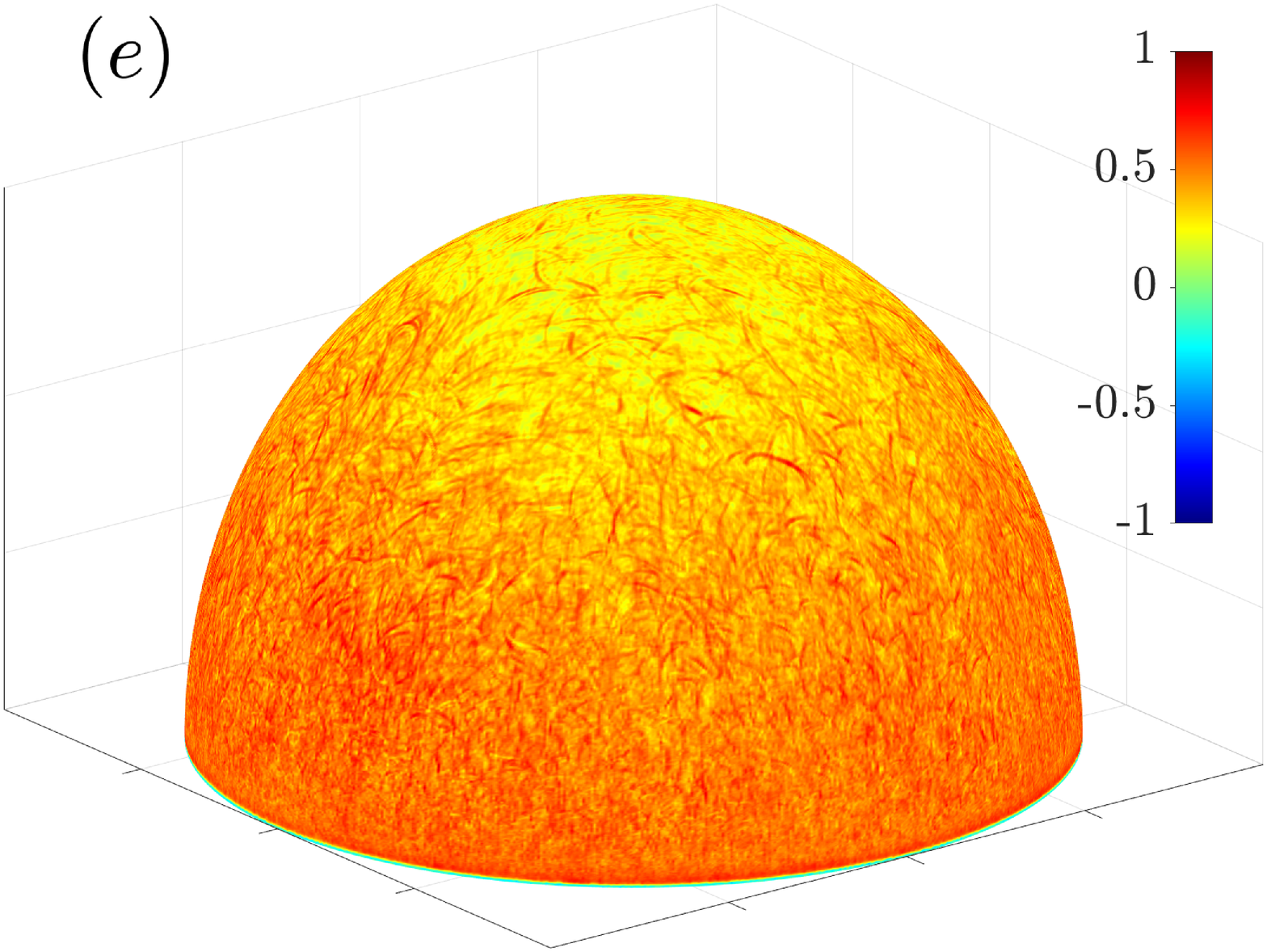}
	\includegraphics[width = 0.24\textwidth]{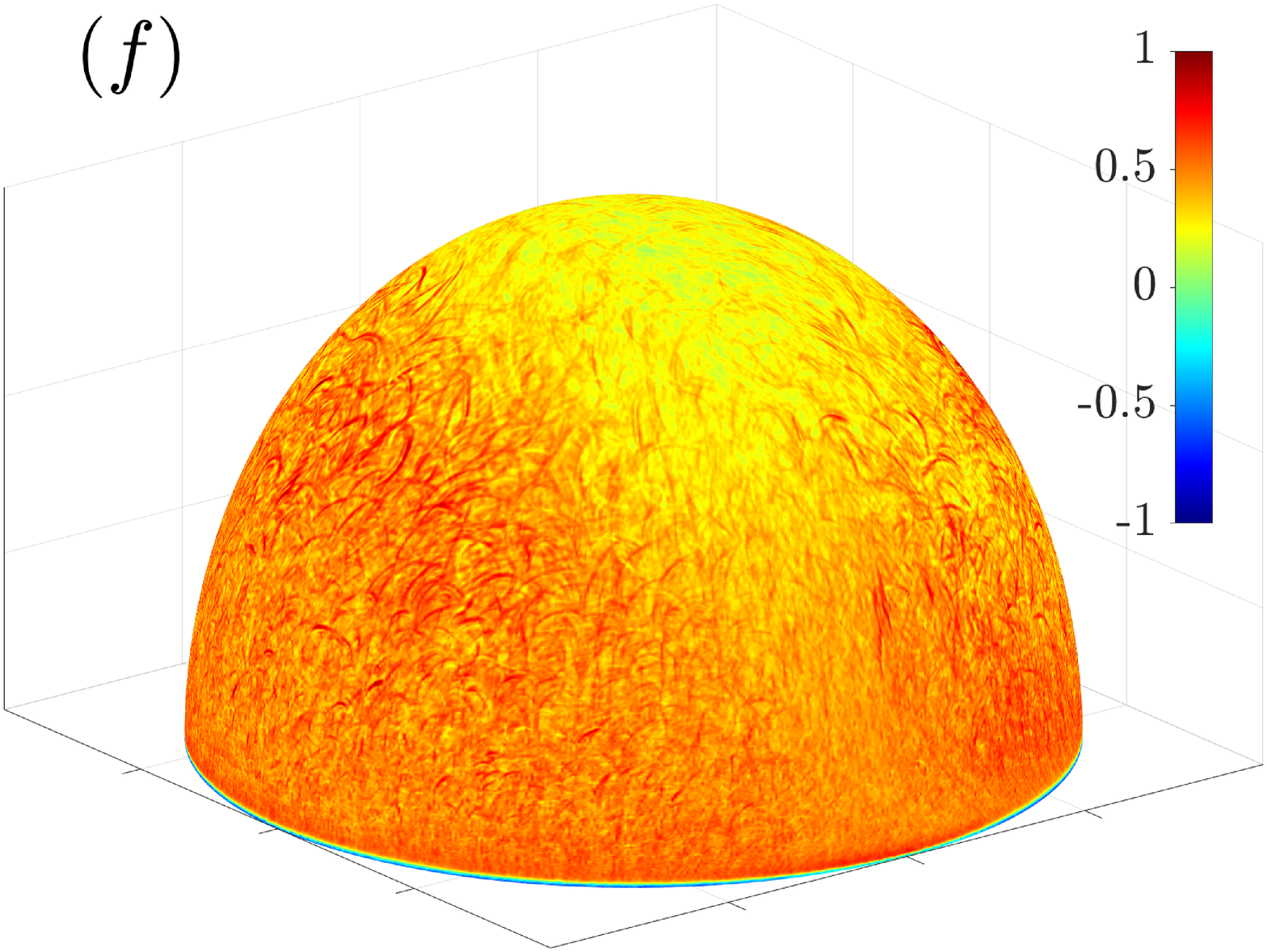}
	\includegraphics[width = 0.24\textwidth]{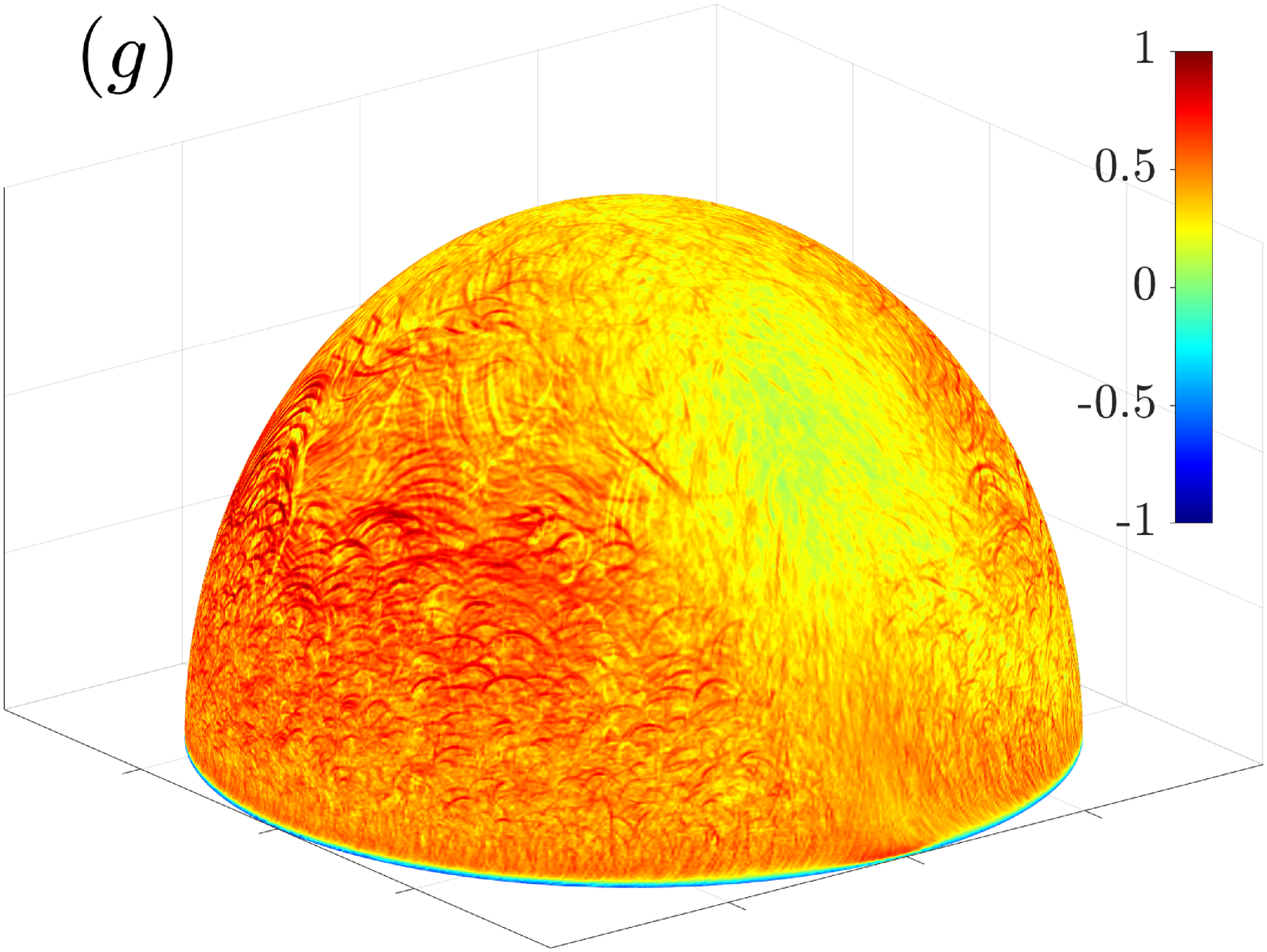}
	\includegraphics[width = 0.24\textwidth]{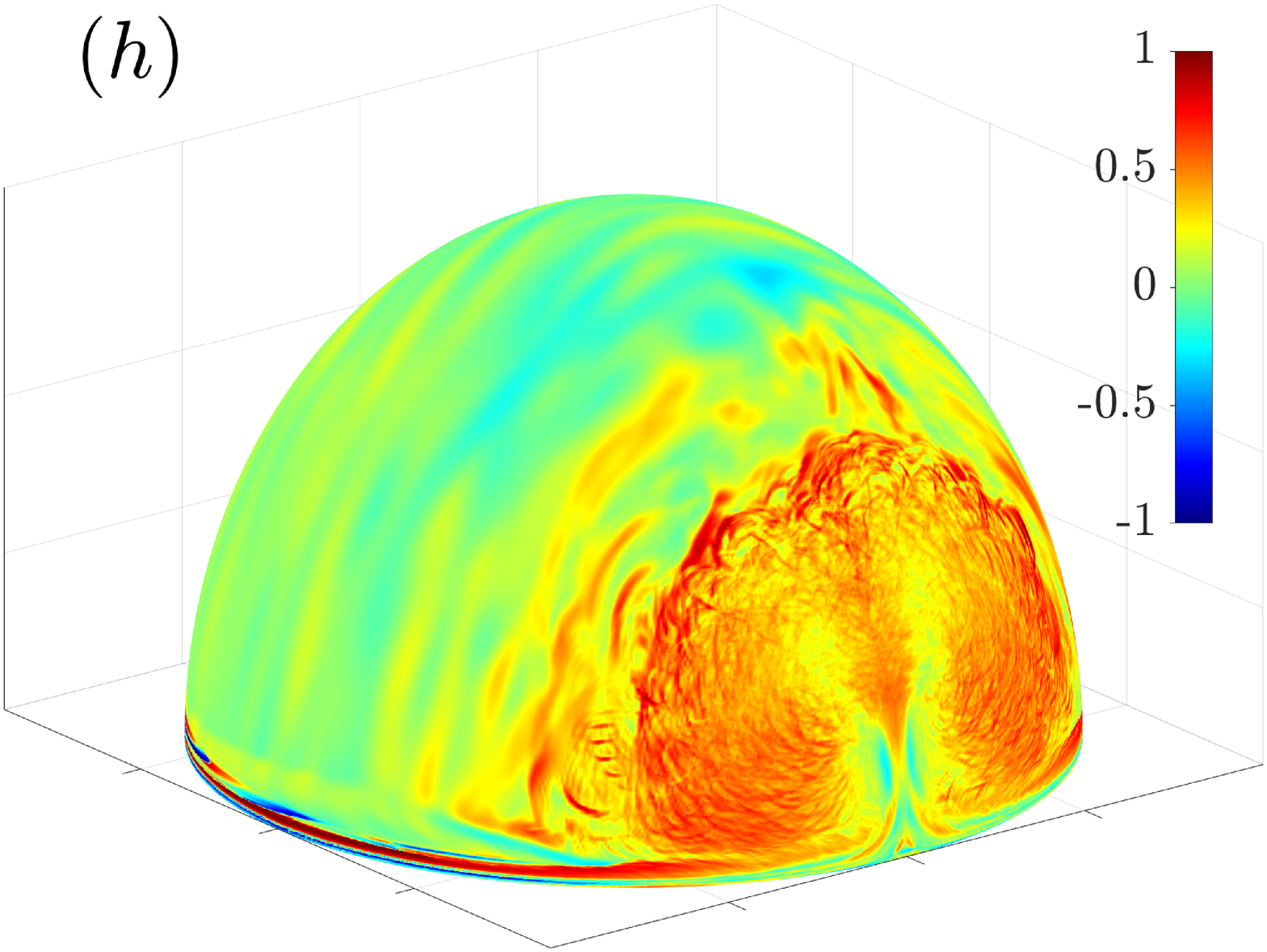}
	\caption{The distribution of $\langle c\rangle$ on the bubble
    for $Ra=3\times10^6$(the upper line: $(a)$ to $(d)$) and 
    $Ra=3\times10^9$(the lower line: $(e)$ to $(h)$) 
    with $\delta=0^{\circ},30^{\circ},60^{\circ},90^{\circ}$(from left to right).} 
	\label{fig:correlation}
\end{figure*}
\begin{figure}
	\includegraphics[width = 0.6\textwidth]{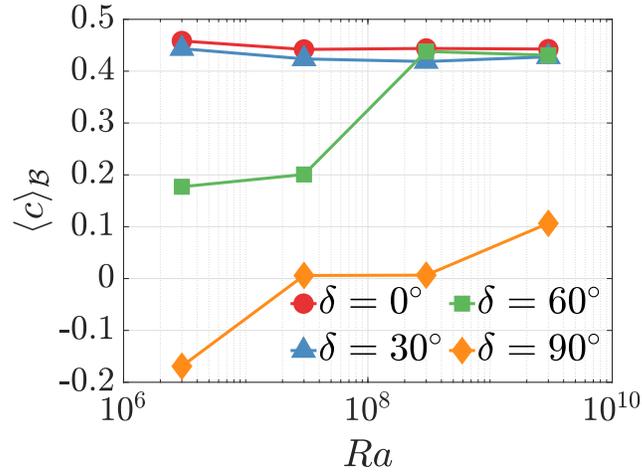}
	\caption{The global correlation coefficients corresponding to different $Ra$ and $\delta$}
	\label{fig:globalCorrelation}
\end{figure}

Figure \ref{fig:correlation} illustrates how $\langle c\rangle$ varies across the surface of the bubble as $\delta$ is varied and the flow transitions between the DPR and SPR.
For the DPR, $\langle c\rangle$ is positive over most of the bubble, and over a considerable part of the surface the correlation is quite high. For the SPR, significant regions of the bubble 
have $\langle c\rangle<0$ when $Ra=3\times 10^6$, indicating significant regions of negative correlation between $\epsilon_{T}$ and $\epsilon_{u}$. However, for $Ra=3\times 10^9$, when $\delta=90^\circ$ and the flow is in the SPR, there is still, however, a significant positive correlation near the lower edge of the bubble where vigorous turbulence still exists. This differing behavior is probably due to the fact that while the SPR for $Ra=3\times 10^9$ is still vigorously turbulent near the lower edge of the bubble, for $Ra=3\times 10^6$ the flow is almost laminar.

It is seen that the SPR are characterized by the filamented and convoluted patches of high $\langle c\rangle$.
The patches are more filamented and convoluted for higher $Ra$.
For the DPR($(a)$, $(b)$, $(e)$, $(f)$ and $(g)$ in figure \ref{fig:correlation}), 
$\langle c\rangle$ decreases in the domain where the stable plume occupies with $\delta$ increasing. 
For the SPR($(c)$, $(d)$ and $(h)$ in figure \ref{fig:correlation}), 
the distribution of $\langle c\rangle$ is more complex.
For relative small $Ra$($(c)$, $(d)$ in figure \ref{fig:correlation}), the dynamic plume disappears on the bubble and there is only the stable plume on the bubble.
Thus the patches of high or low $\langle c\rangle$ have large size and cover the whole surface of the bubble.
But when $Ra$ is enough high($(h)$ in figure \ref{fig:correlation}), the stable and dynamic plumes coexist on the bubble.
$\langle c\rangle$ on the higher edge of the bubble is close to $0$.
In the region near the stable plume, the patch of high $\langle c\rangle$ become filamented and convoluted as in the DPR.

Figure \ref{fig:globalCorrelation} shows the globally averaged correlation coefficient $\langle c\rangle_{\mathcal{B}}$ corresponding to all the cases in table \ref{tab:case}. Here,
the global averaging operator is defined for an arbitrary field variable ${a}(\boldsymbol{x},t)$ as
\begin{equation}
    \langle{a}\rangle_{\mathcal{B}} = \frac{\oint_{\mathcal{B}} {a}(\boldsymbol{x},t)\mathrm{d}s}{2\pi R},
\end{equation}
where $\mathrm{d}s$ is the elemental area on the bubble surface $\mathcal{B}$.

For the DPR, $\langle c\rangle_{\mathcal{B}}$ is almost independent of $Ra$ and approximately equal to $0.42$.
This value validates the observation based on the instantaneous flow field that the thermal and kinetic energy dissipation rates
should be correlated to each other since they are both driven by plumes in the flow.
It is also interesting to note that a similar value of $\langle c\rangle_{\mathcal{B}}\approx 0.4$ was obtained 
for Rayleigh-B\'enard convection\cite{ZhangZhouSun2017JFM}.
Once the flow transitions from the DPR to SPR, the magnitude of $\langle c\rangle_{\mathcal{B}}$ reduces significantly,
with $|\langle c\rangle_{\mathcal{B}}|\lesssim 0.2$ in the SPR.

\subsection{Nusselt $Nu$ and Reynold number $Re$}
We now turn to consider the scaling relation of the Nusselt number $Nu$ and Reynolds number $Re$ versus $Ra$ in the DPR and SPR.
For the bubble flow, $Nu$ is defined differently from that in RBC. 
In RBC, thermal energy passes through the layer of fluid and
$Nu$ is defined as the non-dimensional heat flux through the fluid layer in order to quantify the efficiency of the heat transport.
By contrast, in the bubble flow, heat is absorbed by the fluid at the equator and the thermal energy is dissipated entirely within the flow, with no cold boundary through which it can pass.
For the bubble flow, we are therefore interested in the efficiency of the heat transport away from the equator and so $Nu$ is defined as the non-dimensional heat flux across the equator\cite{He2021}
\begin{equation}
    Nu = \frac{Q_{turb}}{Q_0},
    \label{eq:Nu}
\end{equation}
where $Q_{turb}$ is the heat flux at the equator for the turbulent flow and $Q_0$ is the ideal heat flux associated with pure conduction at the equator.
The quantity $Q_{turb}$ is obtained by the temperature field as
\begin{equation}
  Q_{turb} = -\langle\nabla_z T|_{\theta=\pi/2}\rangle.
\end{equation}
The ideal heat flux of pure conduction $Q_0$ is the heat flux in the hypothesis that the fluid is motionless all over the bubble:
\begin{equation}
  Q_0 = \nabla_z T_0|_{\theta=\pi/2},
\end{equation}
where $T_0$ is obtained as the solution to \eqref{eq:T} using $\boldsymbol{u}=\boldsymbol{0}, S=0$ and boundary conditions $T|_{\theta=\pi/2}=1$ and $T|_{\theta=0}=0$.

For evaluating $Re$, the root mean square (r.m.s) velocity $u_{rms}= \sqrt{\langle\|\boldsymbol{u}\|^2\rangle_{\mathcal{B}}}$ 
is usually used as a global measure of the turbulent velocity scale in studies of RBC\cite{ZhangZhouSun2017JFM,SunXia-155},
and using this gives
\begin{equation}
	Re = \sqrt{\frac{Ra}{Pr}}u_{rms}.
	\label{eq:Re}
\end{equation}

\begin{figure}
	\centering
	\includegraphics[width = 0.6\textwidth]{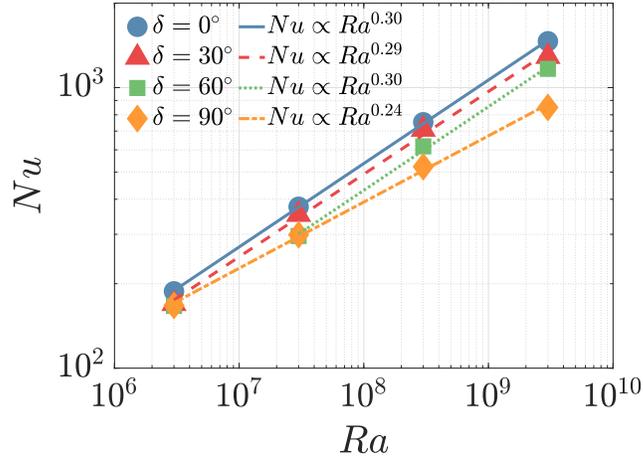}
	\caption{The variation of $Nu$ with $Ra$ and $\delta$}
	\label{fig:Nu}
\end{figure}
\begin{figure}
	\centering
	\includegraphics[width = 0.6\textwidth]{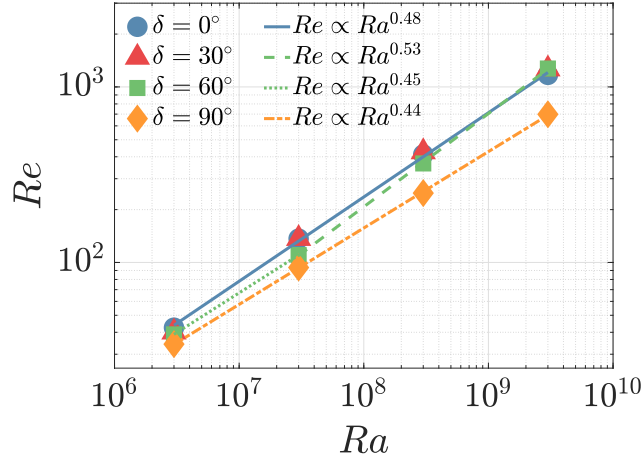}
	\caption{The variation of $Re$ with $Ra$ and $\delta$}
	\label{fig:Re}
\end{figure} 
Figures \ref{fig:Nu} and \ref{fig:Re} show $Nu$ and $Re$ as a function of $Ra$ for different $\delta$,
with power-law fits of the data illustrated by solid or dash lines.
For $\delta=0^\circ$, the scaling relations of $Nu$ and $Re$ are $Nu\propto Ra^{0.30}$ and 
$Re\propto Ra^{0.48}$ respectively, which match those reported by previous studies of RBC\cite{SunXia-155,XuXi2019POF}.
As $\delta$ is increased, we observe that the scaling behaviours are strongly influenced by the flow regime.
When the flow is in the DPR, $Nu$ scales with $Ra$ as a power-law form with scaling exponent close to $0.30$.
As the flow transitions to the SPR, the scaling exponent for $Nu$ decreases from $0.30$ to $0.24$.
Moreover, there is a strong reduction in the actual values of $Nu$ when the flow transitions from the DPR to the SPR, especially for higher $Ra$.

Concerning $Re$, for $\delta=30^{\circ}$ the dependence of $Re$ on $Ra$ is almost identical to that for the case with $\delta=0^\circ$ 
For $\delta=60^{\circ}$, the scaling relation turns into $Re\propto Ra^{0.45}$ for the DPR and 
$Re\propto Ra^{0.53}$ for the SPR.
The data for $\delta=90^{\circ}$ can be described by a single power law $Re\propto Ra^{0.44}$ since all cases are in the SPR. 
There is also a considerable drop in the magnitude of $Re$ when the flow transitions from the DPR to the SPR.
These results show that there is a clear quantitative effect of $\delta$ on both $Nu$ and $Re$ and their dependence on $Ra$, which corresponds
to the transition the flow undergoes when moving from the DPR to the SPR as $\delta$ is increased.

\subsection{Probability density functions (PDFs) of $\epsilon_{T^{\prime}}$ and $\epsilon_{u^{\prime}}$}
\begin{figure*}
	\centering
	\includegraphics[width = 0.45\textwidth]{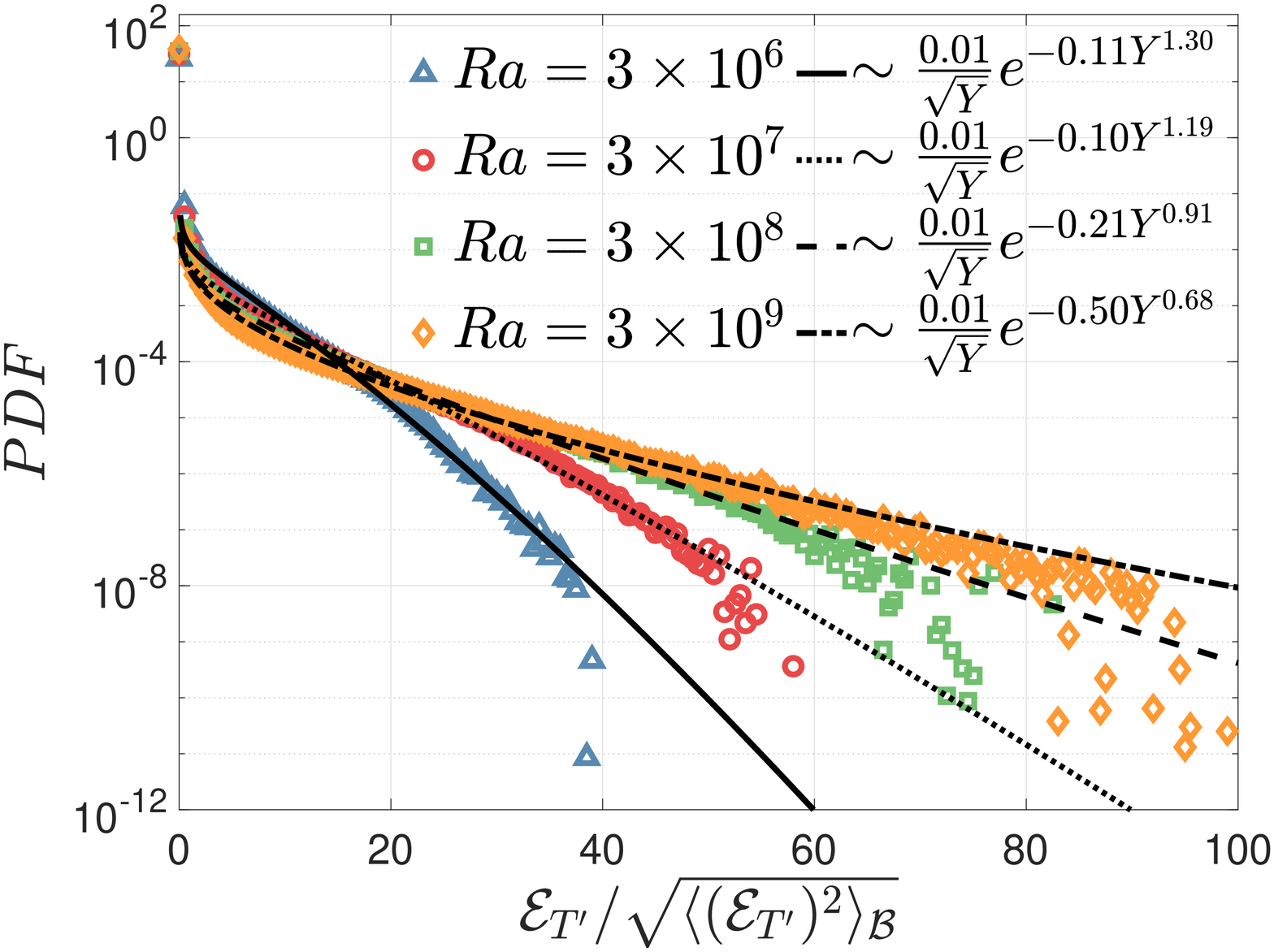}
	\includegraphics[width = 0.45\textwidth]{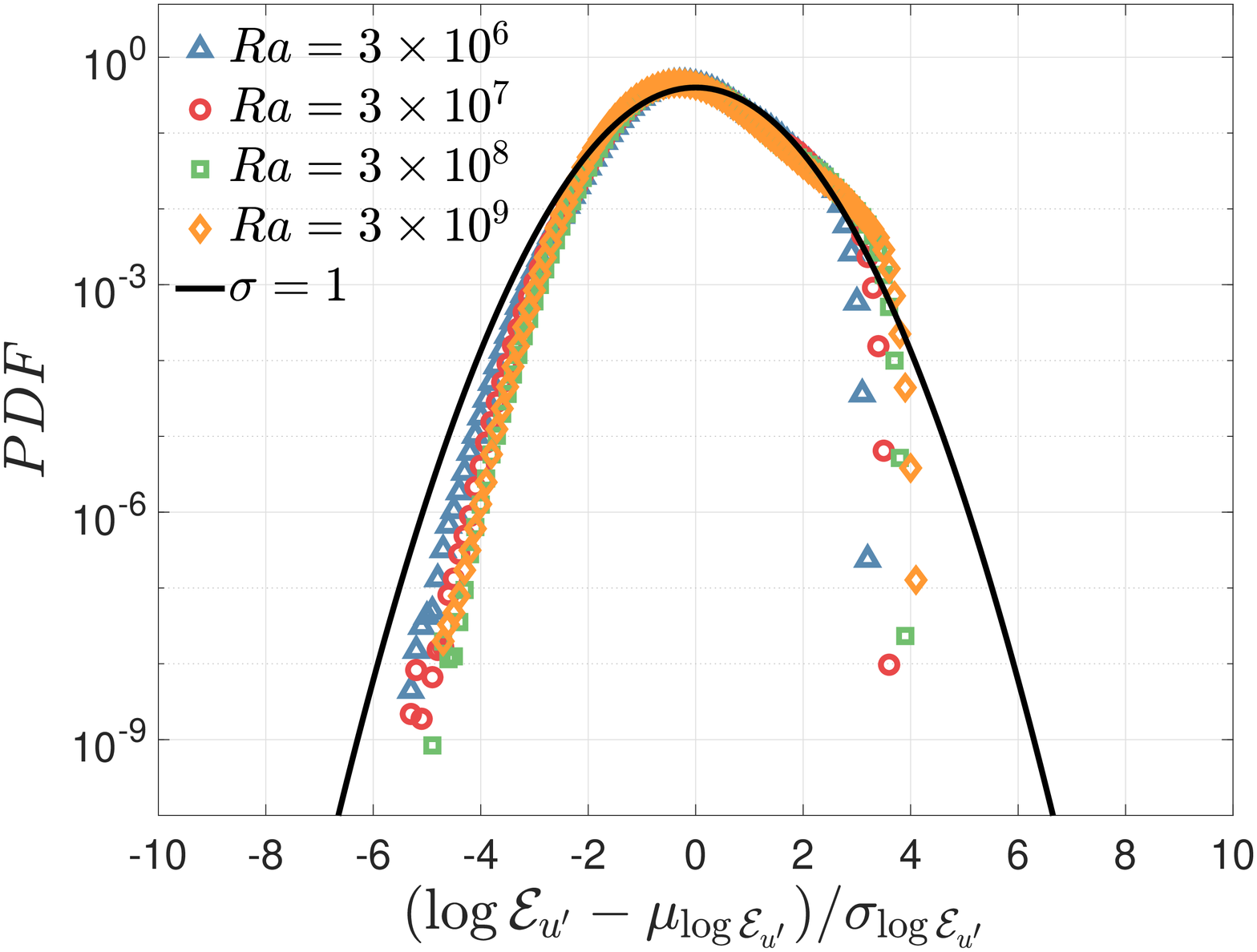}
	\includegraphics[width = 0.45\textwidth]{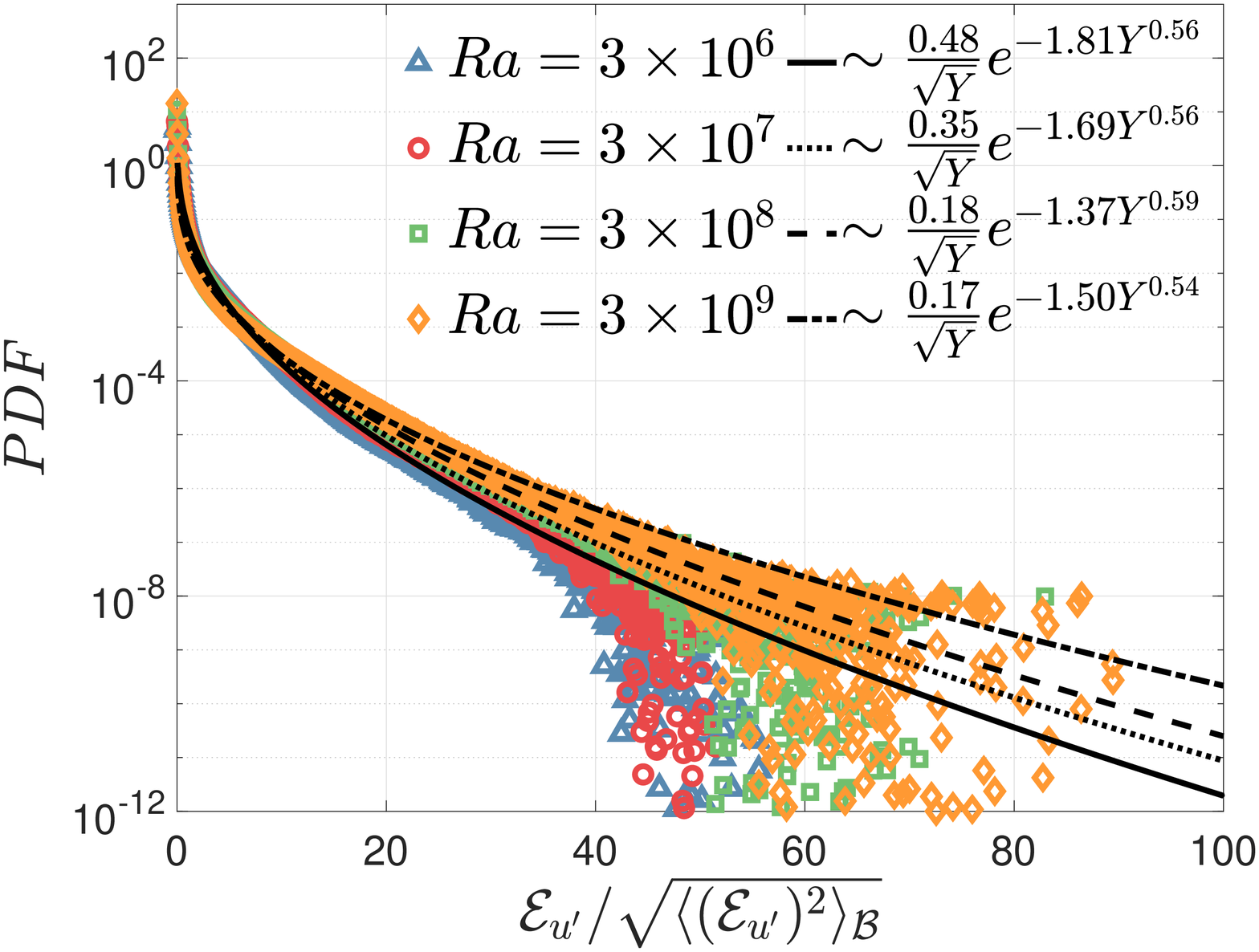}
	\includegraphics[width = 0.45\textwidth]{pic/fig_PDF_logDispTher_VsRa.eps}
	\caption{the PDFs of $\epsilon_{T^{\prime}}$ and $\epsilon_{u^{\prime}}$ with different $Ra$ for $\delta=0^{\circ}$ }
	\label{fig:pdfDispTherKinVsRa}
\end{figure*}

We now turn to consider the statistical characteristics of the turbulent thermal energy dissipation rate $\epsilon_{T^{\prime}}$ and 
kinetic energy dissipation rate $\epsilon_{u^{\prime}}$, which are defined as
\begin{align}
	\epsilon_{T'} &= \epsilon_{T}-\epsilon_{\langle T\rangle},\\
	\epsilon_{T} &= {\kappa}\|\boldsymbol{\nabla}{T}\|^2,\\
	\epsilon_{\langle T\rangle} &= {\kappa}\|\boldsymbol{\nabla}{\langle T\rangle}\|^2,	
\end{align}
and
\begin{align}
	\epsilon_{u^\prime} &= \epsilon_{u}-\epsilon_{\langle u\rangle},\\
	\epsilon_{u} &= \frac{1}{2}{\nu}\|\boldsymbol{\nabla}{\boldsymbol{u}} +\boldsymbol{\nabla}{\boldsymbol{u}}^\top \|^2,\\
	\epsilon_{\langle u\rangle} &=\frac{1}{2}{\nu}\|\langle\boldsymbol{\nabla}{\boldsymbol{u}} +\boldsymbol{\nabla}{\boldsymbol{u}}^\top \rangle\|^2,
\end{align}
Figure \ref{fig:pdfDispTherKinVsRa}
show the PDFs of $\epsilon_{T^{\prime}}$ and $\epsilon_{u^{\prime}}$
for different $Ra$ with $\delta=0^{\circ}$.
As is common in  RBC studies\cite{EmranSchumacher2008JFM,ZhangZhouSun2017JFM,XuXi2019POF},
$\epsilon_{T^{\prime}}$ and $\epsilon_{u^{\prime}}$ are normalized by their global root-mean-square (r.m.s) which are
$\sqrt{\langle\epsilon_{T^{\prime}}\rangle}_{\mathcal{B}}$ and $\sqrt{\langle\epsilon_{u^{\prime}}\rangle}_{\mathcal{B}}$, respectively.
We also plot $\log(\epsilon_{T^{\prime}})$ and $\log(\epsilon_{u^{\prime}})$ with their local mean values $\mu_{\log\epsilon_{T^{\prime}}}=\langle\log\epsilon_{T^{\prime}}\rangle$, 
$\mu_{\log\epsilon_{u^{\prime}}}=\langle\log\epsilon_{u^{\prime}}\rangle$ subtracted, and normalized by their local standard deviations $\sigma_{\epsilon_{T^{\prime}}}=\sqrt{\langle(\log\epsilon_{T^{\prime}}-\mu_{\log\epsilon_{T^{\prime}}})^2\rangle}$, 
$\sigma_{\epsilon_{u^{\prime}}}=\sqrt{\langle(\log\epsilon_{u^{\prime}}-\mu_{\log\epsilon_{u^{\prime}}})^2\rangle}$, in order to consider how close the random variables are to being log-Normally distributed. In the figures for the logarithmic variables, the solid lines show a standardized Gaussian PDF for the reference.
%

The results show that the PDFs of $\epsilon_{T^{\prime}}$ and $\epsilon_{u^{\prime}}$ have increasingly wider tails 
as $Ra$ is increased. This indicates increasing small-scale intermittency in the fields $\|\boldsymbol{\nabla}{T}\|^2$ 
and $\|\langle\frac{1}{2}\left( \boldsymbol{\nabla}{\boldsymbol{u}} +\boldsymbol{\nabla}{\boldsymbol{u}}^\top \right)\rangle\|^2$ that
occurs as an increase in $Ra$ leads to an increase in $Re$.
The presence of intermittency is also clearly seen in the PDFs of logarithmic variables,
which show that the PDFs of these logarithmic variables clearly depart from a Gaussian PDF.

For two dimensional turbulent convection with large $Pr$, Chertkov et al.\cite{Chertkov1998} showed analytically
that the PDF for the gradients of a passive scalar field can be described by stretched exponential functions 
\begin{equation}
	PDF(Y) = \frac{c}{\sqrt{Y}}e^{\left(-m Y^{\alpha}\right)},
	\label{eq:fitShape}
\end{equation}
where the sample-space variable $Y$ is conjugate to the random variable (gradient of scalar) normalized by its modal value, $c$, $m$ and $\alpha$ are fitting parameters, and
$\alpha$ is deduced to be $1/3$ for a passive scalar.
Figure \ref{fig:pdfDispTherKinVsRa} shows that with appropriate choices for $c$, $m$ and $\alpha$, the PDFs of $\epsilon_{T^{\prime}}$ and $\epsilon_{u^{\prime}}$ can also be well described
by such stretched exponential functions (illustrated by the black lines in the figure), with some deviations in the far tails of the PDFs.
The fitting exponent $\alpha$ for $\epsilon_{T^{\prime}}$ decreases from $1.30$ to $0.68$ as $Ra$ is increased.
By contrast, $\alpha$ for $\epsilon_{u^{\prime}}$ has only a slight dependence on $Ra$.

These results share much in common with those acquired from standard RBC\cite{EmranSchumacher2008JFM,ZhangZhouSun2017JFM,XuXi2019POF,HeTongXia2007PRL,HeTong2009PRE}, but there are also some differences. 
He et al.\cite{HeTongXia2007PRL,HeTong2009PRE} measured the local thermal energy dissipation rate
in RBC at the cell center and close to the vertical wall by means of experiments.
They found that the PDFs of $\epsilon_{T^{\prime}}$, scaled by its local r.m.s value $\sqrt{\langle\epsilon^{2}_{T^{\prime}}\rangle}$, 
are well described by stretched exponential functions.
Moreover, regardless of $Ra$, they found $\alpha=0.35$ in the cell center and $\alpha=0.44$ close to the vertical wall, values
which are smaller than those we find for the bubble flow. 
It should also be noted that $Ra$ in the experiments of He et al.\cite{HeTongXia2007PRL,HeTong2009PRE} ranges from $1.7\times10^{9}$ to
$8.2\times10^9$ covering one order. 
These include larger values of $Ra$ than our study, and so some of the differences in the measured $\alpha$ may be due to different $Ra$, as well as the fundamental differences between the canonical RBC they considered, and
the convective bubble flow we are considering. Numerous DNS and experimental studies of RBC have also found that the PDFs of the dissipation rates are well described by stretched exponential functions\cite{EmranSchumacher2008JFM,kaczorowski2009,ZhangZhouSun2017JFM,XuXi2019POF}. The values they find for the fitting parameters do vary somewhat between the studies, which may be due to differences in the RBC geometry, the values of $Ra, Pr$ explored, as well as the approach used to perform the averaging operations when constructing the statistics.

\begin{figure*}
	\centering
	\includegraphics[width = 0.45\textwidth]{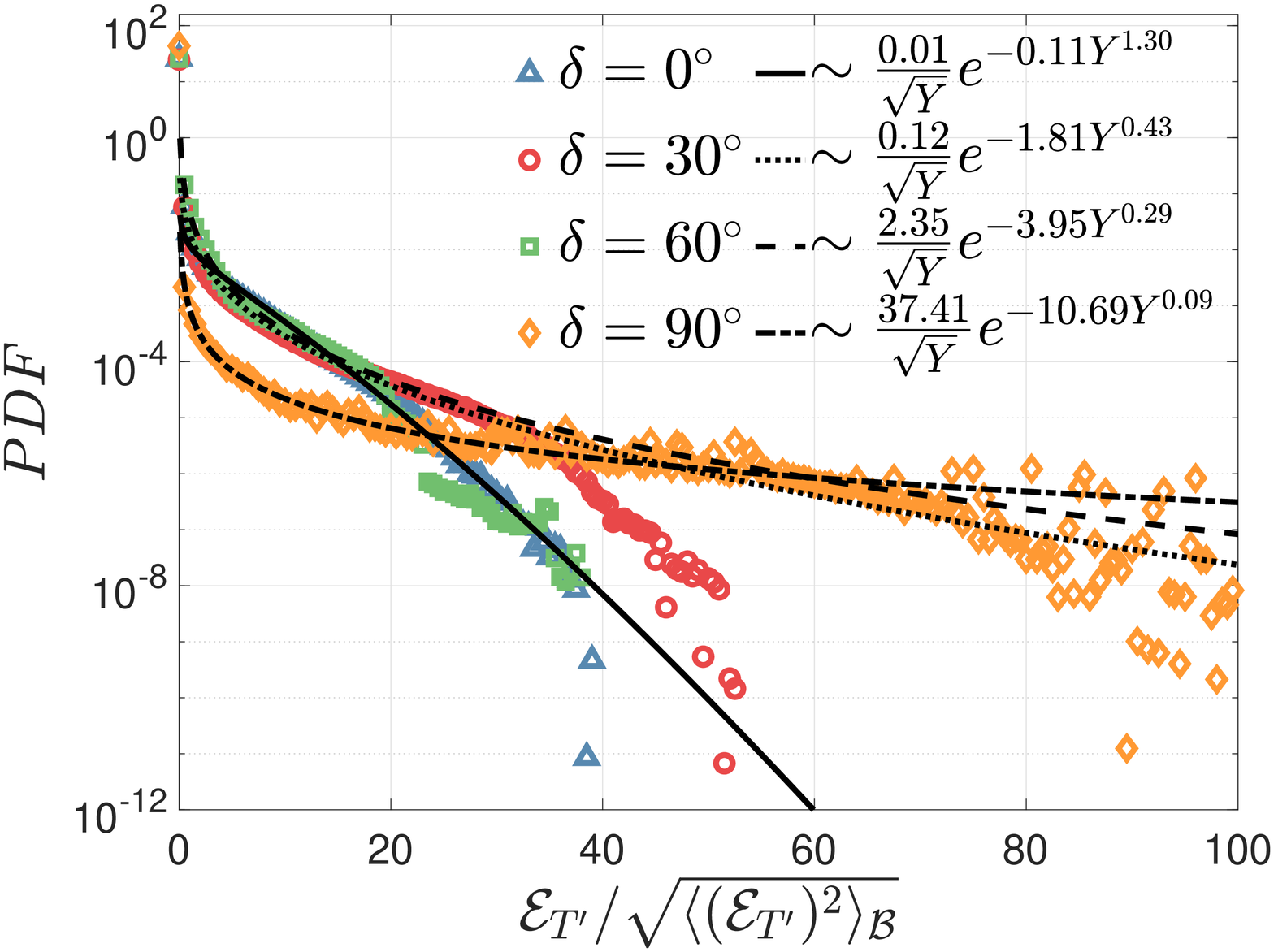}
	\includegraphics[width = 0.45\textwidth]{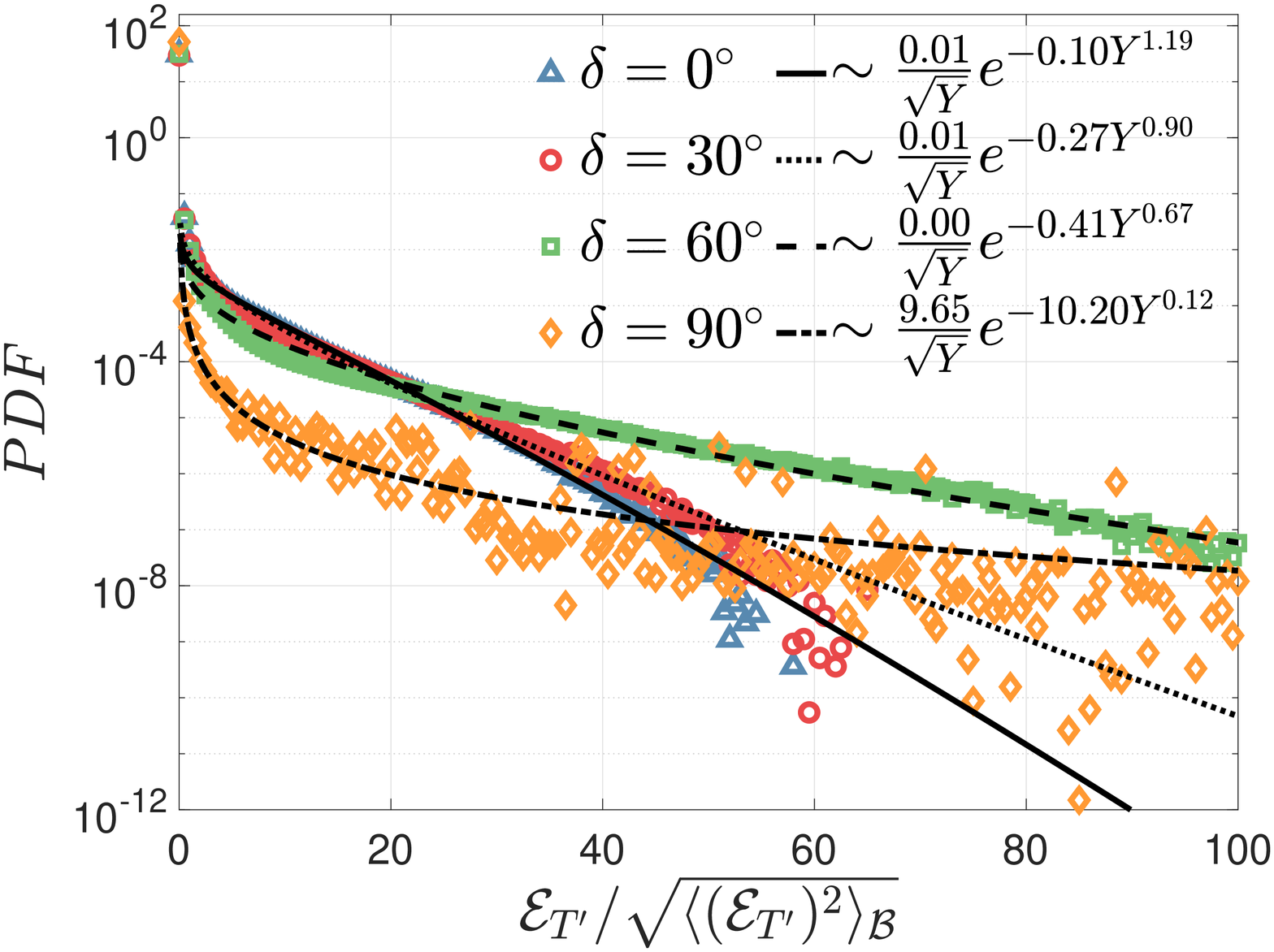}
	\includegraphics[width = 0.45\textwidth]{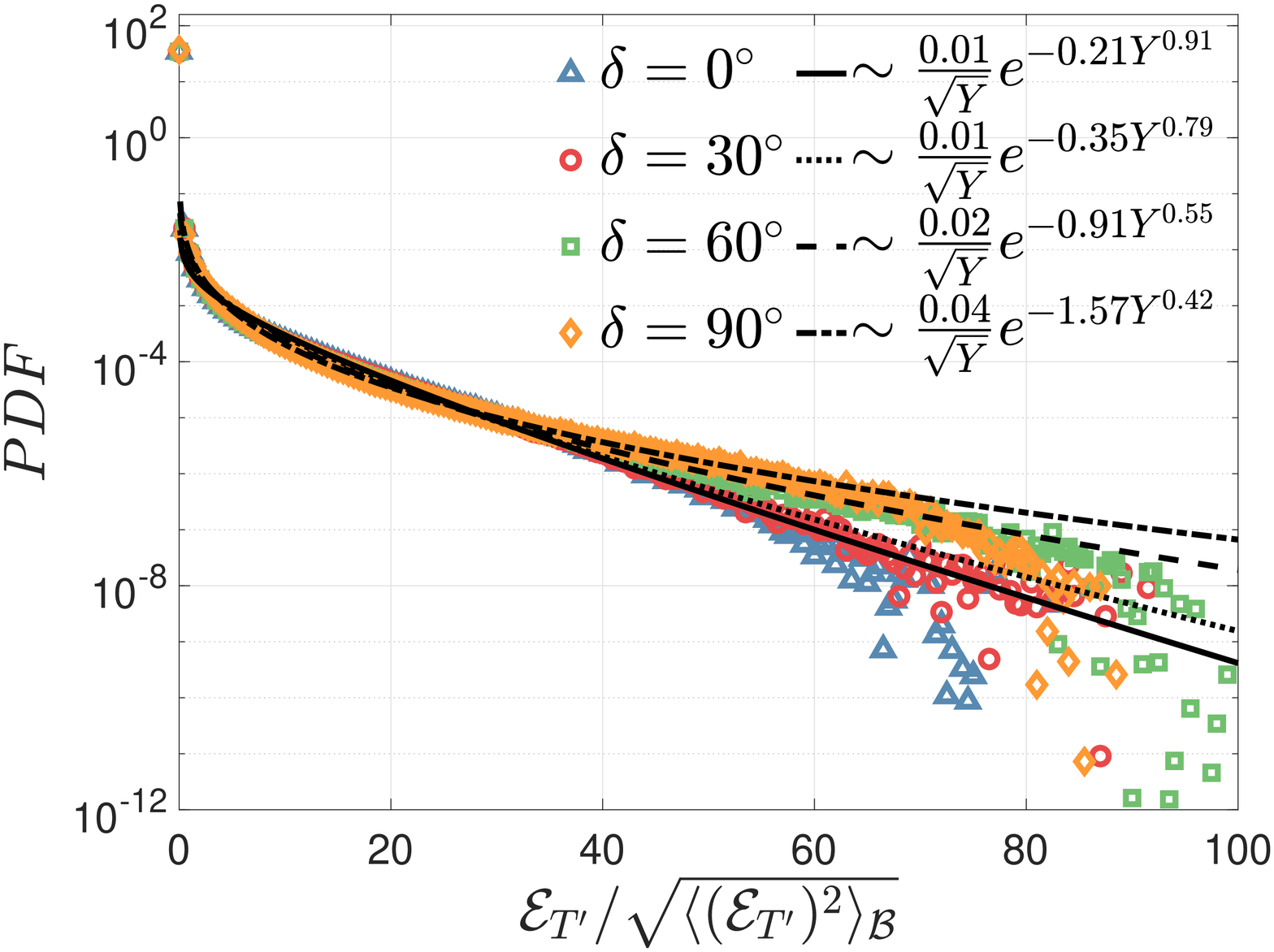}
	\includegraphics[width = 0.45\textwidth]{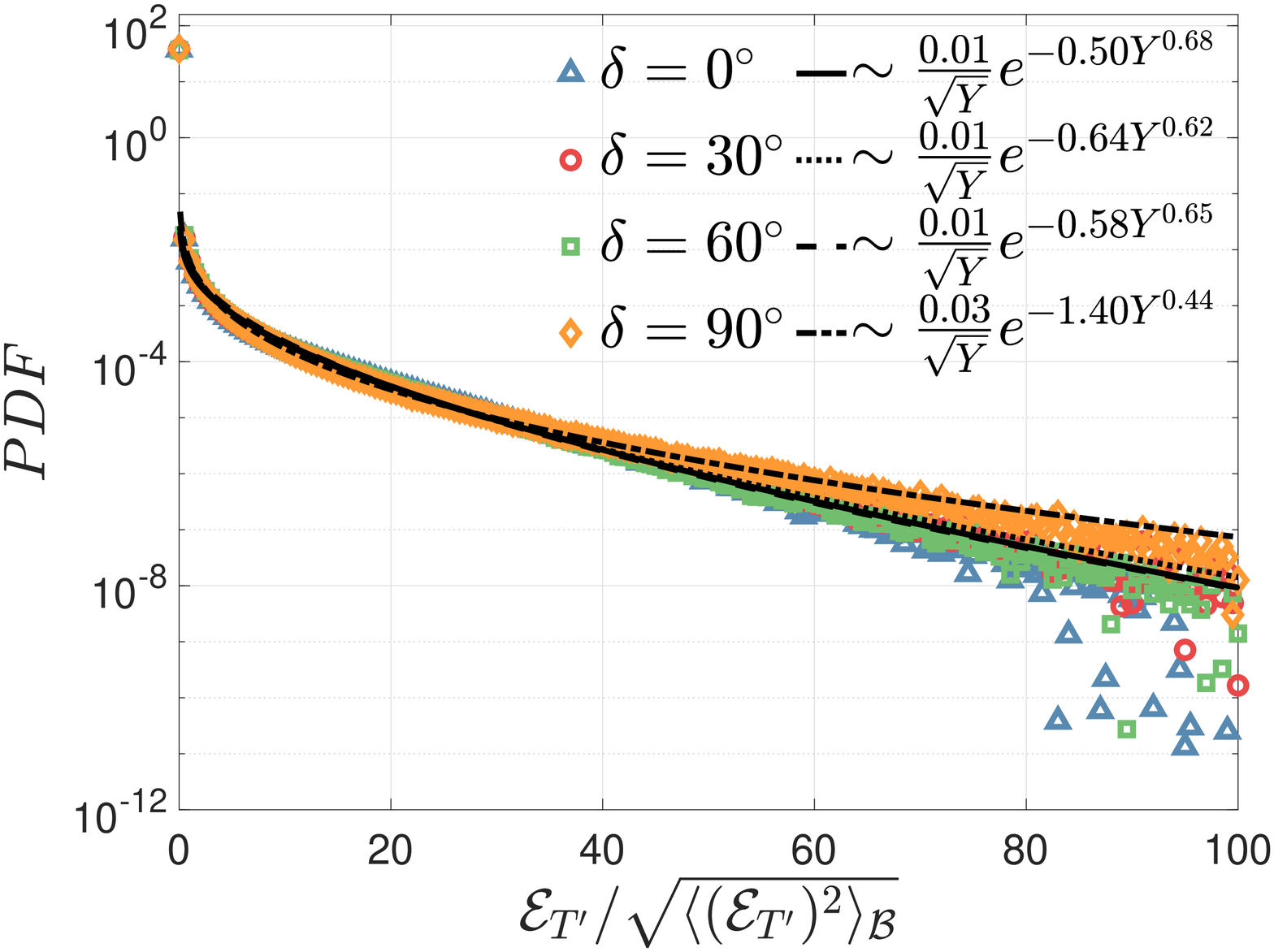}
    \caption{The PDFs of $\epsilon_{T^{\prime}}$ with different $\delta$ and $Ra$.
    The first line, from left to right: $Ra=3\times10^6$, $Ra=3\times10^7$.
    The second line, from left to right: $Ra=3\times10^8$, $Ra=3\times10^9$.}
	\label{fig:pdfEpsilonTVsDelta}
\end{figure*}
%

\begin{figure*}
    \centering
	\includegraphics[width = 0.45\textwidth]{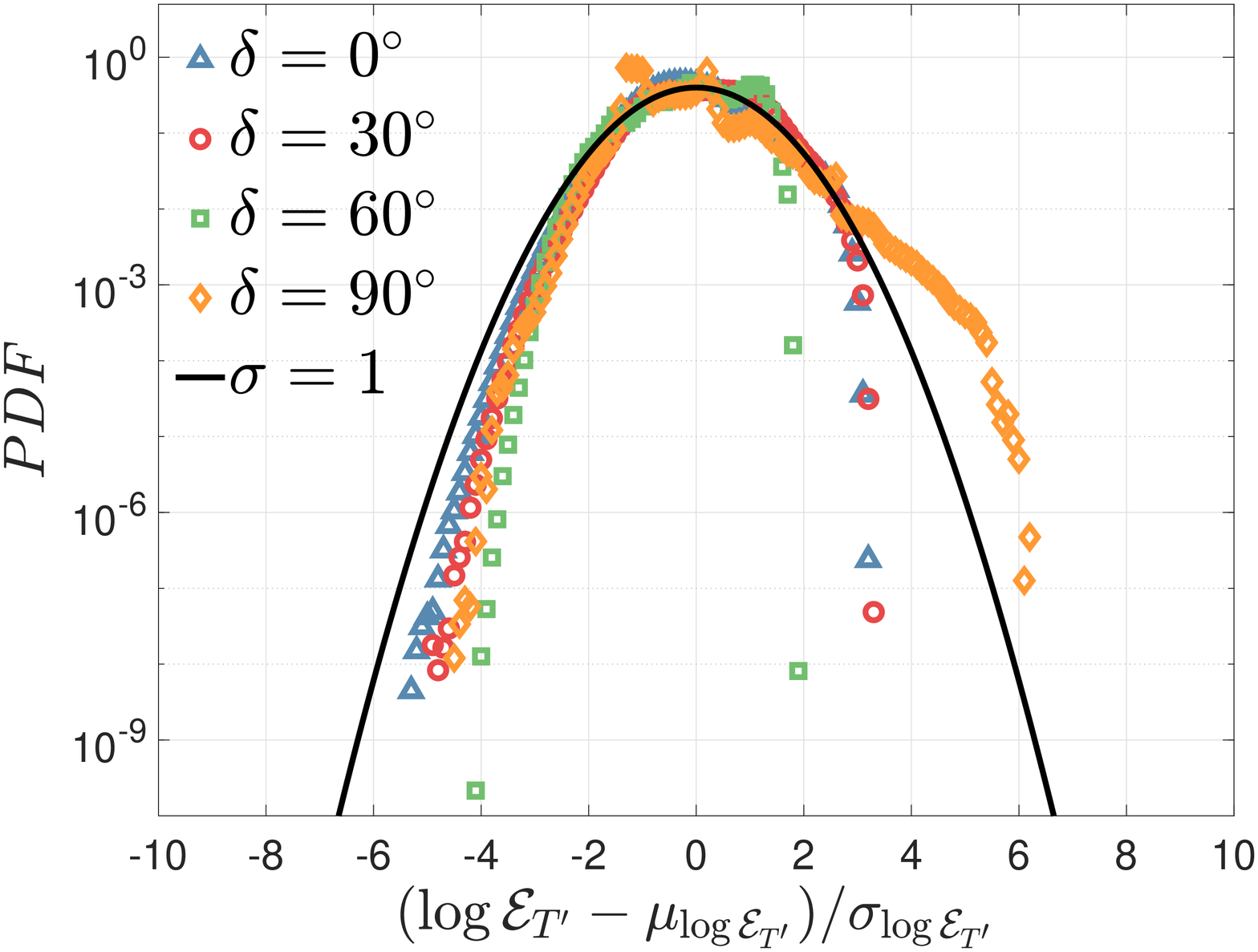}
	\includegraphics[width = 0.45\textwidth]{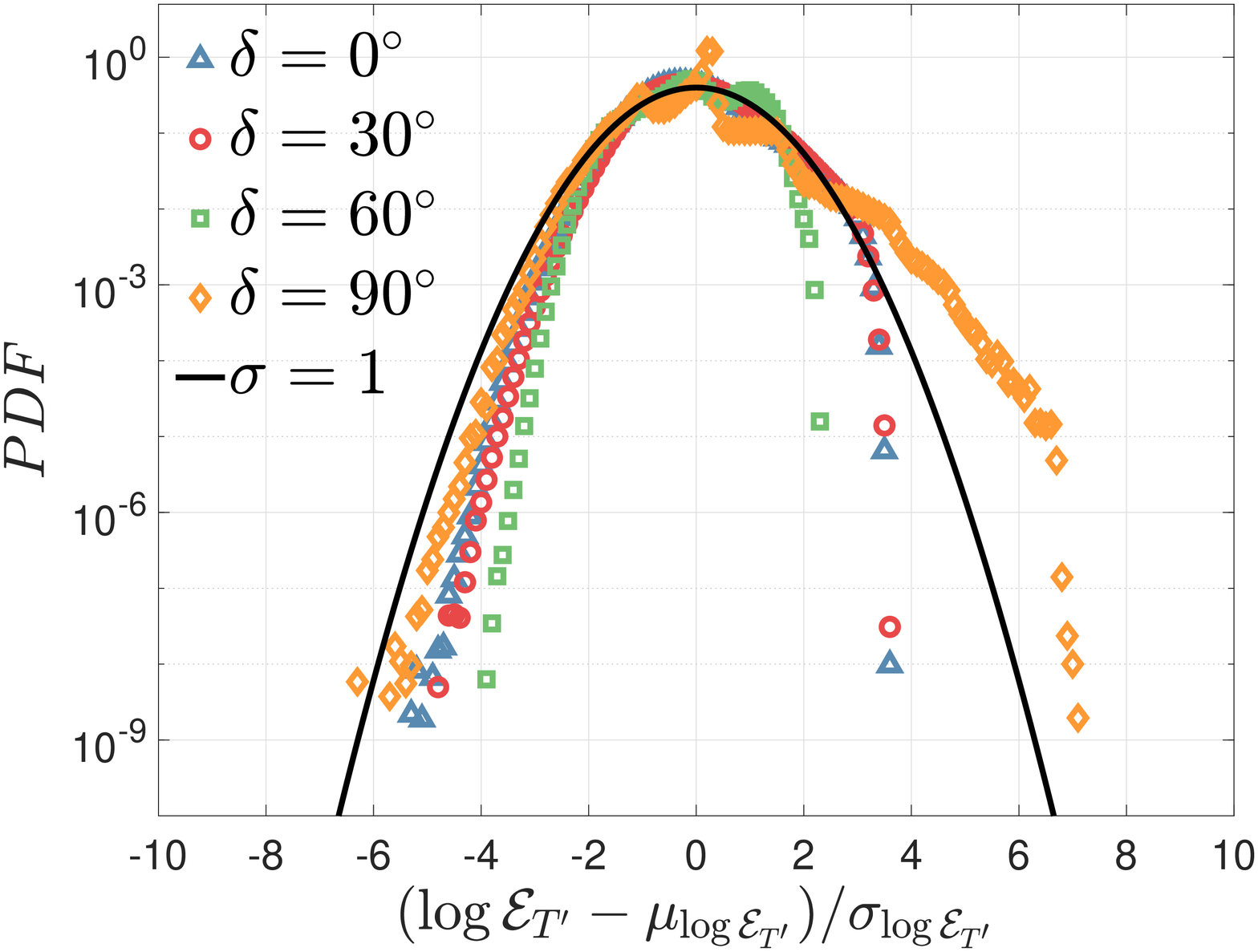}
	\includegraphics[width = 0.45\textwidth]{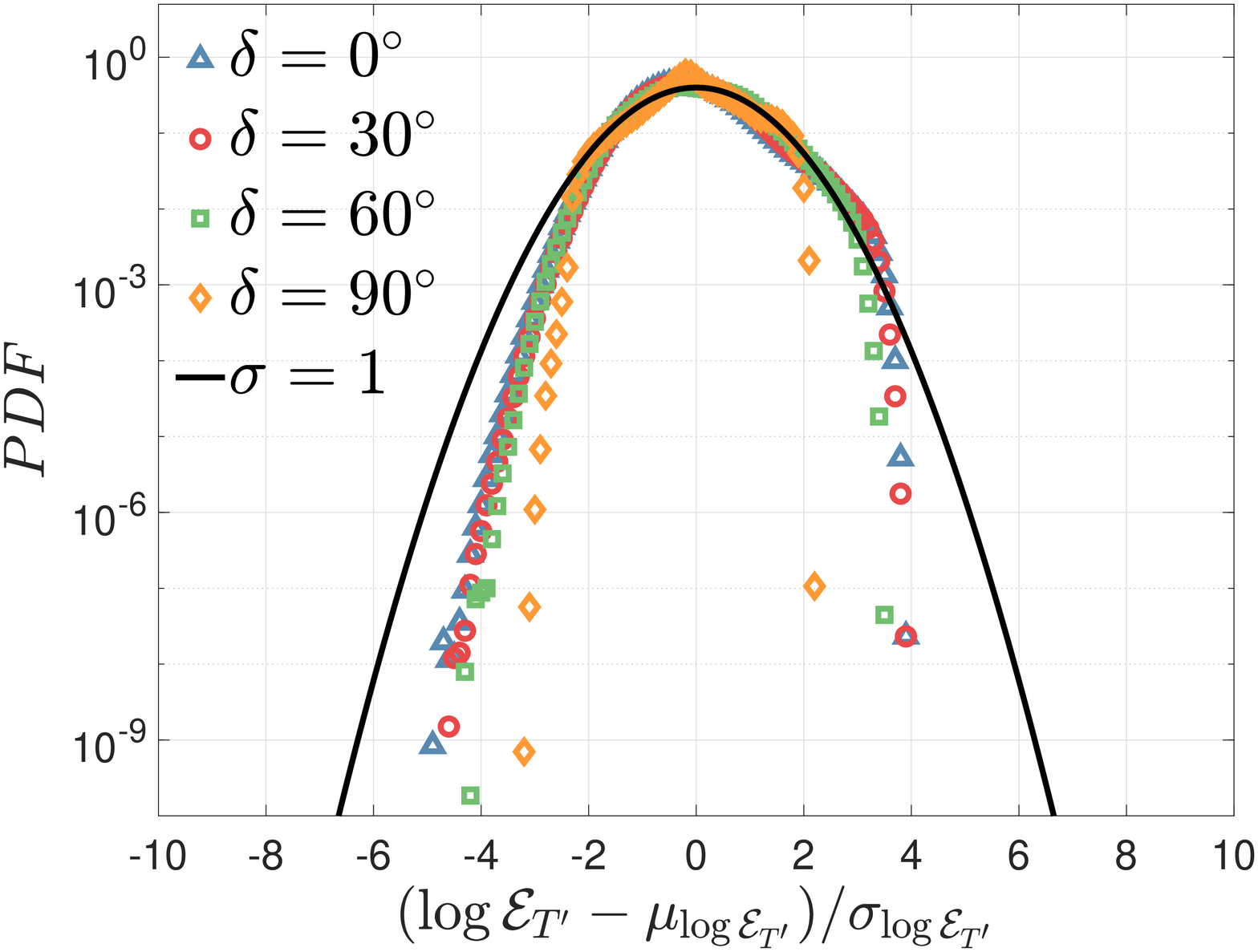}
	\includegraphics[width = 0.45\textwidth]{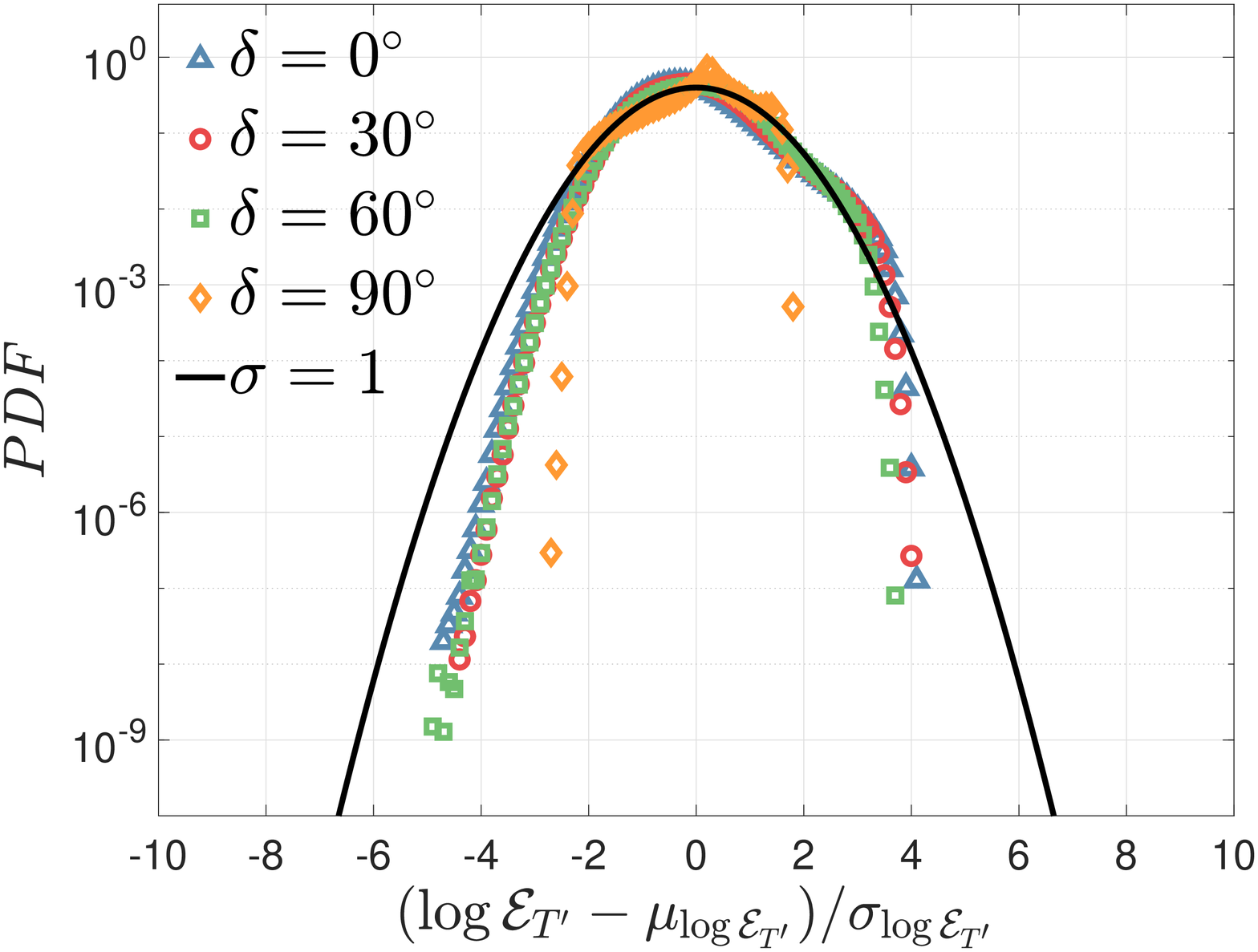}
	\caption{The PDFs of $\log\epsilon_{T^{\prime}}$ with different $\delta$ and $Ra$
    The first line, from left to right: $Ra=3\times10^6$, $Ra=3\times10^7$.
    The second line, from left to right: $Ra=3\times10^8$, $Ra=3\times10^9$.}
	\label{fig:pdfLogEpsilonTVsDelta}
\end{figure*}
%

\begin{figure*}
	\centering
	\includegraphics[width = 0.45\textwidth]{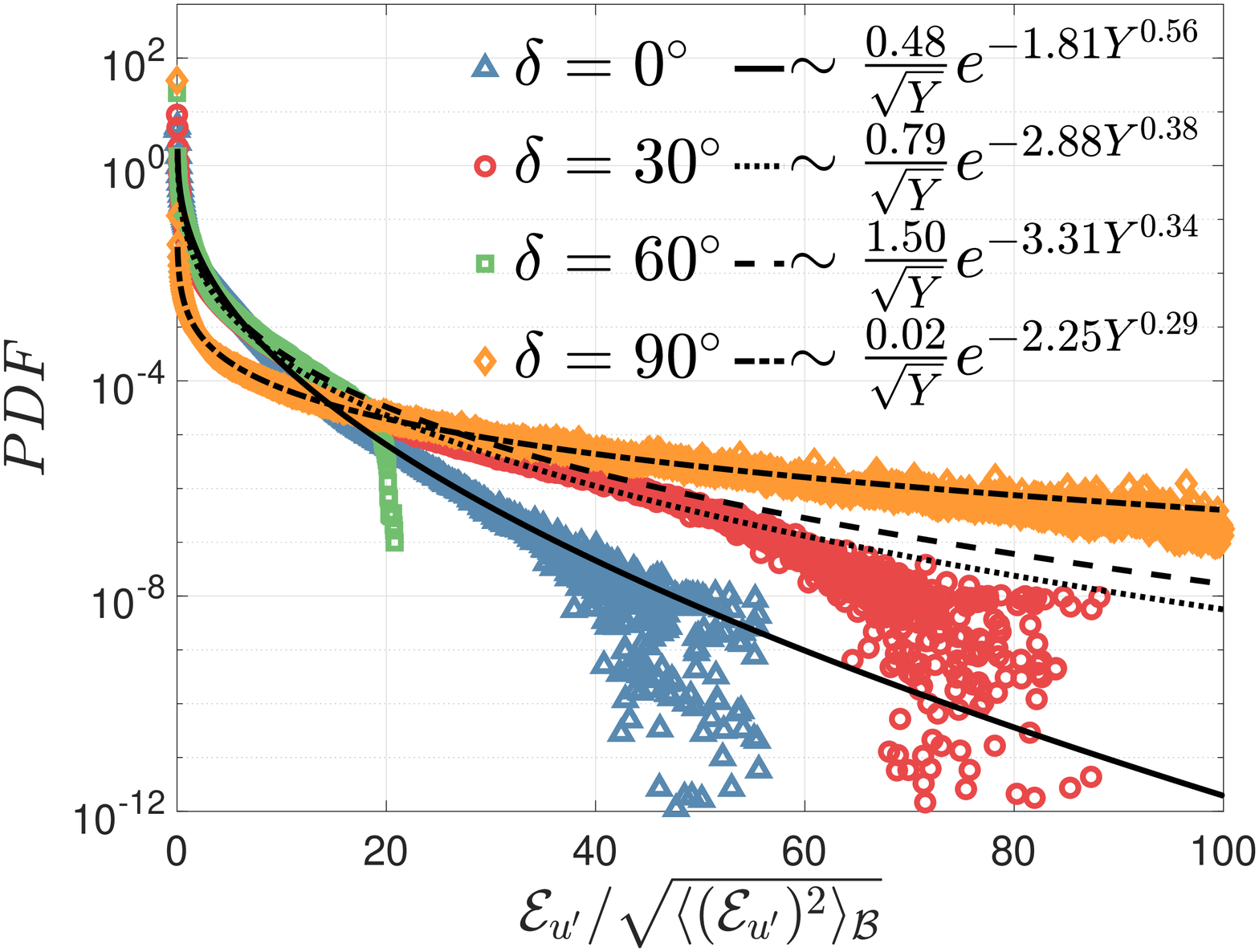}
	\includegraphics[width = 0.45\textwidth]{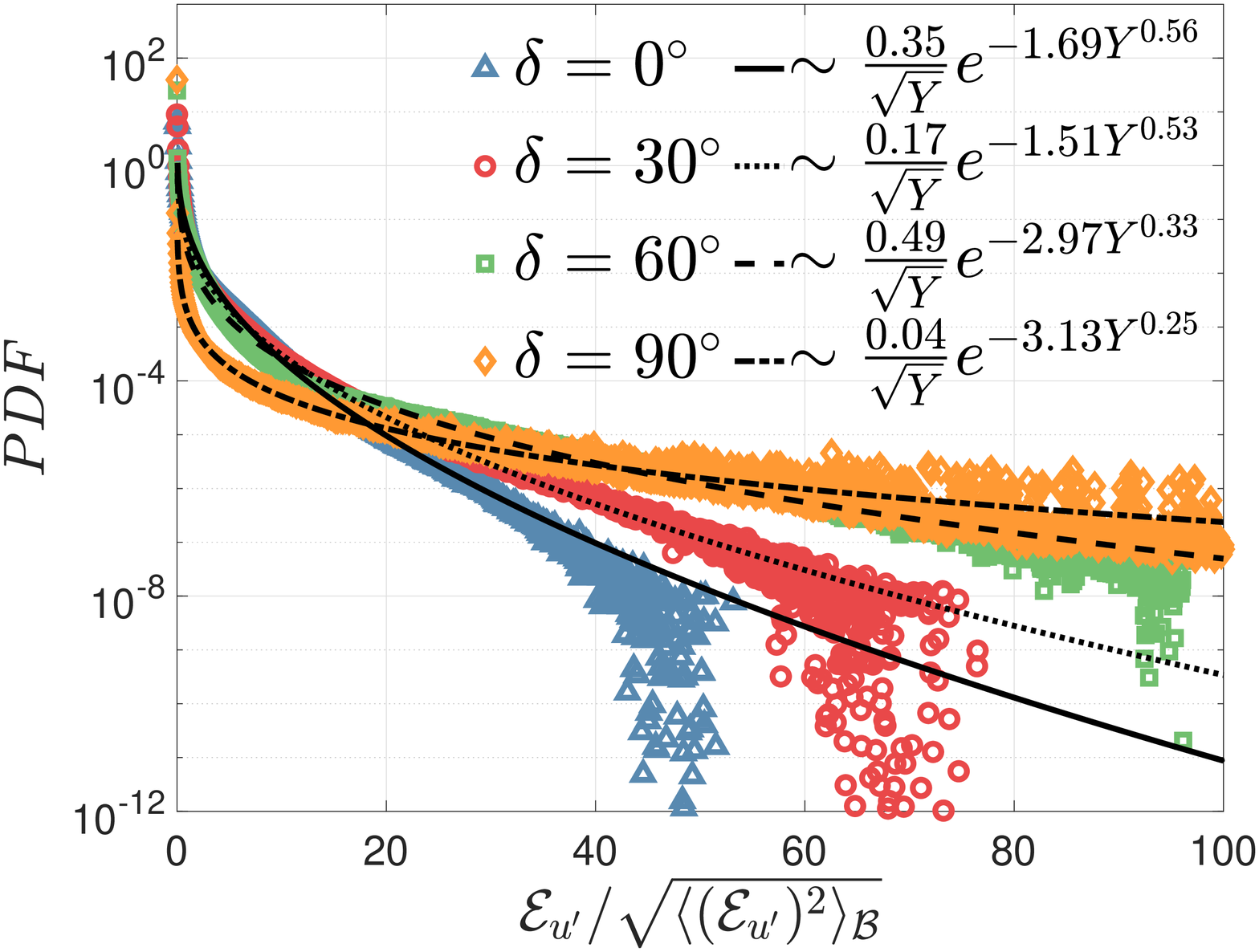}
	\includegraphics[width = 0.45\textwidth]{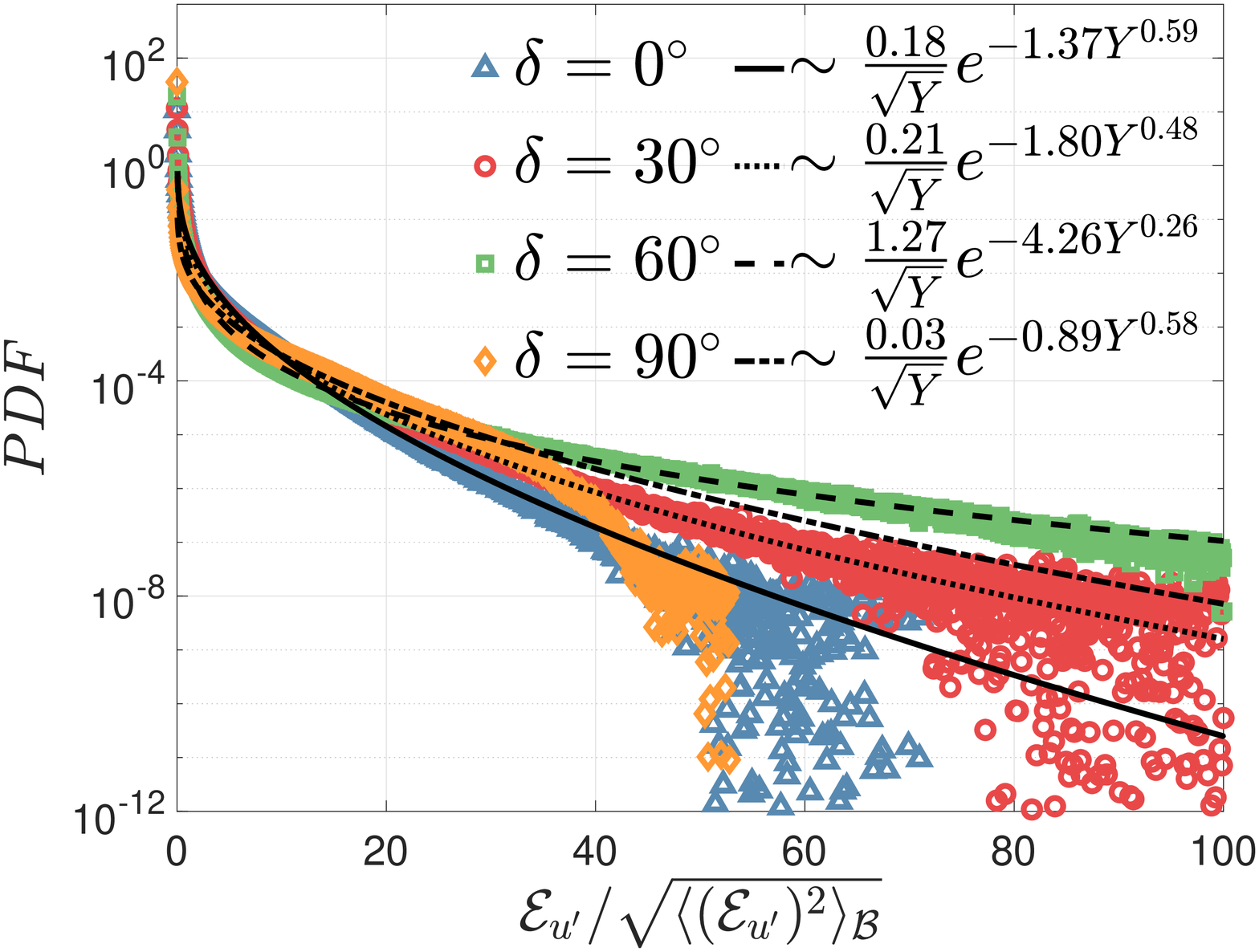}
	\includegraphics[width = 0.45\textwidth]{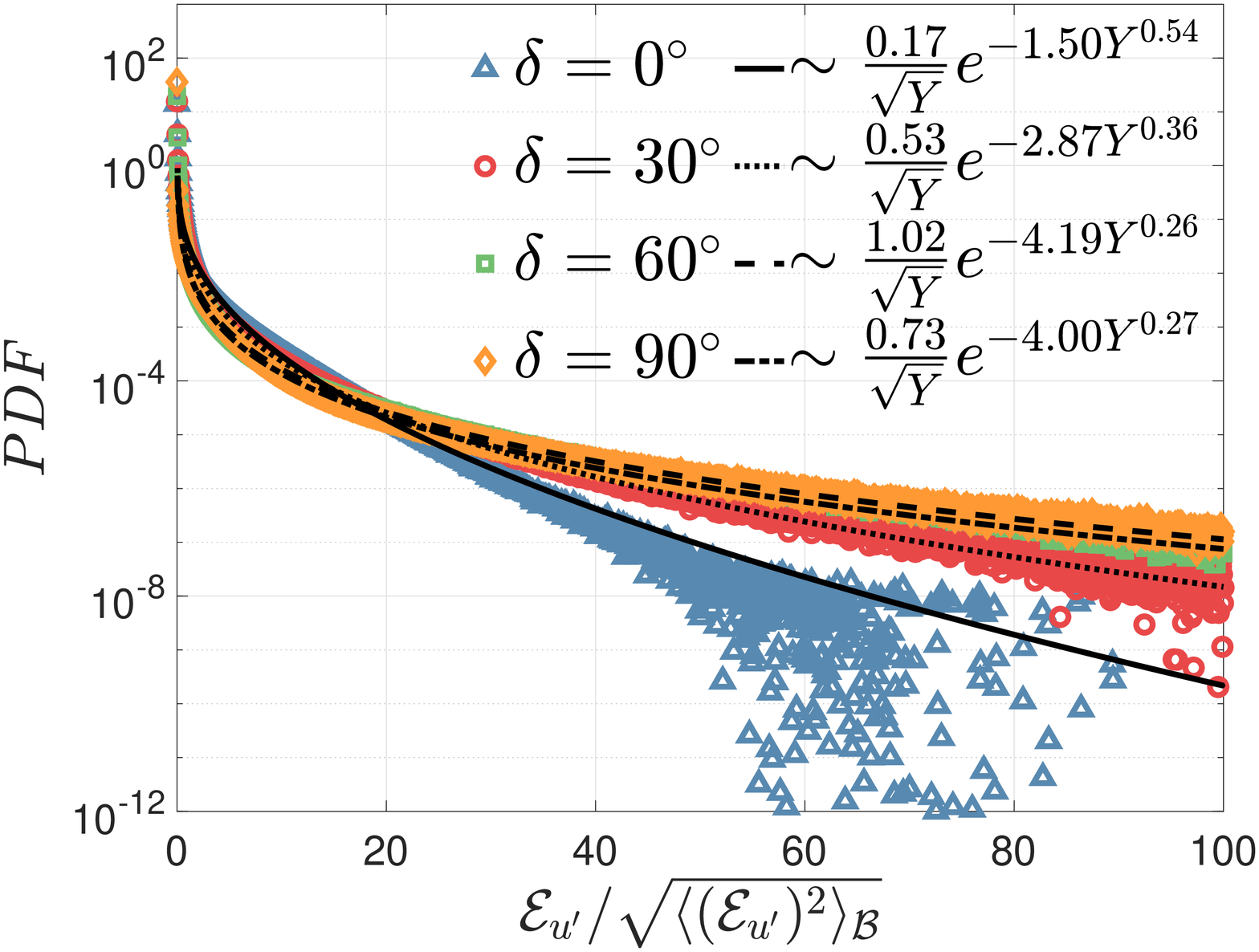}
	\caption{The PDFs of $\epsilon_{u^{\prime}}$ with different $\delta$ and $Ra$.
    The first line, from left to right: $Ra=3\times10^6$, $Ra=3\times10^7$.
    The second line, from left to right: $Ra=3\times10^8$, $Ra=3\times10^9$.}
	\label{fig:pdfEpsilonUVsDelta}
\end{figure*}
%

\begin{figure*}
	\centering
	\includegraphics[width = 0.45\textwidth]{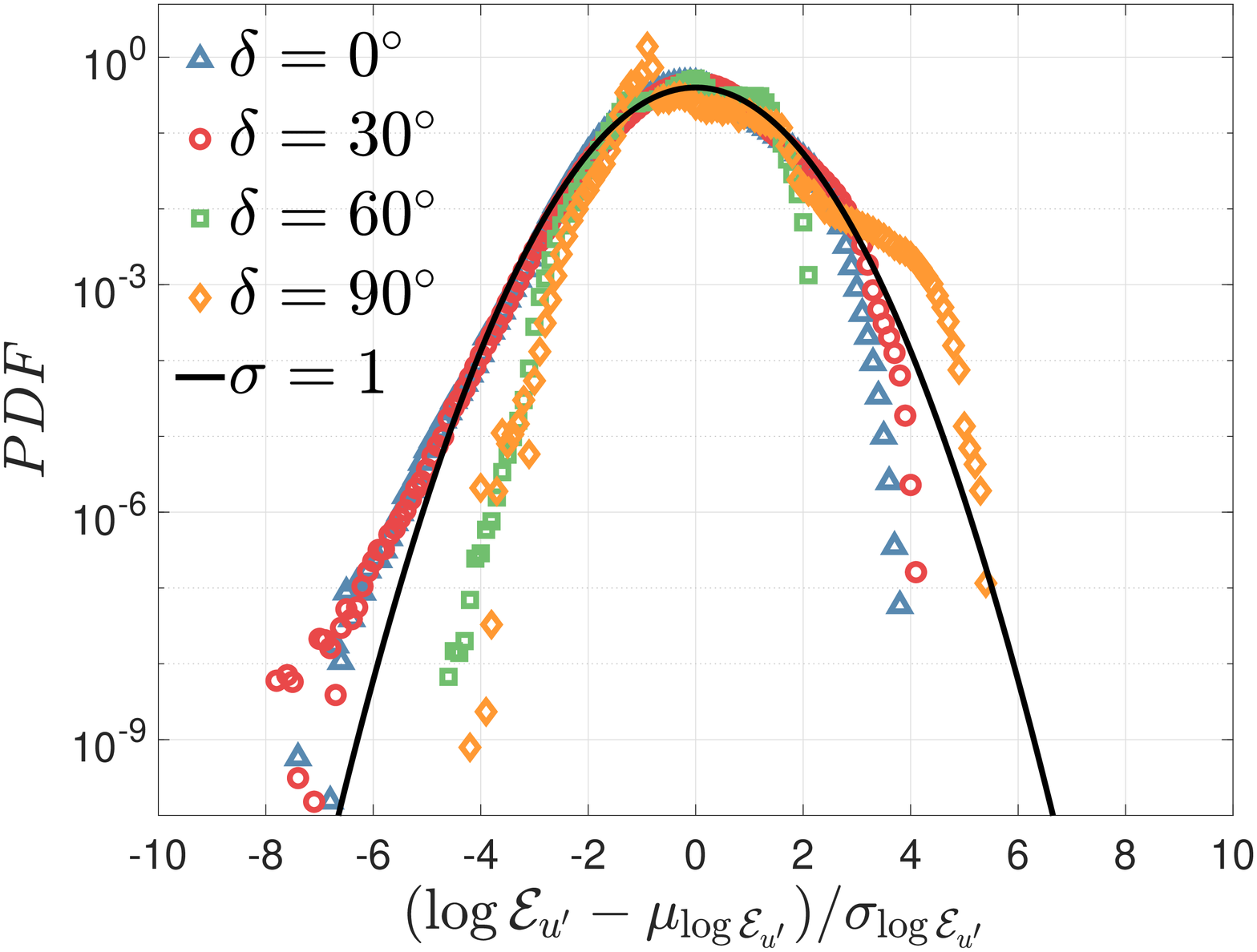}
	\includegraphics[width = 0.45\textwidth]{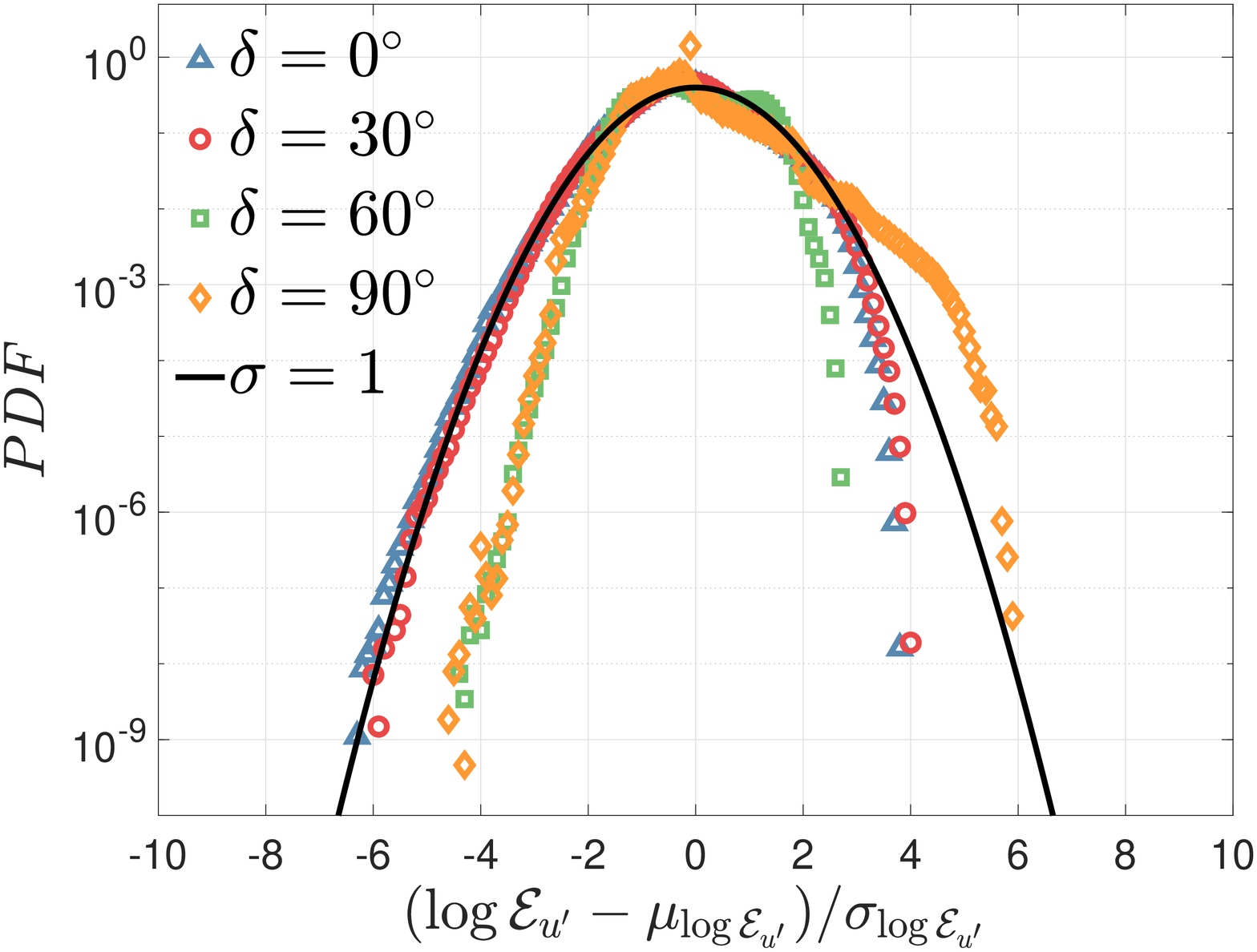}
	\includegraphics[width = 0.45\textwidth]{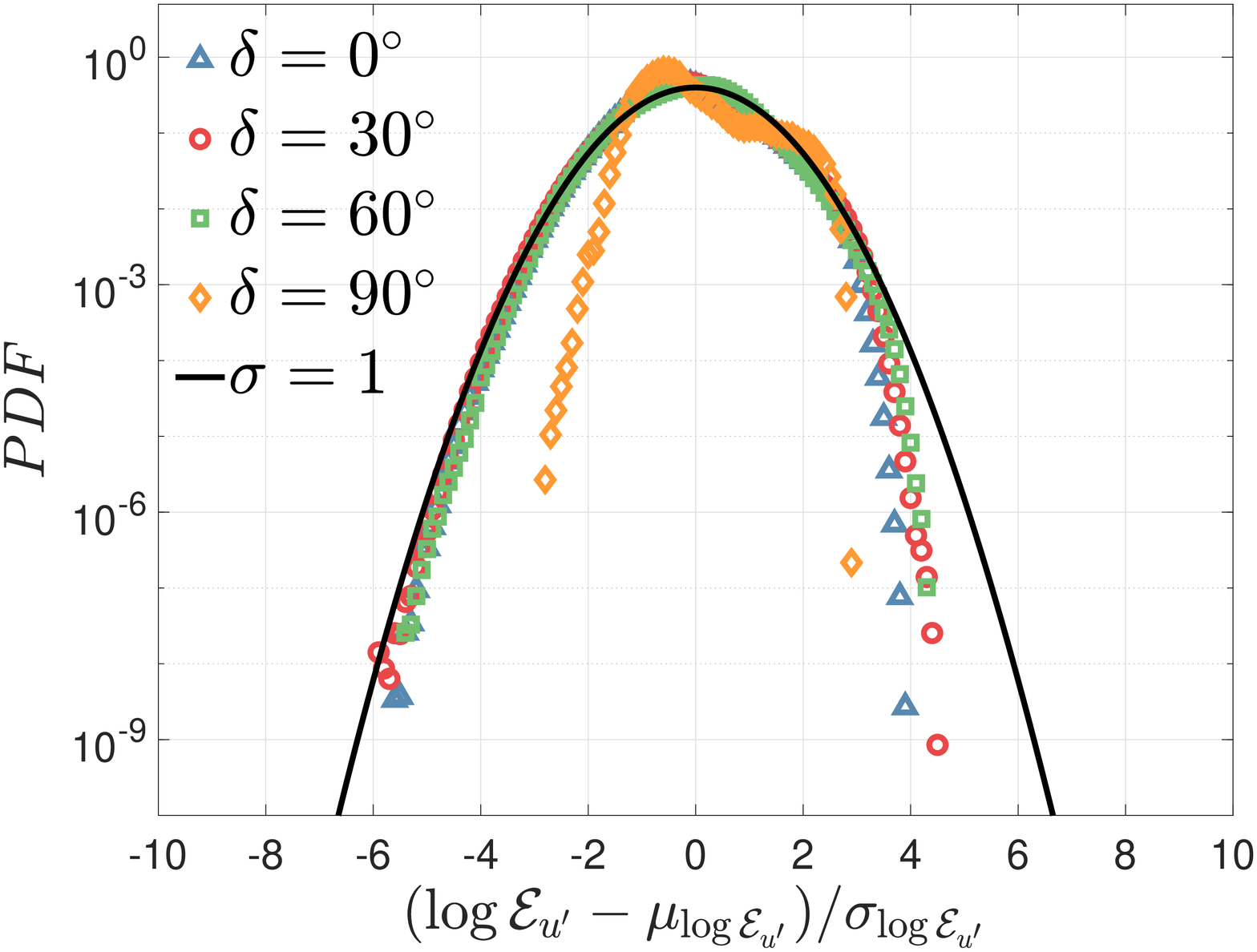}
	\includegraphics[width = 0.45\textwidth]{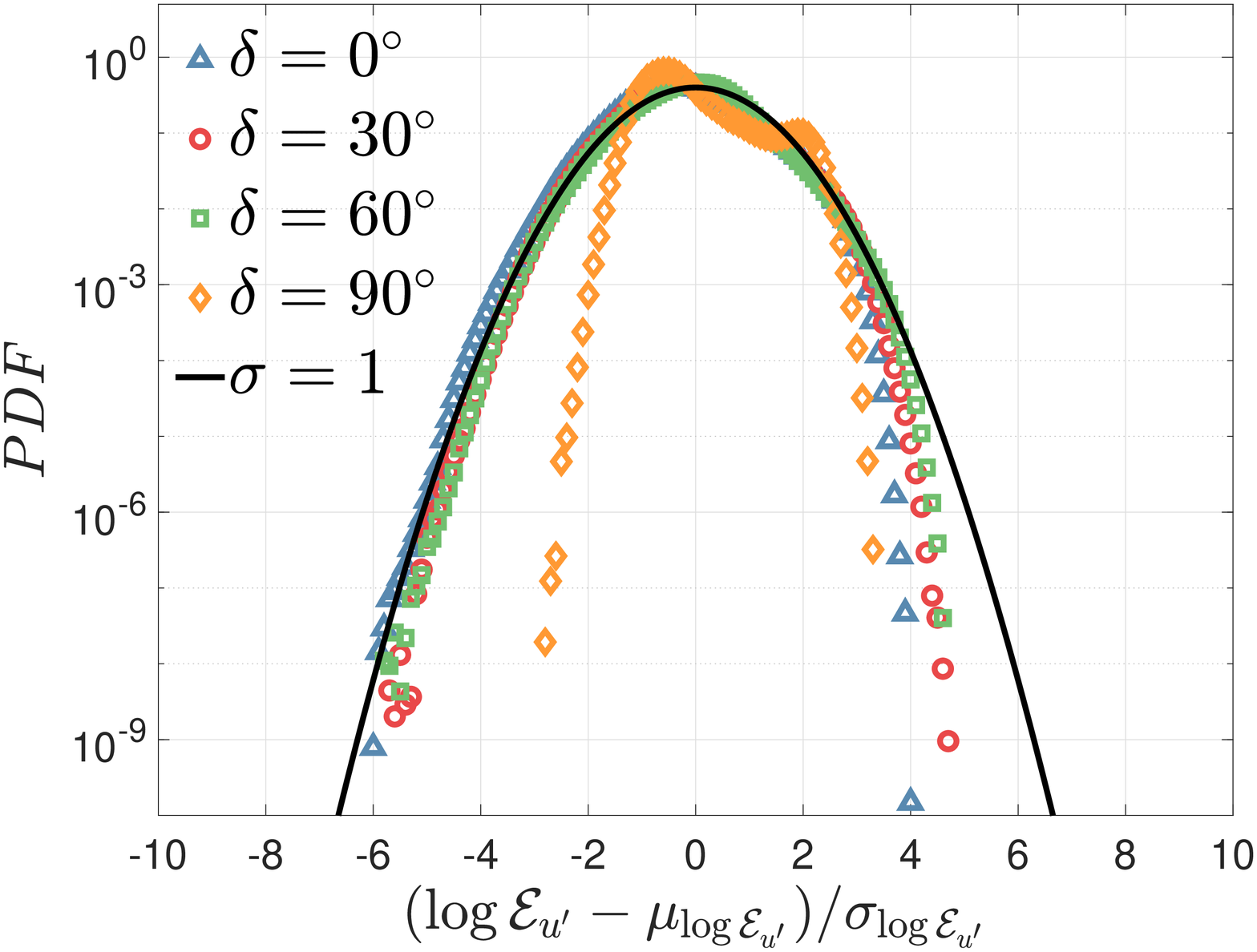}
	\caption{The PDFs of $\log\epsilon_{u^{\prime}}$ with different $\delta$ and $Ra$.
    The first line, from left to right: $Ra=3\times10^6$, $Ra=3\times10^7$.
    The second line, from left to right: $Ra=3\times10^8$, $Ra=3\times10^9$.}
	\label{fig:pdfLogEpsilonUVsDelta}
\end{figure*}
%

Next, we consider the effect of $\delta$ on the PDFs of the dissipation.
The PDFs of $\epsilon_{T^{\prime}}$ for different $\delta$ are shown in
figure \ref{fig:pdfEpsilonTVsDelta} for $Ra=3\times10^6$, $Ra=3\times10^7$, $Ra=3\times10^8$
and $Ra=3\times10^9$. The results show that increasing $\delta$ has a dramatic effect on
the PDFs. For $Ra=3\times10^6$, $\alpha$ decreases monotonically as $\delta$ increases, indicating that the tails of the PDFs
are decaying more slowly. This enhanced intermittency is because as $\delta$ is increased, the turbulence becomes localized
to the lower edge of the bubble, and hence while in this region there is turbulence and dissipation, over vast
portions of the bubble, the flow is almost quiescent.

As $Ra$ is increased, the impact of increasing $\delta$ on the PDFs of $\epsilon_{T^{\prime}}$ becomes much less dramatic, with $\alpha$ still decreasing with increasing $\delta$, but the effect of $\delta$ on $\alpha$ becoming 
much weaker as $Ra$ is increased. Indeed, the effect of tilt on the PDFs is quite weak for $Ra=3\times 10^9$. This reduced effect of $\delta$ as $Ra$ is increased is because with larger $Ra$, the turbulence
produced at the lower edge of the bubble is sill vigorous and dominates the behavior of $\epsilon_{T^{\prime}}$, in contrast to the case of lower $Ra$ where the turbulence at the lower edge is strongly suppressed as $\delta$ is increased.

Figure \ref{fig:pdfLogEpsilonTVsDelta}
shows the PDFs of the logarithm of $\epsilon_{T^{\prime}}$ in normalized form, with a Gaussian distribution
plotted as a solid black line for reference.
Once again, we see that the deviation of the logarithmic PDFs from the Gaussian distribution is enhanced as $\delta$ is increased, but thus enhancement becomes weaker as $Ra$ increases.
Plotting the PDFs in this form also helps reveal a significant difference between the DPR and SPR, namely, that while the logarithmic PDF is approximately Gaussian for small fluctuations of the dissipation in the DPR, it
is far from a Gaussian in the SPR even for small fluctuations. Moreover, the results show that for $\delta=90^\circ$, the logarithmic PDF has a right tail that is heavier than a Guassian for the lower $Ra$ cases, but becomes lighter than a Guassian as $Ra$ increases. 
Moreover, for $Ra=3\times10^8$ and $Ra=3\times10^9$, the PDFs are weakly dependent on $\delta$ for $\delta\leq 60^\circ$.

The PDFs of $\epsilon_{u^{\prime}}$ are plotted in figure \ref{fig:pdfEpsilonUVsDelta}
for different $\delta$ and fixed $Ra$. These PDFs are again well described by stretched exponential functions, as was the case for the PDFs of $\epsilon_{T^{\prime}}$. The most striking difference compared to the PDFs of $\epsilon_{T^{\prime}}$
are that the PDFs of $\epsilon_{u^{\prime}}$ are much more sentitive to $\delta$, and remain sensitive to $\delta$ even for
the largest $Ra$ considered. This difference can also be observed by considering the normalized PDFs of the logarithm of $\log\epsilon_{u^{\prime}}$ which are shown in figure \ref{fig:pdfLogEpsilonUVsDelta}.
As with the normalized PDFs of the logarithm of $\log\epsilon_{T^{\prime}}$, 
those of $\log\epsilon_{u^{\prime}}$ shown in figure \ref{fig:pdfLogEpsilonUVsDelta} show clear differences depending on whether the flow is
in the DPR or the SPR. While the logarithmic PDF is approximately Gaussian for small fluctuations of the dissipation in the DPR, it
is far from a Gaussian in the SPR even for small fluctuations. Indeed, for $Ra=3\times10^8$ and $Ra=3\times10^9$ the PDF for $\delta=90^\circ$ becomes bi-modal.
The results also show that for $\delta=90^\circ$, the logarithmic PDF has a right tail that is heavier than a Guassian for the lower $Ra$ cases, but becomes lighter than a Guassian as $Ra$ increases. 

\subsection{The Global Dissipation Scaling Rules}

Having considered the PDFs of the dissipation rates, which quantify the local fluctuations of the dissipation rates in the flow, we now turn to consider
the globally averaged dissipation rates, both due to the mean-fields and due to the fluctuating fields.

\begin{figure}
	\centering
	\includegraphics[width = 0.6\textwidth]{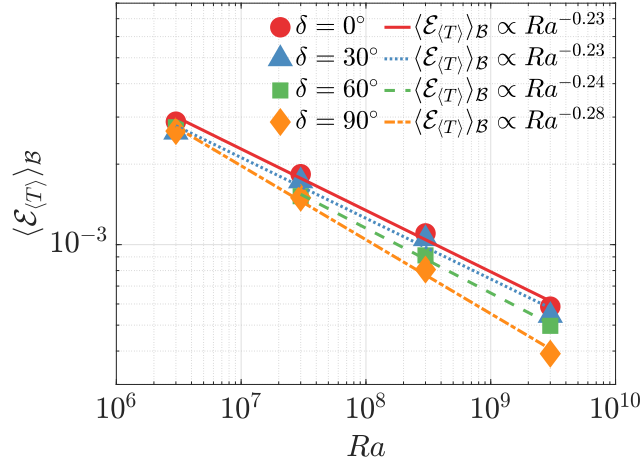}
	\caption{The variation of $\langle\epsilon_{\langle T\rangle}\rangle_{\mathcal{B}}$ with $Ra$ and $\delta$ }
	\label{fig:e_MeanT}
\end{figure}

\begin{figure}
	\centering
	\includegraphics[width = 0.6\textwidth]{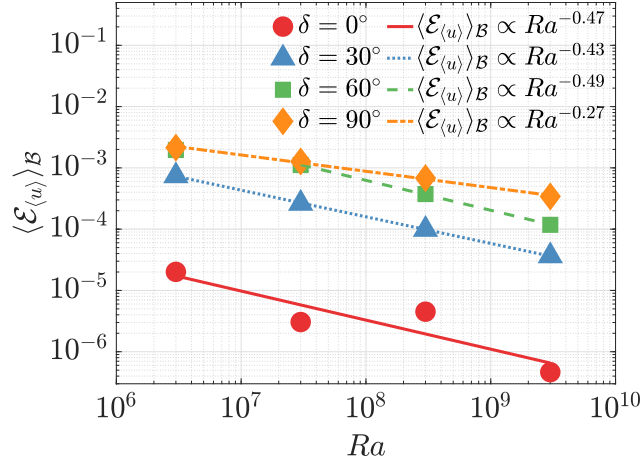}
	\caption{The variation of $\langle\epsilon_{\langle u\rangle}\rangle_{\mathcal{B}}$ with $Ra$ and $\delta$ }
	\label{fig:e_MeanU}
\end{figure}

The globally averaged thermal and kinetic energy dissipation rates due to the mean-fields are denoted by $\langle\epsilon_{\langle T\rangle}\rangle_{\mathcal{B}}$
and $\langle\epsilon_{\langle u\rangle}\rangle_{\mathcal{B}}$. Figures \ref{fig:e_MeanT} and \ref{fig:e_MeanU} show $\langle\epsilon_{\langle T\rangle}\rangle_{\mathcal{B}}$
and $\langle\epsilon_{\langle u\rangle}\rangle_{\mathcal{B}}$ as function of $Ra$ for different $\delta$, with power-law fits illustrated
by solid or dash lines. It is interesting that while the results show that $\langle\epsilon_{\langle T\rangle}\rangle_{\mathcal{B}}$ decreases with increasing $\delta$,
$\langle\epsilon_{\langle u\rangle}\rangle_{\mathcal{B}}$ increases strongly with increasing $\delta$. This is due to the fact that as $\delta$ increases, the stable plume that
dominates in the SPR creates two symmetric vortices, and these enhance the mean-shear in the flow.


Power-law fits to the data yield a scaling law of $\langle\epsilon_{\langle T\rangle}\rangle_{\mathcal{B}}\propto Ra^{-0.23}$ in the DPR 
and $\langle\epsilon_{\langle T\rangle}\rangle_{\mathcal{B}}\propto Ra^{-0.28}$ in the SPR. 
For $\langle\epsilon_{\langle u\rangle}\rangle_{\mathcal{B}}$, when $\delta=0^\circ$ the mean flow field is very weak, 
and $\langle\epsilon_{\langle u\rangle}\rangle_{\mathcal{B}}\propto Ra^{-0.47}$ is found, but the fitting errors are considerable.
When $\delta=30^{\circ}$ and the flow is still in the DPR, the scaling law becomes $\langle\epsilon_{\langle u\rangle}\rangle_{\mathcal{B}}\propto Ra^{-0.43}$ with small negligible fitting error.
The scaling law turns into $\langle\epsilon_{\langle u\rangle}\rangle_{\mathcal{B}}\propto Ra^{-0.49}$ for $\delta=60^{\circ}$ in the DPR.
In the SPR, the behavior becomes $\langle\epsilon_{\langle u\rangle}\rangle_{\mathcal{B}}\propto Ra^{-0.27}$.
The differing scaling behaviour of $\langle\epsilon_{\langle T\rangle}\rangle_{\mathcal{B}}$ and 
$\langle\epsilon_{\langle u\rangle}\rangle_{\mathcal{B}}$ in the DPR and SPR provide further evidence of the quantitative differences in the flow in these two distinct regimes.

Finally, we consider the globally averaged turbulent thermal and kinetic energy dissipation rates which are denoted by
$\langle\epsilon_{T^{\prime}}\rangle_{\mathcal{B}}$ and $\langle\epsilon_{u^{\prime}}\rangle_{\mathcal{B}}$.
\begin{figure}
	\centering
	\includegraphics[width = 0.6\textwidth]{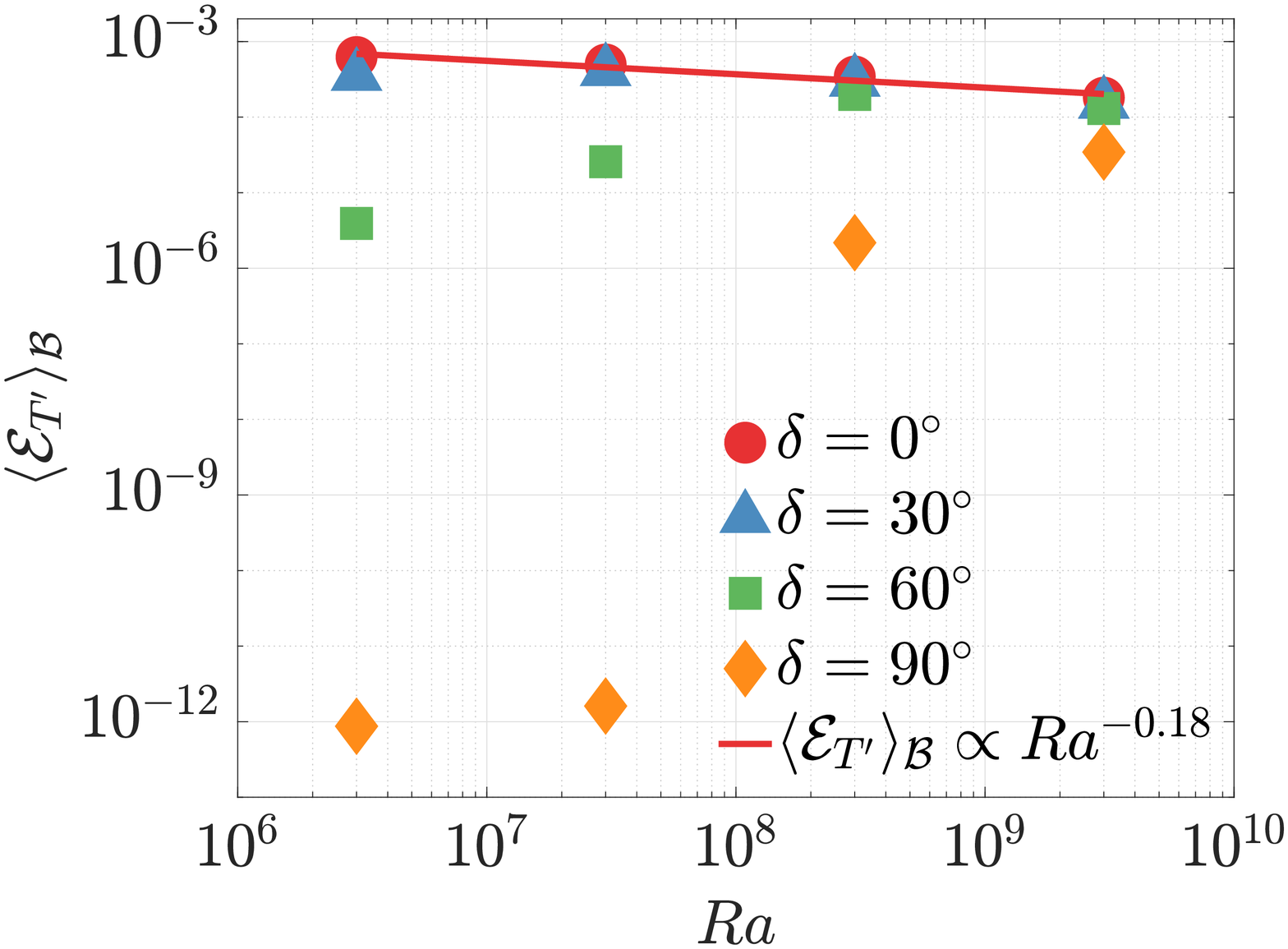}
	\caption{The variation of $\langle\epsilon_{T^{\prime}}\rangle_{\mathcal{B}}$ with $Ra$ and $\delta$}
	\label{fig:e_TurbT}
\end{figure}
\begin{figure}
	\centering
	\includegraphics[width = 0.6\textwidth]{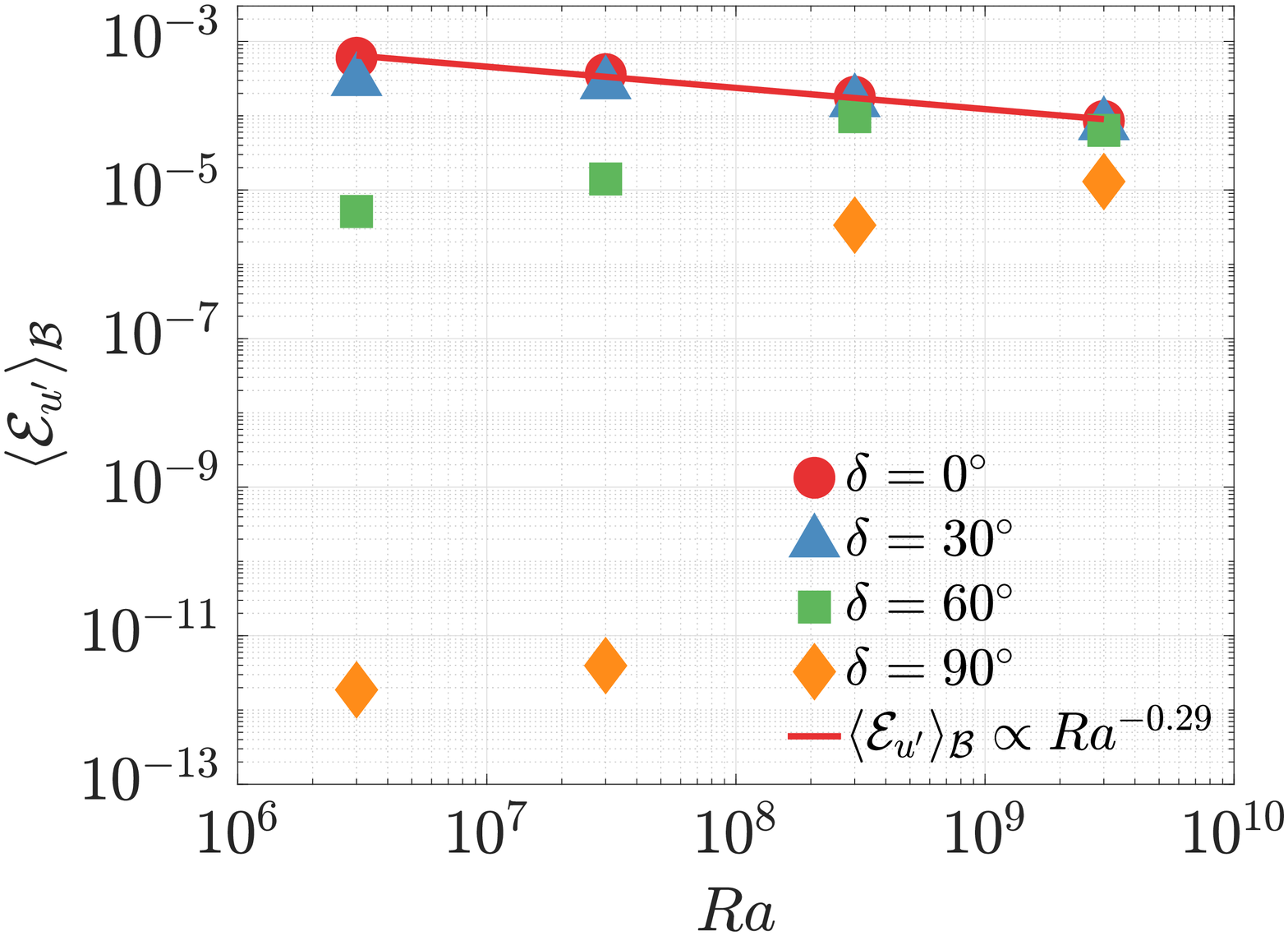}
	\caption{The variation of $\langle\epsilon_{u^{\prime}}\rangle_{\mathcal{B}}$ with $Ra$ and $\delta$}
	\label{fig:e_TurbU}
\end{figure} 

Figures \ref{fig:e_TurbT} and \ref{fig:e_TurbU} show $\langle\epsilon_{T^{\prime}}\rangle_{\mathcal{B}}$ 
and $\langle\epsilon_{u^{\prime}}\rangle_{\mathcal{B}}$ plotted as a function of $Ra$ for different $\delta$.
The data shows an enormous influence of $\delta$ at lower $Ra$, which gradually reduces as $Ra$ is increased. More particularly,
$\langle\epsilon_{T^{\prime}}\rangle_{\mathcal{B}}$ and $\langle\epsilon_{u^{\prime}}\rangle_{\mathcal{B}}$ dramatically reduce as $\delta$ is increased when $Ra$ is relatively small, e.g. $Ra=3\times10^6$ and $Ra=3\times10^7$, due to the strong suppression of convection and turbulence due to the tilting of the bubble.
For relatively large $Ra$, e.g. $Ra=3\times10^8$ and $Ra=3\times10^9$, the reduction of $\langle\epsilon_{T^{\prime}}\rangle_{\mathcal{B}}$ and 
$\langle\epsilon_{u^{\prime}}\rangle_{\mathcal{B}}$ due to increasing $\delta$ are smaller, but still considerable.
When the flow is in DPR, the scaling behaviours of $\langle\epsilon_{T^{\prime}}\rangle_{\mathcal{B}}$ and 
$\langle\epsilon_{u^{\prime}}\rangle_{\mathcal{B}}$ are $\langle\epsilon_{T^{\prime}}\rangle_{\mathcal{B}}\propto Ra^{-0.18}$ and 
$\langle\epsilon_{u^{\prime}}\rangle_{\mathcal{B}}\propto Ra^{-0.29}$.
In contrast, the scaling of $\langle\epsilon_{T^{\prime}}\rangle_{\mathcal{B}}$ and 
$\langle\epsilon_{u^{\prime}}\rangle_{\mathcal{B}}$ in the SPR cannot be described by a power-law function that spans over the range of $Ra$ considered.
This is another quantitative difference between the behavior of the flow in the DPR and SPR.

\section{Conclusion}
In this paper, we have used DNS to explore the effect of tilt on the turbulent thermal convection taking place on a bubble.
Visualizations of the flow reveal that as the tile angle $\delta$ is varied, the flow patterns fall into
one of two regimes. When $\delta$ is relatively small, the flow is dominated by dynamic plumes that 
detach from the boundary layer at random locations on the equator. Turbulent thermal convection 
occurs on the bubble, associated with the continual generation and dissipation of these plumes.
This flow pattern is referred to as the dynamic plume regime (DPR).
On the other hand, when $\delta$ becomes sufficiently large, a single, large stable plume prevails on the bubble, 
emanating from the lower edge of the bubble. The stable large plume arises from an essentially fixed location on the equator and is persistent in time.
This flow pattern is referred to as the stable plume regime (SPR). These qualitatively different flow regimes arise due to the geometric effect that tilting the bubble has on the direction of the local buoyancy force that drives the flow, and
the way in which this depends upon location on the bubble surface.

The first quantitative difference between two regimes explored concerns the scaling behaviours of $Nu$ and $Re$.
Concerning $Nu$, our result shows $Nu\propto Ra^{0.3}$ in the DPR, with a weak dependency of the exponent on $Ra$ and $\delta$.
In the SPR, the scaling changes significantly and becomes $Nu\propto Ra^{0.24}$.
For $Re$, the scaling in the DPR lies between $Re\propto Ra^{0.48}$ and $Re\propto Ra^{0.53}$ depending on $Ra$ and $\delta$,
while in the SPR, the scaling lies between $Re\propto Ra^{0.44}$ and $Re\propto Ra^{0.45}$.

We then explored the behavior of the thermal and kinetic energy dissipation rates in the flow. The standardized PDFs of $\log\epsilon_{T^{\prime}}$ and $\log\epsilon_{u^{\prime}}$ have very different shapes
in the DPR and SPR. For $\log\epsilon_{T^{\prime}}$ and $\log\epsilon_{u^{\prime}}$, the shape of the PDFs are close to a Guassian PDFs for 
small values, but deviates from it for large values in the DPR.
In the SPR, the PDFs of $\log\epsilon_{T^{\prime}}$ and $\log\epsilon_{u^{\prime}}$ depart considerably
from a Gaussian PDF for both small and large values, and the PDF of $\log\epsilon_{u^{\prime}}$ has a bi-modal shape at small values.

The globally averaged thermal energy dissipation rate due to the mean temperature field was
shown to exhibit the scaling $\langle\epsilon_{\langle T\rangle}\rangle_{\mathcal{B}}\propto Ra^{-0.23}$ in the DPR,
and $\langle\epsilon_{\langle T\rangle}\rangle_{\mathcal{B}}\propto Ra^{-0.28}$ in the SPR.
The globally averaged kinetic energy dissipation rate due to the mean velocity field
was shown to exhibit the scaling $\langle\epsilon_{\langle u\rangle}\rangle_{\mathcal{B}}\propto Ra^{-0.47}$ in the DPR (the exponent reduces from $0.47$ to $0.43$ as $\delta$ is increased up to $30^\circ$).
In the SPR, the behavior changes considerably to $\langle\epsilon_{\langle u\rangle}\rangle_{\mathcal{B}}\propto Ra^{-0.27}$.
For the turbulent dissipation rates, the results indicate the scaling $\langle\epsilon_{T^{\prime}}\rangle_{\mathcal{B}}\propto Ra^{-0.18}$ and 
$\langle\epsilon_{u^{\prime}}\rangle_{\mathcal{B}}\propto Ra^{-0.29}$ in the DPR.
However, the dependencies of $\langle\epsilon_{T^{\prime}}\rangle_{\mathcal{B}}$ and $\langle\epsilon_{u^{\prime}}\rangle_{\mathcal{B}}$
on $Ra$ cannot be described by power-laws in the SPR.

Taken together, these results show that the two-dimensional flow on the half-soap bubble undergoes dramatic changes, both qualitative and quantitative, as the bubble is tilted relative to the direction of gravity. This has significant impacts for understanding convective flows in natural and engineered contexts where mean temperature gradients in the flow are often not aligned with gravity, and where the flow may take place on (or in) curved geometries.


%
%

%


\bibliography{refCinq}

\end{document}